%% file: OTOQuantumDissv4.tex
\documentclass[a4paper,11pt]{article}
\usepackage{jheppub}
    \usepackage{etoolbox}
    \makeatletter
    \patchcmd{\maketitle}{\@fpheader}{}{}{}
    \makeatother

\usepackage[normalem]{ulem}
\usepackage{xcolor}
\usepackage{tikz, pgf}
\usetikzlibrary{shapes.misc}
\usetikzlibrary{shapes}
\usetikzlibrary{fit}
\usetikzlibrary{arrows}
\usepackage{array,tikz-cd}

\usepackage{subfigure}
\usepackage{braket}
\usepackage{enumitem}

\include{oto-macros}

\title{Out of Time Ordered  Quantum Dissipation}

\author[a]{Bidisha Chakrabarty,}
\author[a]{Soumyadeep Chaudhuri}
\author[a]{R. Loganayagam}
\affiliation[a]{International Centre for Theoretical Sciences,
Tata Institute of Fundamental Research,
Shivakote, Hesaraghatta,
Bangalore 560089, India}
\emailAdd{bidisha.chakrabarty@icts.res.in}
\emailAdd{soumyadeep.chaudhuri@icts.res.in}
\emailAdd{nayagam@icts.res.in}

\abstract{We consider a quantum Brownian particle interacting with two harmonic baths, which is then perturbed by a 
cubic coupling linking the particle and the baths. This cubic coupling induces non-linear dissipation and noise terms
in the influence functional/master equation of the particle. Its effect on the Out-of-Time-Ordered Correlators (OTOCs) of the particle
cannot be captured by the conventional Feynman-Vernon formalism.We derive the generalised influence functional 
which correctly encodes the physics of OTO fluctuations, response, dissipation and decoherence.   We examine an example 
where Markovian approximation is valid for the OTO dynamics. 

If the original cubic coupling has a definite time-reversal parity,
 the leading order OTO influence functional is completely determined by the couplings in the usual master equation via OTO generalisation of
Onsager-Casimir relations. New OTO fluctuation-dissipation
relations connect the non-Gaussianity of the thermal noise to the thermal jitter in the damping constant of the Brownian particle.}

\newcommand{\cubicfunc}{Q}

\begin{document} 
\maketitle

\section{Motivation}
The dynamics of a Brownian particle interacting with a thermal bath is a topic that has been studied for over a hundred years. A systematic understanding of a 
quantum Brownian particle emerged in the 1960s with the works of Schwinger\cite{Schwinger:1960qe} and Feynman-Vernon\cite{Feynman:1963fq}. In these
 works, an effective theory was derived for a quantum Brownian particle by tracing out the thermal bath's degrees of freedom. 

These analyses were later extended by Caldeira and Leggett \cite{Caldeira:1982iu} to a concrete model of a particle linearly coupled to a harmonic bath. 
The bath degrees of freedom can be exactly integrated out to get a non-local, non-unitary theory describing the evolution of the particle. In this simple 
model, Caldeira and Leggett managed to show the following interesting result:
First, if the distribution of the bath oscillator frequencies is chosen appropriately then the  bath correlators decay with time. Consequently, 
at sufficiently long time-scales, one obtains a local effective theory for the particle which in classical limit reduces to the standard Langevin 
dynamics. This model of Caldeira-Leggett and its generalisations \cite{PhysRevD.45.2843,PhysRevD.47.1576} (see 
\cite{BRE02, weiss2012quantum,schlosshauer2007decoherence} for textbook level discussion) have been crucial in understanding dissipation and decoherence in quantum systems.

In this work, we seek to generalise  Caldeira-Leggett like models. We extend such models in two directions : one extension is to go beyond the
linear Langevin description to non-linear generalisations of Langevin equation. While the non-linear versions have indeed been considered in different contexts before (see references above), the authors of this work
are not aware of a systematic classification of all possible terms at the leading level of non-linearity for a Brownian particle coupled to a generic bath. Such a 
systematic study along with the concomitant generalisation of the famous fluctutation dissipation theorems and Onsager's reciprocal relations is one of our objectives. 

The power of Caldeira-Leggett model lies in its ability  to relate the effective couplings of a dissipative/open quantum system 
description to the underlying microscopic physics. At a superficial level, the idea is the one familiar in usual effective theories : 
one computes correlators in the effective description and in the microscopic description and then matches the two. But, in open quantum systems, 
the integrated out degrees of freedom are not quite the heavy degrees of freedom. Thus, the integrated degrees of freedom can go on-shell
resulting in the non-unitarity of the effective description. For this reason, the matching is not merely one of comparing low frequency behaviour. 

At high temperatures, a whole band of frequency domain contributes right upto the thermal scale. Further, this contribution is 
modulated by how effectively the particle couples to the various states of the bath. We will take these aspects  into
account by writing the effective couplings as frequency integrals over  complex contours with appropriate $i\epsilon$ prescription 
that picks up the correct causal response. The integrands would be the \emph{spectral functions} of the bath which  encode the 
effective number of  bath states accessible to the particle in a given process. From a mathematical point of view, we will write down
expressions that relate the effective couplings of the Langevin description to generalised discontinuities of the appropriate (in general, out of time ordered)
bath correlators.

A second related objective is the generalisation of Caldeira-Leggett like models to take into account  out of time order
correlations(OTOCs) and how they get transmitted from the bath to the Brownian particle. Two of us\cite{Chaudhuri:2018ihk}
recently considered the general formalism to tackle the question of  how the OTOCs of a probe records information about
 the OTOCs of a system. This work is motivated by trying to find a concrete realisation of the ideas considered in \cite{Chaudhuri:2018ihk}. 

The perspective provided by extending the ideas about open systems to include  out of time order correlations is crucial to our work.
As we shall show explicitly, the fluctuation-dissipation theorems and Onsager's reciprocal relations relate parameters governing 
time-ordered correlations to those controlling  out of time ordered correlations. The straightforward way to understand the relations between
time-ordered couplings is via the relations between out of time-ordered correlations and time-ordered correlations. This is the macroscopic counterpart to the 
observation that including out of time ordered correlations simplify the structure of thermal correlators in a general quantum field theory\cite{2017arXiv170102820H,Haehl:2017eob,Chaudhuri:2018ymp}. 

We illustrate our results about non-linear corrections to Langevin theory by considering a simple  extension of Caldeira-Leggett model. Our model consists of a Brownian
particle coupled to two sets of oscillators at the same temperature. Apart from the usual linear coupling, we will also turn on  weak three-body interactions involving 
the particle and two other bath oscillators, one drawn from each set. We show that under appropriate distribution of couplings, the bath continues to be Markovian. We derive the non-linear corrections to Langevin equation that result from such an interaction in the Markovian limit and check that these couplings indeed satisfy the correct generalisations of fluctuation-dissipation/Onsager reciprocal relations.

In the following sections, we will elaborate on these ideas and summarise our results. We will begin by reviewing the model by Caldeira-Leggett in section~\S\ref{sec:CaldLeg}. This is followed by a detailed description of a non-linear generalisation of Langevin theory in section~\S\ref{sec:nonLinearLangevin}, where we also summarise our main results on such non-linear corrections for a general bath. This is followed by a description of the particular model we work with. A complete specification of the qXY model is given in section~\S\ref{sec:qXY}. The derivation of the non-linear  Langevin theory from qXY model is the subject of section~\S\ref{sec:qXYtoLangevin}. We give a general analysis of both the generalised Onsager and  fluctuation-dissipation relations in the  non-linear  Langevin theory in the  section~\S\ref{sec:Relations}, before our concluding remarks in  the  section~\S\ref{sec:discussion}. 

Appendix~\ref{app:DimensionalAnalysis} enumerates the dimension of various physical quantities that appear in this work.  In Appendix~\ref{a1}, we give a brief  review of the rules 
that constrain the form of effective theories that can arise by integrating out the bath. This is followed by  Appendix~\ref{app:poles}, where we have summarised various contour integrals that are useful in computation of  the effective Langevin couplings from our microscopic model.

\section{Review of Caldeira-Leggett model}\label{sec:CaldLeg}
We will now begin with a brief review of the salient features of the quantum Brownian particle and later, how these get generalised in the context of our work. We will sketch the 
basic structural features postponing detailed derivations for later. The reader familiar with this material is encouraged to quickly skim over this description, making note of our notation.

\subsection{Review of Langevin theory and Fluctuation-Dissipation theorem }
Consider, for definiteness, a Brownian particle  whose evolution is described by Langevin equation :
\begin{equation}
\begin{split} 
\frac{d^2q}{dt^2} +\gamma \frac{dq}{dt} =\langle f^2\rangle\ \mathcal{N}(t)\ . \end{split}
\end{equation}
Here $\gamma$ is the damping constant of the particle. The term $\langle f^2\rangle \mathcal{N}(t)$ is the fluctuating force (`the noise term') which is  commonly approximated to be Gaussian and delta-correlated, viz., its only 
non-zero cumulant is  taken to be of the form
\be\begin{split} \langle \mathcal{N}(t) \mathcal{N}(t')\rangle_{noise} \equiv \frac{1}{\langle f^2\rangle}\ \delta(t-t') . \end{split}\ee
Here $\mathcal{N}(t)$ is normalised to have the inverse dimension of velocity for later convenience, so that $\langle f^2\rangle$ measures the statistical variance of the fluctuating force per unit mass (and has the dimension of acceleration$^2\times$time). The cumulant statement above can also be recast into a statement about the probability 
distribution governing the noise  ensemble :
\be
\begin{split} 
P[\mathcal{N}] \propto \exp\left\{-\frac{\langle f^2\rangle }{2 }\int dt\  \mathcal{N}^2\right\}\ . 
  \end{split}
\ee
The significance of Caldeira-Leggett model is that it shows how such a system of equations can arise from an underlying quantum mechanical model, under appropriate approximations,  via integrating out the effects of a harmonic bath.

Another triumph of Caldeira-Leggett model is that it can reproduce the well-known fluctuation-dissipation relation\cite{PhysRev.32.97,PhysRev.32.110,PhysRev.83.34,BERNARD:1959zz}  between 
the parameter $ \langle f^2\rangle$ characterising thermal fluctuations and the parameter $\gamma$ characterising dissipation. Let us remind the 
reader the classical argument why such a relation should be expected : the  thermal fluctuations in the bath which source the noise also cause the dissipative 
effects of the bath. On one hand, the rate of damping $\gamma$ is roughly determined by the effective number of degrees of freedom of the bath with which it interacts. 
On the other hand, the thermal noise is produced by the energy  fluctuations in the bath which is roughly $k_B T$ times the number of degrees of freedom (here, $T$ 
denotes the temperature of the bath and $k_B$ is the Boltzmann constant). This suggests a relationship of the form $ \langle f^2\rangle \sim  \gamma \frac{k_B T}{m_p} $ between the fluctuation and the dissipation parameters. Here, we have introduced the mass of the particle $m_p$ to match the dimensions.

A more precise argument is as follows : at the level of classical stochastic dynamics, one can integrate the Langevin equation to obtain 
\be\begin{split} \frac{dq}{dt} = e^{-\gamma(t-t_0)} \left(\frac{dq}{dt}\right)_{t_0} + \langle f^2\rangle \int_{t_0}^t dt'  e^{-\gamma(t-t')}\mathcal{N}(t')\ .  \end{split}\ee
We square this equation and  average over the noise under the assumption that the initial velocity distribution and the fluctuating force are 
uncorrelated. At large times, this yields for the long-time average kinetic energy for a particle of mass $m_p$ :
\begin{equation}\begin{split}
\lim_{t\to \infty} \Big\langle\frac{1}{2}m_p\left(\frac{dq}{dt}\right)^2\Big\rangle_{noise} 
&= \frac{m_p\langle f^2\rangle }{4\gamma}\ . 
\end{split}\end{equation}
Demanding that this average kinetic energy approach the thermal equipartition value $\frac{1}{2} k_B T$, we obtain the fluctuation-dissipation relation
\begin{equation}\begin{split}\label{eq:FDTIntro}
 \langle f^2\rangle = 2\gamma\frac{k_B T}{m_p}\equiv 2\gamma v_{th}^2 \ ,
 \end{split}\end{equation}
 where we have introduced the rms thermal velocity $v_{th}$ of the Brownian particle defined via
 \begin{equation}\begin{split}
v_{th}^2 \equiv \frac{k_B T}{m_p}\ .
 \end{split}\end{equation}
The physics behind such a  classical stochastic argument is clear : the kicks of the fluctuating force should, on average, replenish the  energy lost due to dissipation 
so that the eventual balance is achieved at the Maxwell-Boltzmann value for the average kinetic energy.

As pointed out by  Kubo, Martin and Schwinger \cite{Kubo:1957mj,Martin:1959jp}, 
 the origins of this relation can be traced back, in the underlying quantum description, to the mathematical structure of thermal correlators.
In the linear response theory, both the fluctuating force and damping rate felt by the Brownian particle can be computed from a characteristic \emph{spectral function}
of the bath denoted by $\rho(\omega)$, related to the Fourier domain commutator of bath operators that couple to the particle (see below).
We then have the frequency  integrals (often termed  the \emph{sum rules}):
\begin{equation}\label{eq:2ptSumIntro}
\begin{split}
m_p^2 \langle f^2\rangle &=\int_{-\infty-i\epsilon}^{\infty-i\epsilon}\frac{d\omega}{2\pi i}\frac{\rho(\omega)}{\omega^2}\  \left(1+2 \bose_\omega\right)\hbar \omega\ \ , \\
m_p\gamma&=\int_{-\infty-i\epsilon}^{\infty-i\epsilon}\frac{d\omega}{2\pi i}\frac{\rho(\omega)}{\omega^2}\  \ ,
\end{split}
\end{equation}
where $\beta\equiv \frac{\hbar}{k_B T}$ is the  periodicity of the thermal circle and $\bose_\omega\equiv \frac{1}
{e^{\beta\omega}-1}$ is the Bose-Einstein distribution. These relations make precise the afore-mentioned intuition that the same effective degrees of freedom (encoded in the single function $\rho(\omega)$ of the bath)
 determine both $ \langle f^2\rangle$ and $\gamma$.

Note the factor $\left(1+2\bose\right)$ whose three terms describe spontaneous emission, stimulated emission and absorption of the fluctuation of frequency $\omega$, all of which add incoherently into the variance of noise. At high temperature limit (i.e., small $\beta$) one then recovers Eq.\eqref{eq:FDTIntro}. The $i\epsilon$ prescription  for the frequency integral is the frequency domain analogue of step functions that appear in time domain retarded Green functions. For example, the formula above for $\gamma$ can also be written in the form
\begin{equation}
\begin{split}
m_p\gamma=\int d\tau\ (\tau-t)\Theta(t-\tau)\int\frac{d\omega}{2\pi i}\rho(\omega) e^{-i\omega(t-\tau)}\  \ ,
\end{split}
\end{equation}
which can be thought of as coming in turn from the approximation of a non-local retarded Green function :
\begin{equation}
\begin{split}
m_p\gamma\dot{q}(t)&\in \int d\tau\ \left\{\Theta(t-\tau)\int\frac{d\omega}{2\pi i}\rho(\omega) e^{-i\omega(t-\tau)}\right\}\ q(\tau)\  \ .
\end{split}
\end{equation}
While such time domain expressions clearly exhibit the causality properties, we will find it convenient to work in frequency domain where the structure of thermal
correlators can be examined clearly. All our frequency domain contour integrals can, if needed, be readily converted into time domain integrals with appropriate 
step functions. 

Let us examine in  some more detail how this works : consider a probe particle coupled to a bath in a general time-independent state and assume that at long enough time scales, a local autonomous description is still
possible for the particle (i.e., Markovian approximation can be justified and any effect of the memory of the bath can be ignored). Under these assumptions, the effect of the bath can be integrated out following the method of Schwinger\cite{Schwinger:1960qe} and Feynman-Vernon\cite{Feynman:1963fq} to get a local Schwinger-Keldysh effective action (or
equivalently a local Feynman-Vernon influence functional). We can match this local effective action  against the generating functional for the Langevin correlators following the method of
Martin-Siggia-Rose\cite{1973PhRvA...8..423M}-De Dominicis-Peliti\cite{1978PhRvB..18..353D}-Janssen\cite{1976ZPhyB..23..377J} (See \cite{2017JPhA...50c3001H} for a recent review). This results in a general  expression for the Langevin couplings :
\begin{equation}\label{eq:2ptSumIGeneral}
\begin{split}
 m_p^2\langle f^2\rangle &=\int_{-\infty-i\epsilon}^{\infty-i\epsilon}\frac{d\omega_1}{2\pi i}\int_{-\infty+i\epsilon}^{\infty+i\epsilon}\frac{d\omega_2}{2\pi}\ \frac{\rho[12_+]}{\omega_1}\ \ , \\
m_p\gamma&=\int_{-\infty-i\epsilon}^{\infty-i\epsilon}\frac{d\omega_1}{2\pi i}\int_{-\infty+i\epsilon}^{\infty+i\epsilon}\frac{d\omega_2}{2\pi}\ \frac{\rho[12]}{\omega_1^2}\  \ .
\end{split}
\end{equation}
Here the integrands are the Fourier transformed expectation values of the commutator/anti-commutator 
\begin{equation}\begin{split}
\rho[12] &\equiv \frac{1}{\hbar}\int dt_1 \int dt_2\  e^{i(\omega_1t_1+\omega_2t_2)}\ \langle [\mathcal{O}(t_{1}),\mathcal{O}(t_{2})] \rangle_B ,\\
\rho[12_+] &\equiv\int dt_1 \int dt_2\  e^{i(\omega_1t_1+\omega_2t_2)} \langle \{\mathcal{O}(t_{1}),\mathcal{O}(t_{2})\} \rangle_B\ .
\end{split}\end{equation}
In the above, $\mathcal{O}$  is the  bath operator that couples linearly to the 
Brownian particle position $q$ and the expectation values are evaluated in an appropriate time-independent state of the bath.\footnote{In the following, we 
stick to the  convention that every commutator is divided by $\hbar$ in the definition of spectral functions.  This has the advantage that the classical limit of the spectral function
$\rho[12]$ is just the Fourier transform of the Poisson bracket. For example, Eq.\eqref{eq:2ptSumIGeneral} is also valid in classical statistical mechanics, provided $\rho[12]$  is 
computed via Poisson brackets.} 

The reader should also note the specific energy-conserving $i\epsilon$ prescription needed to write the sum-rules above :
\begin{equation}
\begin{split}
 \int_{\mathcal{C}_2} \equiv \int_{-\infty-i\epsilon}^{\infty-i\epsilon}\frac{d\omega_1}{2\pi}\int_{-\infty+i\epsilon}^{\infty+i\epsilon}\frac{d\omega_2}{2\pi}\    \ .
\end{split}
\end{equation}
The above expressions relating the microscopic dynamics of the bath degrees of freedom to the effective couplings can be obtained by standard procedures of 
effective theory : for instance  by matching the correlators predicted by  the effective theory against the microscopic computations. The $i\epsilon$ prescription
then naturally appears when comparing appropriately retarded correlators. 

Another important property of the complex contour $\mathcal{C}_2$ is the following :
it remains invariant under a simultaneous complex conjugation and the reversal of frequencies. It follows then that, if the integrands are similarly invariant (as the above integrands indeed are), the resultant answers are real.

The expressions above imply that, from the point of view of thermal correlators, whereas the damping constant  $\gamma$ is related to the \emph{commutator} of bath operators, the fluctuation $ \langle f^2\rangle$
is related to the \emph{anti-commutator}. If the bath state is thermal, the Kubo-Martin-Schwinger(KMS) conditions relate these frequency domain functions as
\begin{equation}\begin{split}\label{eq:KMSIntro}
\rho[12_+] = \hbar(1+2\bose_1)\rho[12]=-\hbar(1+2\bose_2)\rho[12] \ .
\end{split}
\end{equation}
Using time translation invariance to write $\rho[12]= \rho(\omega_1)\ 2\pi\delta(\omega_1+\omega_2)$, we  then recover Eq.\eqref{eq:2ptSumIntro}.
The Eq.\eqref{eq:KMSIntro} can be motivated by the following relation  between thermal averages in a harmonic oscillator of frequency $\omega$ :
\begin{equation}\begin{split}
 \langle \{a,a^\dag\}\rangle_\beta = \hbar \left(1+2 \bose_\omega\right) \frac{1}{\hbar}\langle [a,a^\dag] \rangle_\beta \ . 
 \end{split}
\end{equation}
KMS showed  that such a relation continues to hold true for a general quantum system, thus giving rise to Eqs. \eqref{eq:2ptSumIntro} and \eqref{eq:FDTIntro} under very general assumptions.   

To summarise, the microscopic justification of fluctuation-dissipation theorem lies in the following steps :
\begin{itemize}
\item In the first step, one identifies the relevant effective couplings of the system (here $ \langle f^2\rangle$ and $\gamma$ of the Brownian particle)
and connects it with the appropriate bath correlators (via sum rules like the ones in Eq.\eqref{eq:2ptSumIGeneral}). This step already assumes the emergence of 
a Markovian description which can be checked explicitly in a simple model like that of Caldeira-Leggett. 
\item In the next step, one uses thermality to derive KMS relations akin to Eq.\eqref{eq:KMSIntro}. Note that this step, by itself, does not 
immediately result in a simple relation between the effective couplings of the local Markovian description.

KMS condition merely says that two couplings are related to two \emph{different moments} of the same spectral function (see Eq.\eqref{eq:2ptSumIntro}). These two moments are a priori two independent numbers provided by  the theory behind the bath. Thus, the KMS condition alone is of limited utility to an experimentalist probing the local dynamics of the Brownian particle.\footnote{One could however imagine a fine-grained experiment sensitive to non-Markovian/memory effects of the bath (such experiments are within the realm of possibility\cite{2011PhRvL.106w3601M,2013arXiv1305.6942G,2017Sci...355..156M,2018NatCo...9..904P}) and think of the integrands in Eq.\eqref{eq:2ptSumIntro} as part of the memory functions of the bath. For this reason, many authors (see, for example, Stratonovich \cite{stratonovich2012nonlinear,stratonovich2013nonlinear}) refer to equations like Eq.\eqref{eq:KMSIntro} as \emph{non-Markovian} fluctuation-dissipation theorems.}
\item In the last step, we take a high temperature limit to get a fluctuation-dissipation relation of the form Eq.\eqref{eq:FDTIntro}. It is in this step that one obtains the  fluctuation-dissipation relation 
between the effective  couplings.
\end{itemize}

Apart from the fluctuation-dissipation relations, there are other useful relations that can be derived when one has more than one Brownian degree of freedom and
when one can assume additional symmetries. An example of  such a symmetry is the microscopic time-reversal invariance whose  consequences were explored by Onsager \cite{PhysRev.37.405, PhysRev.38.2265} and Casimir\cite{RevModPhys.17.343} (we refer the reader to the monograph by  Stratonovich\cite{stratonovich2012nonlinear} for extensions and a detailed exposition). Say we had many degrees of freedom denoted by the coordinate $q_A$,  which undergo coupled Langevin dynamics governed by a matrix of 
Langevin couplings  $\gamma_{_{AB}}$ and $\langle f^2_{_{AB}}\rangle$. In that case, the sum rules of the discussion above generalise to
\begin{equation}
\begin{split}
m_p^2 \langle f^2_{_{AB}}\rangle &=\int_{-\infty-i\epsilon}^{\infty-i\epsilon}\frac{d\omega_1}{2\pi i}\int_{-\infty+i\epsilon}^{\infty+i\epsilon}\frac{d\omega_2}{2\pi}\ \frac{\rho[1_A2_{B+}]}{\omega_1}\ \ , \\
m_p\gamma_{_{AB}}&=\int_{-\infty-i\epsilon}^{\infty-i\epsilon}\frac{d\omega_1}{2\pi i}\int_{-\infty+i\epsilon}^{\infty+i\epsilon}\frac{d\omega_2}{2\pi}\ \frac{\rho[1_A2_B]}{\omega_1^2}\  \ .
\end{split}
\end{equation}
Here  $\mathcal{O}_A$ is the bath operator that couples to the coordinate $q_A$ and we have used a matrix of spectral functions 
\begin{equation}\begin{split}
\rho[1_A2_B] &\equiv \frac{1}{\hbar}\int dt_1 \int dt_2\  e^{i(\omega_1t_1+\omega_2t_2)}\ \langle [\mathcal{O}_A(t_{1}),\mathcal{O}_B(t_{2})] \rangle ,\\
\rho[1_A2_{B+}] &\equiv\int dt_1 \int dt_2\  e^{i(\omega_1t_1+\omega_2t_2)} \langle \{\mathcal{O}_A(t_{1}),\mathcal{O}_B(t_{2})\} \rangle\ .
\end{split}\end{equation}

Say the dynamics and the initial state of the bath are microscopic time-reversal invariant and we will assume that the Hermitian operators $\{\mathcal{O}_A\}$ have definite time-reversal parities $\{\eta_A\}$. The microscopic time-reversal then acts by the simultaneous exchange of $\omega_1$ and $\omega_2$ along with complex conjugation.\footnote{We note that 
the contour $\mathcal{C}_2$ is left invariant under this operation. Hence, the symmetry property of integrands under this operation is inherited by the effective couplings. } One then  obtains the famous \emph{Onsager reciprocal relations} \cite{PhysRev.37.405, PhysRev.38.2265} :
\begin{equation}
\begin{split}
 \langle f^2_{_{AB}}\rangle &=\eta_{_A}\eta_{_B}\  \langle f^2_{_{BA}}\rangle  \ , \quad
\gamma_{_{AB}}=\eta_{_A}\eta_{_B}\ \gamma_{_{BA}}\  \ .
\end{split}
\end{equation}
In this work, we will mainly confine ourselves to studying the dynamics of a single degree of freedom, which can in principle couple to many bath operators $\{\mathcal{O}_A\}$
with different couplings $g_{_A}$. Then, the  Onsager reciprocal relations can be interpreted as statements relating the contributions proportional to $g_{_A}g_{_B}$ to the damping constant and the noise variance.  We will later meet an example of similar relations,  when we study non-linear corrections to the dynamics of a single degree of freedom (see the discussion in section~\ref{time-reversal odd}).

\subsection{Model of the harmonic bath in Caldeira-Leggett like models }
We will now move from the general description  to particular models of the bath. The  equation \eqref{eq:2ptSumIGeneral} for the damping constant $\gamma$ naively suggests the following:  if we can engineer a bath of harmonic oscillators  with a linear spectral function of the form\footnote{We note that, for Markovian approximation to hold, the spectral function should be sufficiently analytic near real axis of the frequency domain\cite{PhysRevD.45.2843,PhysRevD.47.1576,2018PhRvB..97j4306C}.} 
\begin{equation}
\begin{split}
\rho[12]\equiv \rho(\omega_1) 2\pi\delta(\omega_1+\omega_2)\ \sim m_p \gamma\omega_1\ 2\pi\delta(\omega_1+\omega_2)\ ,
\end{split}
\end{equation}
 a naive residue integral seems to pickup the required contribution. This is however misleading. In fact, the integral for $\gamma$ with the linear 
spectral function is UV divergent and needs to be regulated appropriately. A simple and commonly used regulator is to assume
a Lorentz-Drude form for the spectral function, with the linear growth at low frequencies :
 \begin{equation}\label{eq:LorRho}
\begin{split}
\rho(\omega) = 2\ \text{Im}\left\{ m_p\gamma \frac{i \Omega^2 }{\omega+i\Omega}\right\}=2m_p \gamma\omega \frac{\Omega^2}{\omega^2+\Omega^2}\ ,
\end{split}
\end{equation}
which does give back $\gamma$ after the contour integral is performed. 

How can such a  distribution be obtained from  the underlying microscopic dynamics of the harmonic bath ? Say we are interested in the thermal harmonic bath where $q(t)$ of the Brownian particle  is coupled linearly to the bath oscillators via  
\be\begin{split} \mathcal{O}(t)= \sum_i g_{x,i}\ X_i(t)\ .\end{split}\ee
We want to now examine what  set of couplings $g_{x,i}$ and masses $m_{x,i}$ for the bath oscillators will result in a Lorentz-Drude spectral function. Computing the commutator of the bath operator 
in the thermal state and Fourier transforming yields
\begin{equation}\begin{split}
\rho[12] &=\frac{1}{2\omega_1}\ 2\pi \delta(\omega_1+\omega_2)  \sum_i    \frac{g^2_{x,i}}{m_{x,i}} 2\pi\delta(|\omega_1|-\mu_i) \\
&\equiv  2\pi \delta(\omega_1+\omega_2) \times \int_0^\infty \frac{d\mu_x}{2\pi}\  (2\pi)\ \text{sgn}(\omega_1) \delta(\omega_1^2-\mu_x^2)\ \Big\langle \Big\langle \frac{g_x^2}{m_x}\Big\rangle\Big\rangle\ ,
\end{split}\end{equation}
where we have defined a function of $\mu_x$ :
\begin{equation}\begin{split}
\Big\langle \Big\langle \frac{g_x^2}{m_x} \Big\rangle\Big\rangle &\equiv \sum_i    \frac{g^2_{x,i}}{m_{x,i}} 2\pi\delta(\mu_x-\mu_i)\ .
\end{split}\end{equation}

A spectral function $\rho(\omega)$ as in Eq.\eqref{eq:LorRho} is then obtained, if we take a continuum of oscillators whose couplings
add  up to  give
\begin{equation}
\begin{split}
\Big\langle \Big\langle \frac{g_x^2}{m_x}\Big\rangle\Big\rangle = m_p  \gamma  \frac{4\mu_x^2\Omega^2}{\mu_x^2+\Omega^2}\ .
\end{split}
\end{equation}
Thus,  the continuum approximation with an infinite set of oscillators gives  the required smooth form for the spectral function.
Only in this limit can the set of bath oscillators  be idealised as a perfect thermal bath into which the Brownian particle can 
irreversibly dissipate into. It is also only in this limit that the information about the particle is quickly forgotten by the thermal 
bath, thus allowing us to ignore any memory effect. This last point can be explicitly checked by computing the bath correlators
and confirming that they indeed decay at time scales set by $\Omega^{-1}$. Thus, we expect a local description to be good when 
there is a  hierarchy of frequency scales :
\[ \gamma \ll \Omega \ll \frac{k_BT}{\hbar}\ . \]
The continuum approximation and the resultant irreversibility are good approximations at time scales much smaller than the inverse of the typical gap in the bath spectrum.

This ends our brief review of the standard Langevin theory. One main goal of this work is to see  many of these ideas and expressions generalise to higher point functions,
out of time order correlations  and to non-linear Langevin dynamics.

\section{ Introduction to Non-linear Langevin equation}\label{sec:nonLinearLangevin}
It is a natural question to ask how these results generalise once we go beyond the linear Langevin description. A particular
non-linear generalisation is of relevance to this work, which we shall now describe. Consider a non-linear generalisation of Langevin theory described by the following stochastic equation :
\begin{equation}\begin{split}
\mathcal{E}[q]\equiv \frac{d^2q}{dt^2}+(\gamma+\zeta_\gamma \mathcal{N})  \frac{dq}{dt} + (\bar{\mu}^2+\zeta_{\mu} \mathcal{N})\ q +\left(\overline{\lambda}_3-\overline{\lambda}_{3\gamma}\ \frac{d}{dt}\right)\frac{q^2}{2!} -F =\langle f^2\rangle \mathcal{N}\ \ . 
  \end{split}
  \label{NLD dynamics}
\end{equation} 
Here, we will take $\mathcal{N}$ to be a random noise drawn from the non-Gaussian probability distribution
\begin{equation}
\begin{split} 
P[\mathcal{N}] \propto \exp\left\{-\frac{1}{2 \langle f^2\rangle}\int dt\ \Bigl(\langle f^2\rangle \mathcal{N}-\zeta_N \mathcal{N}^2\Bigr)^2-\frac{1}{2}Z_I\int dt\ \dot{\mathcal{N}}^2\right\}\ . 
  \end{split}
\end{equation} 
We will assume that the corrections to the Langevin equation are small : this is equivalent to assuming that the  parameters
$\{\zeta_\gamma,\zeta_{\mu} ,\overline{\lambda}_3,\overline{\lambda}_{3\gamma},\zeta_N,Z_I\}$ are small. 

The physical meaning of these  non-linear parameters should be evident : $\zeta_\gamma,\zeta_{\mu}$ characterise the thermal jitter in the damping constant $\gamma$ 
and the (renormalised) natural frequency $\bar{\mu}^2$ of the Brownian oscillator. The parameters $\overline{\lambda}_3,\overline{\lambda}_{3\gamma}$ control the anharmonicity in the model whereas $\zeta_{_N}$
characterises the non-Gaussianity of the thermal noise. The above equation includes all terms upto quadratic in amplitudes of $q$ and $\mathcal{N}$ and upto one time derivative
acting on $q$ (except for the inertial term $\frac{d^2q}{dt^2}$ which was already present before the bath came into picture). In this sense, this is indeed 
the most generic leading non-linear correction to the linear Langevin theory.

Another equivalent way to define these couplings would be to state how they occur in the long-time three point cumulants of the Brownian particle which starts off from the 
harmonic oscillator vacuum at an initial time $t=t_0$. While specification in terms of cumulants does 
not have the immediate intuitive appeal of the description above, it is a very useful characterisation for  computing these couplings from a microscopic 
model. Consider long enough time scales such that the memory effects of the bath can be ignored and the Markovian approximation is valid. We will however be interested in  the time scales much smaller than $\gamma^{-1}$, the time scale at which damping effects become substantial. In this time window,
one can write down universal expressions for the vacuum cumulants of the Brownian particle in terms of the effective Langevin couplings. In the semi-classical limit (i.e., ignoring
$O(\hbar)$ terms ), they take the following  form  \cite{Chaudhuri:2018ihk} : for $t_1>t_2>t_3$, we get
\be
\begin{split}
\frac{1}{\hbar^2}\langle[q(t_1)q(t_2)q(t_3)]\rangle_c&=\Big(\overline{\lambda}_3-\overline{\lambda}_{3\gamma}\frac{\partial}{\partial t_1}\Big)\cubicfunc_{123}+O(\hbar)\ ,\\
\frac{1}{\hbar^2}\langle[q(t_3)q(t_2)q(t_1)]\rangle_c&=\Big(\overline{\kappa}_3+\overline{\kappa}_{3\gamma}\frac{\partial}{\partial t_3}\Big)\cubicfunc_{321}+O(\hbar)\ ,\\
\frac{i}{\hbar}\langle[q(t_1)q(t_2)q(t_3)_+]\rangle_c&=O(\hbar)
,\\
\frac{i}{\hbar}\langle[q(t_3)q(t_2)q(t_1)_+]\rangle_c
&=
2m_p\Big[\zeta_\mu+2\zeta_{\gamma}\frac{\partial}{\partial t_1}-\frac{2}{3}\widehat{\kappa}_{3\gamma}\Big(\frac{\partial}{\partial t_1}-\frac{\partial}{\partial t_2}\Big)\Big]\cubicfunc_{321}+O(\hbar)\ ,\\
\frac{i}{\hbar}\langle[q(t_1)q(t_3)q(t_2)_+]\rangle_c
&=
-2 m_p\Big[\zeta_\mu+2\zeta_\gamma\frac{\partial}{\partial t_2}+\frac{2}{3}\widehat{\kappa}_{3\gamma}\Big(\frac{\partial}{\partial t_1}-\frac{\partial}{\partial t_2}\Big)\Big]\cubicfunc_{321}+O(\hbar)\ ,\\
\langle[q(t_1)q(t_2)_+q(t_3)_+]\rangle_c
&=\frac{2}{\overline{\mu}^4}\zeta_N\Big[2\cos(\overline{\mu}(t_{13}+t_{23}))+6\cos(\overline{\mu} t_{12})+\cos(\overline{\mu}(t_{10}+t_{20}+t_{30}))\\
&\quad-3\cos(\overline{\mu}(t_{12}-t_{30}))-3\cos(\overline{\mu}(t_{23}-t_{10}))-3\cos(\overline{\mu}(t_{31}-t_{20}))\Big]\\
&\quad+O(\hbar)\ .
\end{split}
\ee
In the above, we have defined the function 
\be
\cubicfunc_{ijk}\equiv\frac{1}{6 (m_p\overline{\mu}^2)^2}\Big\{\cos[\overline{\mu}(t_{ij}+t_{ik})]-\cos[\overline{\mu}(t_{ji}+t_{jk})]
-3\cos[\overline{\mu} t_{ik}]+3\cos[\overline{\mu} t_{jk}]\Big\}\ ,
\ee
with $t_{ij}\equiv t_i-t_j$ and we have used the square bracket notation to indicate nested commutators (with a $+$ subscript indicating anticommutator). For example,
\be
\begin{split}
\langle[q(t_1)q(t_2)q(t_3)]\rangle_c&\equiv \langle[[q(t_1),q(t_2)],q(t_3)]\rangle_c\ ,\\
\langle[q(t_1)q(t_2)q(t_3)_+]\rangle_c&\equiv \langle\{[q(t_1),q(t_2)],q(t_3)\}\rangle_c\ ,\\
\langle[q(t_1)q(t_2)_+q(t_3)_+]\rangle_c
&\equiv \langle\{\{q(t_1),q(t_2)\},q(t_3)\}\rangle_c\  .
\end{split}
\ee
The six correlators given above cover all possible time orderings with three operators. In the above, we have divided by a factor of $\hbar$ for every commutator in LHS, so that the commutators smoothly go to Poisson brackets in the classical limit.

The reader should also note that apart from the non-linear Langevin/1-OTO couplings\footnote{These labels characterise how much out of time ordered a particular correlator/coupling actually is.
The numbers here represent the number of minimum number of time-folds required to define   a particular  correlator/coupling\cite{2017arXiv170102820H}.}, three new `out of time ordered' 2-OTO couplings $\overline{\kappa}_3,\overline{\kappa}_{3\gamma}$ and $\widehat{\kappa}_{3\gamma}$ are needed
to fit the long-time behaviour of arbitrarily ordered cumulants. On general grounds, we expect correlators with four out of the six time orderings to be computable via
standard Schwinger-Keldysh/Feynman-Vernon influence functionals : these are the 1-OTO correlators in the classification of \cite{2017arXiv170102820H}. A basis of 
such correlators is provided by 
\be
\begin{split}
\langle[q(t_1)q(t_2)q(t_3)]\rangle_c\ ,\quad \langle[q(t_1)q(t_2)q(t_3)_+]\rangle_c\ ,\quad\langle[q(t_1)q(t_2)_+q(t_3)_+]\rangle_c\ ,\\
\langle[q(t_1)q(t_2)_+q(t_3)]\rangle_c= -\langle[q(t_3)q(t_2)q(t_1)_+]\rangle_c+ \langle[q(t_1)q(t_3)q(t_2)_+]\rangle_c\ ,
\end{split}
\ee
all of which can be written down in terms of the standard non-linear Langevin/1-OTO couplings, as is evident from the expressions above. The two other remaining 
time-orderings are however genuinely out of time ordered which can neither be captured by the standard Schwinger-Keldysh/Feynman-Vernon influence functionals 
nor can they be written solely in terms of the standard non-linear Langevin couplings.

A historical aside may be in order : many authors  have studied  non-linear generalisation of Langevin equations in a variety of contexts
(See, for example \cite{1971JSP.....3..245B,1973JSP.....9..215Z,1979PhLA...69..313S,Brun:1993qj}). However, as far as the authors are aware,
there is no systematic discussion in the literature including all leading nonlinear corrections allowed on general grounds, nor a
microscopic model within which Markovian approximation is justified and all couplings derived. Similarly,
despite a long and rich literature on non-linear fluctuation-dissipation/Onsager relations \cite{BERNARD:1959zz, stratonovich2012nonlinear, stratonovich2013nonlinear, efremov1969fluctuation,stratonovich1970izv,stratonovich1970contribution,efremov1972gf,bochkov1977general,sitenko1978sov,gordon1978fluctuation,klyshko2018photons},
the relations we derive here for non-linear Langevin theory are new, as far as we know.

In the rest of this section, we will give a detailed summary of our results listing the integrals that relate the couplings that appear in the above equation to 
the bath correlators in the microscopic theory. The reader desirous of a briefer summary is encouraged to consult the subsection \S\S\ref{ssec:summaryNonLin}
at the end of this section.

\subsection{Linear Langevin couplings}
We will now summarise our results regarding how the above coefficients could be computed starting from a non-linear generalisation of Caldeira-Leggett like setup.
Assume the Langevin degree of freedom $q(t)$ is still linearly coupled to some bath operator $\mathcal{O}(t)$.\footnote{We assume that the thermal 1-point function of this bath operator is zero.} As in the discussion above Eq.\eqref{eq:2ptSumIGeneral},
take  the  bath to be in a general time-independent state and assume a Markovian description  at long enough time scales.

 Under these assumptions, a general  expression can be written for the   Langevin couplings in leading order causal perturbation theory in particle-bath coupling (or equivalently by matching the local influence functional to MSRDPJ action of the above stochastic equation to leading order in perturbation theory) :
 \begin{equation}
\begin{split}
m_p\gamma&= \int_{\mathcal{C}_2}\frac{\rho [12] }{i\omega_1^2}\ ,\quad  m_p \Delta \bar{\mu}^2= - \int_{\mathcal{C}_2}\frac{\rho [12] }{\omega_1}\ ,\quad
\Delta m_p =\int_{\mathcal{C}_2}\frac{\rho [12]}{\omega_1^3}\ ,\quad\\
m_p^2\langle f^2\rangle &= \int_{\mathcal{C}_2}\frac{\rho [12_+]}{i\omega_1}=\int_{\mathcal{C}_2}\frac{\hbar}{i\omega_1}(1+2\bose_1)\rho [12]\ ,\\
m_p^2 Z_I &=-\int_{\mathcal{C}_2}\frac{\rho [12_+]}{i\omega_1^3}=-\int_{\mathcal{C}_2}\frac{\hbar}{i\omega_1^3}(1+2\bose_1)\rho [12]\ ,
\end{split}
\end{equation}
where as before, we have an integral over the causal contour
\begin{equation}
\begin{split}
 \int_{\mathcal{C}_2} \equiv \int_{-\infty-i\epsilon}^{\infty-i\epsilon}\frac{d\omega_1}{2\pi}\int_{-\infty+i\epsilon}^{\infty+i\epsilon}\frac{d\omega_2}{2\pi}\    \ .
\end{split}
\end{equation}
Here $\Delta m_p$ and $\Delta \bar{\mu}^2$ denote the renormalisation of mass and frequency of the Brownian oscillators.

The spectral functions appearing in the integrand are defined via Fourier integrals
\begin{equation}\begin{split}
\rho[12_+] &\equiv \int dt_1 \int dt_2\ e^{i(\omega_1t_1+\omega_2t_2)}\ \langle \{\mathcal{O}(t_{1}),\mathcal{O}(t_{2})\} \rangle_B\ ,\\
\rho[12] &\equiv \frac{1}{\hbar}\int dt_1 \int dt_2\ e^{i(\omega_1t_1+\omega_2t_2)}\ \langle [\mathcal{O}(t_{1}),\mathcal{O}(t_{2})] \rangle_B\ .
\end{split}\end{equation}
Further, in writing the second equality for the couplings above, we have assumed that the bath state is thermal and consequently, we have used the Kubo-Martin-Schwinger (KMS) relation to convert the anti-commutator to commutator.

The complex contour $\mathcal{C}_2$ has the following \emph{reality} property : it remains invariant under a simultaneous complex conjugation and the reversal of frequencies ;
\be\label{eq:ContourReality2pt}
\begin{split}
 \mathcal{C}_2^\ast[\omega^c_i\ \to\ -\omega_i]= \mathcal{C}_2[\omega_i],
\end{split}
\ee
where $\omega^c_i$ denotes the complex conjugate of the frequency $\omega_i$. Further, the hermiticity of bath  operator  $\mathcal{O}(t)$ yields the result
\be\label{eq:SpecFnReality2pt}
\begin{split}
 (\rho[12])^\ast_{\omega^c_i\ \to\ -\omega_i}= - \rho[12]\ ,\qquad
  (\rho[12_+])^\ast_{\omega^c_i\ \to\ -\omega_i}=  \rho[12_+]\ .
\end{split}
\ee
Thus, integrating the conjugated, frequency-reversed spectral function over the conjugated, frequency-reversed contour gives the same answer as the original
spectral function integrated over the original contour. Since the relabelling of the integration variables $\omega^c_i\ \to\ -\omega_i$ should not change the value of the integral,
the above assertion is equivalent to  the statement of  reality of the Langevin couplings. 

The complex contour $\mathcal{C}_2$ also has the following  \emph{time-reversal property}\footnote{ We refer the reader to section \S\ref{sec:Relations} for a detailed discussion of time-reversal invariance and its action on  causal contours in the frequency domain.} : it remains invariant under a simultaneous complex conjugation and the exchange of frequencies  $\omega_1$ and $\omega_2$. Say the spectral function has a specific time-reversal parity $\eta_{_{12}}$ inherited from the time parity of operators that define it.
We then have
\be\label{eq:SpecFnTRev2pt}
\begin{split}
 (\rho[12])^\ast_{\omega_1^c \to \omega_2, \ \omega_2^c \to \omega_1}= -\eta_{_{12}} \rho[1 2]\ ,\qquad
  (\rho[12_+])^\ast_{\omega_1^c \to \omega_2, \ \omega_2^c \to \omega_1}= \eta_{_{12}} \rho[12_{+}]\ ,
\end{split}
\ee
thus guaranteeing that the couplings derived above obey Onsager reciprocal relations using a similar argument to the above argument for reality.

\paragraph{A  comment on the force term:}
The force per unit mass $F$ appearing in the particle's equation of motion \eqref{NLD dynamics}  is determined at  leading order in the particle-bath coupling by the 1-point function of the  operator $\mathcal{O}(t)$. For the $qXY$ model introduced in section \ref{sec:qXY}, this thermal 1-point function is zero. This, in turn, leads to the vanishing of the leading order term in $F$. At  sub-leading order, it can receive contributions from the 3-point spectral functions of the bath. We will  ignore such sub-leading corrections to the couplings in this paper.

\subsection{Anharmonicity parameters : time ordered and out of time ordered}
A similar exercise can be carried out for $3$-point functions : for the anhormonic couplings we have 
\be\label{eq:3ptSumIGeneralLambda}
\begin{split}
m_p\overline{\lambda}_3 &=2 \int_{\mathcal{C}_3} \frac{\rho[123]}{\omega_1\omega_3}\ ,\quad 
m_p\overline{\lambda}_{3\gamma} =  \int_{\mathcal{C}_3}\frac{1}{i\omega_1\omega_3}\left(\frac{2}{\omega_1}-\frac{1}{\omega_3}\right)\rho[123]
=
 \int_{\mathcal{C}_3}\frac{(2\omega_3-\omega_1)}{i(\omega_1\omega_3)^2}\rho[123]\ .
\end{split}
\ee
As before, we have defined here the spectral function
\begin{equation}\begin{split}
\rho[123] &\equiv \frac{1}{\hbar^2}\int dt_1 \int dt_2\ \int dt_3\  e^{i(\omega_1t_1+\omega_2t_2+\omega_3t_3)}\ \langle [[\mathcal{O}(t_{1}),\mathcal{O}(t_{2})],\mathcal{O}(t_{3})] \rangle_B\ .
\end{split}\end{equation}
and the frequency domain contour which picks up the causal part of three point  function :
\be
\begin{split}
\int_{\mathcal{C}_3}\equiv \int_{-\infty-i\epsilon_1}^{ \infty-i\epsilon_1} \frac{d\omega_1}{2\pi}\int_{-\infty+i\epsilon_1-i\epsilon_3}^{ \infty+i\epsilon_1-i\epsilon_3} \frac{d\omega_2}{2\pi}
\int_{-\infty+i\epsilon_3}^{ \infty+i\epsilon_3} \frac{d\omega_3}{2\pi}
\end{split}
\ee
where $\epsilon_1,\epsilon_3>0$. We will find it convenient to not fix a particular ordering between $\epsilon_1$ and $\epsilon_3$ : the ordering will not matter, provided
we take care that our integrands do not have poles/branch cuts in $\omega_2$ near the real axis. For example, the above expression for the effective couplings is valid 
provided the  spectral function has no discontinuities or branch cuts in the real axis :
\be
\begin{split}
\text{Disc}_{\omega_2} \rho[123]=  0\ .
\end{split}
\ee
In cases where there are discontinuities, they are known to lead to  long time memory effects which, in turn, lead to a breakdown of Markovian approximation\cite{2018PhRvB..97j4306C}.
Markovian approximation also needs the following constraint on the spectral function :
\be
\begin{split}
\lim_{\omega_3\to0} \rho[123]=  0\ .
\end{split}
\ee

The complex contour $\mathcal{C}_3$ has the following \emph{reality property} : it remains invariant under a simultaneous complex conjugation and the reversal of frequencies (similar to our discussion above for the two-point causal contour $\mathcal{C}_2$) . 
It follows then that, if the spectral function $\rho[123]$ is invariant under this operation, then the resultant answers for the anharmonic couplings are real. 

The line of argument
which establishes this invariance is identical to the argument outlined for two point functions : the hermiticity of
bath  operator  $\mathcal{O}(t)$ guarantees the reality of the double commutator $\langle [[\mathcal{O}(t_{1}),\mathcal{O}(t_{2})],\mathcal{O}(t_{3})] \rangle_B$ and in turn, the invariance of $\rho[123]$
under simultaneous complex conjugation and the reversal of frequencies, viz.,
\be\label{eq:SpecFnReality3}
\begin{split}
 (\rho[123])^\ast_{\omega_i^c\ \to\ -\omega_i}=  \rho[123]\ .
\end{split}
\ee
This along with the invariance of the contour guarantees reality of the couplings.

The complex contour $\mathcal{C}_3$ also has the following \emph{time-reversal property} : it remains invariant under a simultaneous complex conjugation and the exchange of frequencies
$\omega_1$ and $\omega_3$ (with the concomitant exchange of their imaginary parts $\epsilon_1$ and $\epsilon_3$) . If there are no poles/branch cuts in $\omega_2$ near the real axis, we can deform the contour back to  $\mathcal{C}_3$ : it is in this sense that the frequency contour $\mathcal{C}_3$ is time-reversal invariant. This is similar to the 
time-reversal invariance of the two-point causal contour $\mathcal{C}_2$. However, unlike the two point function case, the action on the \emph{integrand} cannot be simply described :
the spectral function $\rho[123]$ gets mapped to a new function $\rho[321]$
\be\label{eq:SpecFnTRev3pt}
\begin{split}
 (\rho[123])^\ast_{\omega_1^c \to \omega_3, \ \omega_2^c \to \omega_2, \ \omega_3^c \to \omega_1}= \eta_{_{123}}  \rho[321]\ ,
 \end{split}
\ee
 where $\eta_{_{123}}$ is the time parity of the spectral function inherited from the time parities of the underlying operators. In the above, the function $\rho[321]$ can be thought of as the time-reversed/out of time order (OTO) spectral function :
\begin{equation}\begin{split}
\rho[321] &\equiv \frac{1}{\hbar^2}\int dt_1 \int dt_2\ \int dt_3\  e^{i(\omega_1t_1+\omega_2t_2+\omega_3t_3)}\ \langle [[\mathcal{O}(t_{3}),\mathcal{O}(t_{2})],\mathcal{O}(t_{1})] \rangle_B\ .
\end{split}\end{equation}
While the expression for $\rho[321]$ looks formally similar to that of $\rho[123]$, note that these are two different functions on the  $\mathcal{C}_3$ contour due to the 
different $i\epsilon$ prescriptions. For a bath with no microscopic time-reversal invariance, these two spectral functions are a priori unrelated. On the other hand, in the 
presence of time-reversal invariance,  the analogue of Onsager relations for cubic couplings relate the Langevin anharmonic couplings to anharmonic couplings in the time-reversed stochastic dynamics.

The above statement can be made precise by introducing the out of time order (or time-reversed) anharmonic couplings 
\be
\begin{split}
m_p\overline{\kappa}_3 &\equiv 2 \int_{\mathcal{C}_3} \frac{\rho[321]}{\omega_1\omega_3}\ ,\quad 
m_p\overline{\kappa}_{3\gamma} \equiv  \int_{\mathcal{C}_3}\frac{1}{i\omega_1\omega_3}\left(\frac{1}{\omega_1}-\frac{2}{\omega_3}\right)\rho[321]=\int_{\mathcal{C}_3}\frac{(\omega_3-2\omega_1)}{i(\omega_1\omega_3)^2}\rho[321]\ .
\end{split}
\ee
Say the bath operators $O_k(t)$ with a definite time-reversal parity $\eta_k$ couple to the single degree of freedom of the Brownian particle. For simplicity, assume that the 
relevant spectral function gets contribution only from the product of three operators $O_{k=1,2,3}$. We can then write down the non-linear,  OTO analogue of the 
Onsager reciprocal relations for the resultant couplings as
\be\boxed{
\begin{split}
\overline{\lambda}_3&= \eta_{_O}\ \overline{\kappa}_3\ ,\qquad
\overline{\lambda}_{3\gamma} = \eta_{_O}\ \overline{\kappa}_{3\gamma} \ ,
\end{split}}
\ee
where $\eta_{_O}\equiv \eta_1 \eta_2 \eta_3$ is the total time-reversal parity of the given spectral function.
Thus, rather than constraining the cubic couplings in the non-linear Langevin dynamics, the time-reversal invariance ends up relating the standard
anharmonic couplings to the OTO anharmonic couplings.

\subsection{Frequency Noise parameter $\zeta_\mu$}

We now move on to quote the results for the other non-linear couplings that appear in the non-linear Langevin equation. We begin with the parameter
$\zeta_\mu$ that governs the thermal jitter in the frequency of the Brownian oscillator: 
\be\label{eq:3ptSumIGeneralZetaMu}
\begin{split}
m_p^2\zeta_\mu
&=   \int_{\mathcal{C}_3} \frac{1}{2i\omega_1\omega_3}\Big(\rho [12_+3]+  \rho[123_+]\Big)\\
&=   \int_{\mathcal{C}_3} \frac{1}{2i\omega_1\omega_3}\Big(\rho [231_+]+ \rho [132_+]+ \rho[123_+]\Big)\\
&=  \int_{\mathcal{C}_3} \frac{\hbar}{i\omega_1\omega_3}\Big\{(1+\bose_1+\bose_2)\rho[321]-(1+\bose_2+\bose_3)\rho[123]\Big\}\  .
\end{split}
\ee
Here, the spectral functions are defined as before by Fourier transforming the appropriate nested commutators/anti-commutators.\footnote{including a factor of inverse $\hbar$ for every commutator to guarantee  smooth classical limit.} For example, we have
\begin{equation}\begin{split}
\rho[12_+3] &\equiv \frac{1}{\hbar}\int dt_1 \int dt_2\int dt_3\  e^{i(\omega_1t_1+\omega_2t_2+\omega_3t_3)} \langle [\{\mathcal{O}(t_{1}),\mathcal{O}(t_{2})\},\mathcal{O}(t_{3})]  \rangle_B\ ,\\
\rho[123_+] &\equiv \frac{1}{\hbar}\int dt_1 \int dt_2\int dt_3\  e^{i(\omega_1t_1+\omega_2t_2+\omega_3t_3)} \langle \{[\mathcal{O}(t_{1}),\mathcal{O}(t_{2})],\mathcal{O}(t_{3})\} \rangle_B\ ,
\end{split}\end{equation}
where the subscript $+$ indicates the anti-commutator. The integral is over the same causal contour $\mathcal{C}_3$ as before. The first line gives the appropriate
causal Green function from which the effective coupling $\zeta_\mu$ can be extracted.

The expression  in the second line above arise from generalised Jacobi identities of the form 
\be\begin{split} \rho [12_+3]=\rho [231_+]+ \rho [132_+]\ ,\qquad \rho [123]+\rho [231]+ \rho [312]=0 \end{split}\ee 
and permutations thereof, and the last line which follows from Kubo-Martin-Schwinger (KMS) relation for three point functions :
\be\begin{split} \rho [123_+]=-\hbar(1+2\bose_3)\rho[123]\ \end{split}\ee
and permutations thereof. Here $\bose_i\equiv \bose(\omega_i)$ are the Bose-Einstein functions of the respective frequencies. Thus, one can express the coupling 
as a causal contour integral over the spectral function $\rho[123]$ and the out of time order spectral function $\rho[321]$ multiplied by the appropriate
Bose-Einstein factors.

The discussion on the reality properties is similar to  before : the causal contour $\mathcal{C}_3$ and the commutator spectral functions are invariant under simultaneous
conjugation and frequency reversal (assuming no $\omega_2$ discontinuities on the real axis). The  anti-commutator spectral functions are odd under simultaneous
conjugation and frequency reversal :
\be\label{eq:SpecFnAntiReality}
\begin{split}
 (\rho[123_+])^\ast_{\omega_i\ \to\ -\omega_i}= - \rho[123_+]\ ,\quad  (\rho[12_+3])^\ast_{\omega_i\ \to\ -\omega_i}= - \rho[12_+3]\ .
\end{split}
\ee

This, then ensures the reality of $\zeta_\mu$. The reality can also be shown
using the expression in terms of Bose-Einstein distributions, however the argument involved is slightly more subtle :  the following property of Bose-Einstein functions
\be\begin{split} \bose(-\omega)= -\left\{1+\bose(\omega)\right\}, \end{split}\ee
should be used along with the assumption that the potential discontinuity due to $\bose_2$ near $\omega_2\to 0$ is cancelled among the two terms.\footnote{One can
explicitly prove that this indeed happens in the Markovian models of the bath we use.}

As discussed in the last subsection, time-reversal acts on the causal contour $\mathcal{C}_3$ via   a simultaneous complex conjugation and the exchange of frequencies
$\omega_1$ and $\omega_3$. If the bath operator $\mathcal{O}(t)$ has a definite time-reversal parity $\eta_{_O}$, we can then write down the non-linear
Onsager reciprocal relation for $\zeta_\mu$ as 
\be\boxed{
\begin{split}
\zeta_\mu&= \eta_{_O}\ \zeta_\mu,
\end{split}}
\ee
i.e., one obtains a trivial relation for time-reversal even bath operators and for time-reversal odd bath operators, $\zeta_\mu=0$ .

At large temperatures (or small $\beta$), one can approximate
\be\begin{split} \hbar\bose(\omega) \approx \frac{\hbar}{\beta\omega} = \frac{m_p v_{th}^2}{\omega}\ . \end{split}\ee
We then get a high temperature formula for $\zeta_\mu$ of the form
\be
\begin{split}
m_p\zeta_\mu
&= v_{th}^2  \int_{\mathcal{C}_3} \frac{1}{i\omega_1\omega_3}\Big\{\left(\frac{1}{\omega_1}+\frac{1}{\omega_2}\right)\rho[321]-\left(\frac{1}{\omega_2}+\frac{1}{\omega_3}\right)\rho[123]\Big\}\\
&= v_{th}^2  \int_{\mathcal{C}_3} \frac{1}{i\omega_2}\Big\{\frac{\rho[123]}{\omega_3^2}-\frac{\rho[321]}{\omega_1^2}\Big\}  ,
\end{split}
\ee
where, in the last line, we have used the fact that the spectral functions are proportional to $\delta(\omega_1+\omega_2+\omega_3)$ for time-independent state of the bath.

We now turn to possible fluctuation-dissipation type relation involving $\zeta_\mu$. While the  integrand which appears in the first line above has structural similarities with the integrands in the sum rules of the couplings $\overline{\lambda}_{3\gamma}$
and $\overline{\kappa}_{3\gamma}$, we have not succeeded in establishing any general fluctuation-dissipation type relation between these couplings. Nevertheless, in many
explicit models with time-reversal invariance\footnote{We will describe one such model in detail in the following section.}, we find the following relation :
\be
\begin{split}
\int_{\mathcal{C}_3} &\frac{1}{i\omega_1\omega_3}\Big\{\left(\frac{1}{\omega_1}+\frac{1}{\omega_2}\right)\rho[321]-\left(\frac{1}{\omega_2}+\frac{1}{\omega_3}\right)\rho[123]\Big\}\\
&= \int_{\mathcal{C}_3}\frac{1}{i\omega_1\omega_3}\left(\frac{2}{\omega_1}-\frac{1}{\omega_3}\right)\rho[123]
= \int_{\mathcal{C}_3}\frac{1}{i\omega_1\omega_3}\left(\frac{1}{\omega_1}-\frac{2}{\omega_3}\right)\rho[321]\ ,
\end{split}
\ee
where the last equality is the relation $\overline{\lambda}_{3\gamma}=\overline{\kappa}_{3\gamma}$ which holds for even time-reversal bath operators.
More generally, we find that the following integral vanishes in many explicit examples :
\be
\begin{split}
&v_{th}^2\left(\overline{\lambda}_{3\gamma}+\overline{\kappa}_{3\gamma}\right)-2\zeta_\mu\\
 &\ =  \frac{v_{th}^2}{m_p} \int_{\mathcal{C}_3}\frac{1}{i\omega_1\omega_3}\left\{\left(\frac{2}{\omega_1}+\frac{2}{\omega_2}+\frac{1}{\omega_3}\right)\rho[123]-\left(\frac{1}{\omega_1}+\frac{2}{\omega_2}+\frac{2}{\omega_3}\right)\rho[321]\right\}=0\ ,
\end{split}
\ee
thus resulting in a relation of the form 
\be\boxed{
\begin{split}
\zeta_\mu=\frac{1}{2}v_{th}^2\left(\overline{\lambda}_{3\gamma}+\overline{\kappa}_{3\gamma}\right)\ .
\end{split}}
\ee

\subsection{Dissipative noise parameter $\zeta_\gamma$ and its OTO counterpart}
We will now move on to the parameter $\zeta_\gamma$ which describes the jitter in the dissipative constant $\gamma$ of the Brownian particle.
The sum rule(s) for computing $\zeta_\gamma$ is given by 
\be
\begin{split}
m_p^2\zeta_\gamma
&=  \int_{\mathcal{C}_3}\frac{1}{(2\omega_1\omega_3)^2}\Bigl((\omega_3-2\omega_1)\ \rho[12_+3]-\omega_2\ \rho[123_+] \Bigr)\\
&=  \int_{\mathcal{C}_3}\frac{1}{(2\omega_1\omega_3)^2}\Bigl((2\omega_1-\omega_3)\ \rho[321_+]+(\omega_3-2\omega_1)\ \rho[132_+]+(\omega_1+\omega_3)\ \rho[123_+] \Bigr)\\
&=  \int_{\mathcal{C}_3}\frac{\hbar}{(2\omega_1\omega_3)^2}\\
&\ \times\Big\{2(\omega_3-2\omega_1)\Big(1+\bose_1+\bose_2\Big)\rho[321] -(\omega_3-2\omega_1)(1+2\bose_2)\rho[123]
+\omega_2(1+2\bose_3)\rho[123]\Big\}\  .
\end{split}
\ee
In the first line, we have the representation in terms of the causal correlator. The second line comes from generalised Jacobi identities and the last line from 
Kubo-Martin-Schwinger (KMS) relations. Following arguments very similar to the ones sketched in the previous subsection, one can argue that $\zeta_\gamma$
produced by the above integral is real.

Next we turn to examine the behaviour of the integrand under  the exchange of frequencies
$\omega_1$ and $\omega_3$ (which would be relevant for the action of time-reversal on this coupling). We have the following identity :
\be
\begin{split}
&(2\omega_1-\omega_3)\ \rho[321_+]+(\omega_3-2\omega_1)\ \rho[132_+]+(\omega_1+\omega_3)\ \rho[123_+] \\
&\quad + (2\omega_3-\omega_1)\ \rho[123_+]+(\omega_1-2\omega_3)\ \rho[312_+]+(\omega_3+\omega_1)\ \rho[321_+] \\
&\qquad= 3  \Bigl( \omega_1\ \rho[321_+]+(\omega_3-\omega_1)\ \rho[132_+]+\omega_3\ \rho[123_+]\Bigr)\  .
\end{split}
\ee
This implies that if the bath operator $\mathcal{O}(t)$ has a definite time-reversal parity $\eta_{_O}$, then the generalised
Onsager reciprocal relation for $\zeta_\gamma$ becomes 
\be\boxed{
\begin{split}
\zeta_\gamma &= \eta_{_O}\ (\widehat{\kappa}_{3\gamma}-\zeta_\gamma)\ , \quad \widehat{\kappa}_{3\gamma} = \eta_{_O}\ \widehat{\kappa}_{3\gamma}\ ,
\end{split}}
\ee
where we have defined a new OTO coupling $\widehat{\kappa}_{3\gamma}$ via 
\be
\begin{split}
m_p^2\ \widehat{\kappa}_{3\gamma}
&\equiv 3\int_{\mathcal{C}_3} \frac{1}{(2\omega_1\omega_3)^2}\Bigl( \omega_1\ \rho[321_+]+(\omega_3-\omega_1)\ \rho[132_+]+\omega_3\ \rho[123_+]\Bigr)\\
&=-3 \int_{\mathcal{C}_3} \frac{1}{(2\omega_1\omega_3)^2}\Bigl( \omega_1\ \rho[12_+3]+\omega_3\ \rho[32_+1]\Bigr)\\
&= -3\int_{\mathcal{C}_3} \frac{\hbar}{(2\omega_1\omega_3)^2}
\Big(\Big\{2\omega_1\Big(1+\bose_1+\bose_2\Big)-\omega_3(1+2\bose_2)\Big\}\rho[321] \Bigr.\\
&\qquad+\Big\{2\omega_3\Big(1+\bose_2+\bose_3\Big)-\omega_1(1+2\bose_2)\Big\}\rho[123]\Big)\ .
\end{split}
\end{equation}
We conclude that if  $\eta_{_O}=1$, one has the constraint $\widehat{\kappa}_{3\gamma}=2\zeta_\gamma$ whereas for $\eta_{_O}=-1$, one has the constraint $\widehat{\kappa}_{3\gamma}=0$. Thus, $\widehat{\kappa}_{3\gamma}$ can be thought of as the part of $\zeta_\gamma$ even under time-reversal, whereas the part odd under time-reversal 
is given by 
\be
\begin{split}
m_p^2\ \left(2\zeta_\gamma-\widehat{\kappa}_{3\gamma}\right)
&\equiv \int_{\mathcal{C}_3} \frac{1}{(2\omega_1\omega_3)^2}\Bigl( 
 (\omega_1-2\omega_3)\ \rho[321_+]+3(\omega_3-\omega_1)\ \rho[132_+]+(2\omega_1-\omega_3)\ \rho[123_+]\Bigr)\\
&= \int_{\mathcal{C}_3} \frac{\hbar}{(2\omega_1\omega_3)^2}
\Big(\Big\{2\omega_2\Big(1+\bose_1+\bose_2\Big)+3\omega_3(1+2\bose_1)\Big\}\rho[321] \Bigr.\\
&\qquad-\Big\{2\omega_2\Big(1+\bose_2+\bose_3\Big)+3\omega_1(1+2\bose_3)\Big\}\rho[123]\Big)\ .
\end{split}
\end{equation}

At high temperature,  these expressions become
\be
\begin{split}
m_p\zeta_\gamma
&= v_{th}^2 \int_{\mathcal{C}_3}\frac{1}{2\omega_1\omega_3}\\
&\ \times\Big\{\left(\frac{1}{\omega_1}-\frac{2}{\omega_3}\right)\Big(\frac{1}{\omega_1}+\frac{1}{\omega_2}\Big)\rho[321] +\left(\frac{2}{\omega_2\omega_3}-\frac{1}{\omega_2\omega_1}+\frac{\omega_2}{\omega_1\omega_3^2}\right)\rho[123]\Big\}\\
&= v_{th}^2 \int_{\mathcal{C}_3}\frac{1}{\omega_1\omega_3}\Big\{\left(\frac{1}{\omega_1^2}+\frac{3}{\omega_1\omega_2}\right)\rho[321] +\left(\frac{3}{\omega_3\omega_2}-\frac{1}{\omega_3^2}\right)\rho[123]\Big\}\  ,
\end{split}
\ee
or equivalently
\be
\begin{split}
m_p\ \left(2\zeta_\gamma-\widehat{\kappa}_{3\gamma}\right)
&= v_{th}^2\int_{\mathcal{C}_3} \frac{1}{\omega_1\omega_3}\Big\{\frac{1}{\omega_1^2}\rho[321] -\frac{1}{\omega_3^2}\rho[123]\Big\}\  ,
\end{split}
\end{equation}
and
\be
\begin{split}
m_p\ \widehat{\kappa}_{3\gamma}
&= -3v_{th}^2\int_{\mathcal{C}_3} \frac{1}{2\omega_1\omega_3}\\
&\ \times
\Big\{ \Big(\frac{1}{\omega_1\omega_3}+\frac{1}{\omega_3\omega_2}-\frac{1}{\omega_1\omega_2}\Big)\rho[321] 
+\Big(\frac{1}{\omega_1\omega_3}+\frac{1}{\omega_1\omega_2}-\frac{1}{\omega_3\omega_2}\Big)\rho[123]\Big\}\\
&= 3v_{th}^2\int_{\mathcal{C}_3} \frac{1}{\omega_1\omega_2\omega_3}
\Big\{\frac{1}{\omega_1}\rho[321] 
+\frac{1}{\omega_3}\rho[123]\Big\}\ .
\end{split}
\end{equation}
In simplifying these expressions, we have used the identity
\be\label{eq:onebyomegasq}
\begin{split}
\frac{1}{\omega_1\omega_3}+\frac{1}{\omega_1\omega_2}+\frac{1}{\omega_3\omega_2}=0\ ,
\end{split}
\end{equation}
which holds because of the $\delta(\omega_1+\omega_2+\omega_3)$ inside the spectral functions.

\subsection{Non-Gaussianity $\zeta_N$ and its fluctuation-dissipation relation}
We finally turn our attention to the last parameter of the non-linear Langevin model which is the non-Gaussianity parameter $\zeta_N$. We find the 
sum rule(s) for computing $\zeta_N$ as 
\be
\begin{split}
m_p^3\zeta_N &=  \int_{\mathcal{C}_3} \frac{\rho[1 2_+ 3_+]}{4\omega_1\omega_3} =  \int_{\mathcal{C}_3} \frac{\hbar^2}{4\omega_1\omega_3}
(1+2\bose_3)\Big\{(1+2 \bose_2) \rho[1 2 3]-2(1+\bose_1+\bose_2) \rho[3 2 1] \Big\}
\ ,
\end{split}
\ee
where we have used the  condition for the nested anti-commutator to write the second equality. The relevant spectral function is
\begin{equation}\begin{split}
\rho[12_+3_+] &\equiv \int dt_1 \int dt_2\int dt_3\  e^{i(\omega_1t_1+\omega_2t_2+\omega_3t_3)} \langle \{\{\mathcal{O}(t_{1}),\mathcal{O}(t_{2})\},\mathcal{O}(t_{3})\}  \rangle_B\ .
\end{split}\end{equation}

 It can be checked that the integral above does give 
a real coupling as an answer using methods described before. In order to study time-reversal, consider the following time-reversed/OTO
counterpart of the spectral function appearing above :
\be
\begin{split}
\rho[3 2_+ 1_+]= \rho[1 2_+ 3_+]+\hbar^2\left(\rho[123]-\rho[321]\right)\ ,
\end{split}
\ee
a relation which can be checked by a simple expansion of the nested commutators/anti-commutators. It follows that the combination of spectral functions with good time-reversal
properties is 
\be
\begin{split}
\rho[1 2_+ 3_+]+\rho[3 2_+ 1_+]= 2\ \rho[1 2_+ 3_+]+\hbar^2\left(\rho[123]-\rho[321]\right)\ .
\end{split}
\ee
With this in mind, we examine the integral 
\be
\begin{split}
  \int_{\mathcal{C}_3} \frac{\rho[1 2_+ 3_+]+\rho[3 2_+ 1_+]}{4\omega_1\omega_3} &=   \int_{\mathcal{C}_3} \frac{2\ \rho[1 2_+ 3_+]+\hbar^2\left(\rho[123]-\rho[321]\right)}{4\omega_1\omega_3}\\
  &=2m_p^3\zeta_N+\frac{\hbar^2}{8}m_p(\overline{\lambda}_3-\overline{\kappa}_3)\ .
\end{split}
\ee
Here, we have used the sum rules quoted before to write a combination of couplings with good time-reversal properties. Microscopic time-reversal
invariance thus implies
\be\boxed{
\begin{split}
2m_p^3\zeta_N+\frac{\hbar^2}{8}m_p(\overline{\lambda}_3-\overline{\kappa}_3)=\eta_{_O}\left\{2m_p^3\zeta_N+\frac{\hbar^2}{8}m_p(\overline{\lambda}_3-\overline{\kappa}_3)\right\}\ ,
\end{split}}
\ee
which is trivially satisfied for $\eta_{_O}=+1$. For $\eta_{_O}=-1$, we get a generalised Onsager condition of the form
\be
\begin{split}
\zeta_N=-\frac{\hbar^2}{16m_p^2}(\overline{\lambda}_3-\overline{\kappa}_3)\ .
\end{split}
\ee
Thus, the non-Gaussianity in thermal noise is quantum suppressed for time-reversal odd coupling to the bath.

At high temperature limit, we get 
\be
\begin{split}
m_p\{\zeta_N +\frac{\hbar^2}{16m_p^2}(\overline{\lambda}_3-\overline{\kappa}_3)\}&= v_{th}^4  \int_{\mathcal{C}_3} \frac{1}{\omega_1\omega_3}
\Big\{\frac{1}{\omega_3\omega_2} \rho[1 2 3]-\Big(\frac{1}{\omega_1\omega_3}+\frac{1}{\omega_3\omega_2}\Big) \rho[3 2 1] \Big\}\\
&=v_{th}^4\int_{\mathcal{C}_3} \frac{1}{\omega_1\omega_3}
\Big\{\frac{1}{\omega_1\omega_2}\rho[321] 
+\frac{1}{\omega_3\omega_2}\rho[123]\Big\} \ .
\end{split}
\ee
where we have used Eq.\eqref{eq:onebyomegasq}. Comparing this against the high temperature limit of $\widehat{\kappa}_{3\gamma}$, we obtain the fluctuation-dissipation relation :
\be\boxed{
\begin{split}
\zeta_N +\frac{\hbar^2}{16m_p^2}(\overline{\lambda}_3-\overline{\kappa}_3)& = \frac{1}{3} \widehat{\kappa}_{3\gamma}v_{th}^2
\ .
\end{split}}
\ee
We note that, in this way of writing both sides of these equations have the same time-reversal property.

\subsection{Summary of relations in non-linear Langevin theory}\label{ssec:summaryNonLin}
We will now summarise all the Onsager type relations between the couplings of the non-linear Langevin theory :
\be
\begin{split}
\overline{\lambda}_3&= \eta_{_O}\ \overline{\kappa}_3\ ,\quad
\overline{\lambda}_{3\gamma} = \eta_{_O}\ \overline{\kappa}_{3\gamma} \ ,\quad
\zeta_\mu= \eta_{_O}\ \zeta_\mu\ ,\quad
\zeta_\gamma = \eta_{_O}\ (\widehat{\kappa}_{3\gamma}-\zeta_\gamma)\ , \quad 
\widehat{\kappa}_{3\gamma} = \eta_{_O}\ \widehat{\kappa}_{3\gamma}\ ,\\
&2m_p^3\zeta_N+\frac{\hbar^2}{8}m_p(\overline{\lambda}_3-\overline{\kappa}_3)=\eta_{_O}\left\{2m_p^3\zeta_N+\frac{\hbar^2}{8}m_p(\overline{\lambda}_3-\overline{\kappa}_3)\right\}\ .
\end{split}
\ee
In addition, we have the consequence of KMS relations :
\be
\begin{split}
\zeta_\mu=\frac{1}{2}v_{th}^2\left(\overline{\lambda}_{3\gamma}+\overline{\kappa}_{3\gamma}\right)\ ,
\qquad \zeta_N  +\frac{\hbar^2}{16m_p^2}(\overline{\lambda}_3-\overline{\kappa}_3)= \frac{1}{3} \widehat{\kappa}_{3\gamma}v_{th}^2\ .
\end{split}
\ee
Note that, in general, each of these conditions relate the time-ordered Langevin couplings to OTO  Langevin couplings. These relations
can be combined to solve for $5$ of the non-linear Langevin couplings in terms of the $3$ remaining couplings
 $\{\overline{\lambda}_3,\overline{\lambda}_{3\gamma},\zeta_\gamma\}$ :
\be\boxed{
\begin{split}
\overline{\kappa}_3&= \eta_{_O}\ \overline{\lambda}_3\ ,\quad
\overline{\kappa}_{3\gamma} = \eta_{_O}\ \overline{\lambda}_{3\gamma} \ ,
\quad \widehat{\kappa}_{3\gamma}  =  \left(1+ \eta_{_O}\right)\zeta_\gamma\ ,\\
\zeta_\mu&=\frac{1}{2}\left(1+ \eta_{_O}\right)\overline{\lambda}_{3\gamma}v_{th}^2\ ,
\qquad \zeta_N  = \frac{1}{3} \left(1+ \eta_{_O}\right)\zeta_\gamma v_{th}^2-\frac{\hbar^2}{16m_p^2}\left(1- \eta_{_O}\right)\overline{\lambda}_3\ .
\end{split}}
\ee
These relations are one of the main results of this work. We collect in Table~\ref{table:couplings3gen}, the Onsager pairs related to each other by time-reversal.

The following tables summarise the integrands  $\mathcal{I}[g]$ associated with each coupling $g$.  The tables~\ref{table:couplings2gen}, \ref{table:couplings2T} and \ref{table:couplings2Thi} 
summarise the integrals for quadratic couplings that are induced, in the leading order in perturbation theory, from a bath in a general time-independent state, by a general thermal bath and by a bath at high temperature
respectively. The couplings are given by expressions of the form 
\be
\begin{split}
g=\int_{\mathcal{C}_2}\mathcal{I}[g] .
\end{split}
\ee
Similarly, the tables~\ref{table:couplings3gen}, \ref{table:couplings3T} and \ref{table:couplings3Thi} summarise the integrals for cubic couplings that are induced, in the leading order in perturbation theory, from a bath in a general time-independent state, by a general thermal bath and by a bath at high temperature
respectively. In this case, the couplings are given by
\be
\begin{split}
g=\int_{\mathcal{C}_3}\mathcal{I}[g] .
\end{split}
\ee

\begin{table}[ht]
\caption{Quadratic Couplings (general environment)}
\label{table:couplings2gen}
\begin{center}
\begin{tabular}{ |c| c|  }
\hline
$g$ & $\mathcal{I}[g]$\\
  \hline
 \hline
 $\Delta m_p$ & $\frac{1}{\omega_1^3}\rho [12]$\\
\hline
$Z_I $ & $-\frac{1}{i m_p^2 \omega_1^3}\rho [12_+]$\\
\hline
$\Delta \bar{\mu}^2$&$-\frac{1}{m_p\omega_1}\rho[12]$\\
\hline
$ \langle f^2\rangle $ & $\frac{1}{im_p^2\omega_1}\rho[12_+]$\\
\hline
$ \gamma$ & $\frac{1}{i m_p\omega_1^2} \rho[12]$\\
\hline
\end{tabular}
\end{center}
\end{table}

\begin{table}[ht]
\caption{Sum rules for Quadratic Couplings (Thermal environment)}
\label{table:couplings2T}
\begin{center}
\begin{tabular}{ |c| c|  }
\hline
$g$ & $\mathcal{I}[g]$\\
  \hline
 \hline
 $\Delta m_p$ & $\frac{1}{\omega_1^3}\rho [12]$\\
\hline
$Z_I $ & $-\frac{\hbar}{i m_p^2 \omega_1^3} (1+2\bose_1)\rho [12]$\\
\hline
$\Delta \bar{\mu}^2$&$-\frac{1}{m_p\omega_1}\rho[12]$\\
\hline
$ \langle f^2\rangle $ & $\frac{\hbar}{im_p^2\omega_1}(1+2\bose_1)\rho [12]$\\
\hline
$ \gamma$ & $\frac{1}{i m_p\omega_1^2} \rho[12]$\\
\hline
\end{tabular}
\end{center}
\end{table}

\begin{table}[ht]
\caption{Sum rules for quadratic Couplings (High Temperature limit)}
\label{table:couplings2Thi}
\begin{center}
\begin{tabular}{ |c| c|  }
\hline
$g$ & $\mathcal{I}[g]$\\
  \hline
 \hline
 $\Delta m_p$ & $\frac{1}{\omega_1^3}\rho [12]$\\
\hline
$Z_I $ & $-\frac{2  v^2_{th}}{ im_p} \frac{1}{\omega_1^4}\rho [12]$\\
\hline
$\Delta \bar{\mu}^2$&$-\frac{1}{m_p\omega_1}\rho[12]$\\
\hline
$ \langle f^2\rangle $ & $\frac{2  v^2_{th}}{i m_p} \frac{1}{\omega_1^2}\rho [12]$\\
\hline
$ \gamma$ & $\frac{1}{i m_p\omega_1^2} \rho[12]$\\
\hline
\end{tabular}
\end{center}
\end{table}

\begin{table}[ht]
\caption{Cubic couplings : doublets/singlets under microscopic time-reversal (general environment)}
\label{table:couplings3gen}
\begin{center}
\begin{tabular}{ |c| c|  }
\hline
$g$ & $\mathcal{I}[g]$\\
  \hline
 \hline
 $\overline{\lambda}_3$ & $\frac{2}{m_p\omega_1\omega_3}\rho [123]$\\
\hline
$\overline{\kappa}_3$ & $\frac{2}{m_p\omega_1\omega_3}\rho [321]$\\
\hline
\hline
$\overline{\lambda}_{3\gamma}$&$\frac{1}{im_p\omega_1\omega_3}\left(\frac{2}{\omega_1}-\frac{1}{\omega_3}\right)\rho[123]$\\
\hline
$ \overline{\kappa}_{3\gamma}$ & $\frac{1}{im_p\omega_1\omega_3}\left(\frac{1}{\omega_1}-\frac{2}{\omega_3}\right)\rho[321]$\\
\hline
\hline
$ \zeta_\gamma$ & $\frac{1}{(2m_p\omega_1\omega_3)^2}\Bigl((2\omega_1-\omega_3)\ \rho[321_+]+(\omega_3-2\omega_1)\ \rho[132_+]+(\omega_1+\omega_3)\ \rho[123_+] \Bigr)$\\
\hline
$\widehat{\kappa}_{3\gamma} - \zeta_\gamma$ & $\frac{1}{(2m_p\omega_1\omega_3)^2}\Bigl((\omega_1+\omega_3)\ \rho[321_+]+(2\omega_3-\omega_1)\ \rho[132_+]+(2\omega_3-\omega_1)\ \rho[123_+] \Bigr)$\\
\hline
\hline
$\widehat{\kappa}_{3\gamma} $ & $\frac{3}{(2m_p\omega_1\omega_3)^2}\Bigl(\omega_1\ \rho[321_+]+(\omega_3-\omega_1)\ \rho[132_+]+\omega_3\ \rho[123_+] \Bigr)$\\
\hline
\hline
 $\zeta_\mu$&$ \frac{1}{2im_p^2\omega_1\omega_3}\Big(\rho[123_+]-\rho [321_+]+ \rho [132_+] \Big)$\\
 \hline
 \hline 
  $ 2\zeta_N+\frac{\hbar^2}{8m_p^2}\left(\overline{\lambda}_3-\overline{\kappa}_3\right)$&$\frac{1}{8m_p^3\omega_1\omega_3}\Bigl(\rho [12_+3_+]+\rho [32_+1_+]\Bigr)$\\
  \hline
 \end{tabular}
\end{center}
\end{table}

\begin{table}[ht]
\caption{Sum rules for Coupling  (Thermal environment)}
\label{table:couplings3T}
\begin{center}
\begin{tabular}{ |c| c|  }
\hline
$g$ & $\mathcal{I}[g]$\\
  \hline
 \hline
 $\overline{\lambda}_3$ & $\frac{2}{m_p\omega_1\omega_3}\rho [123]$\\
\hline
$\overline{\kappa}_3$ & $\frac{2}{m_p\omega_1\omega_3}\rho [321]$\\
\hline
$\overline{\lambda}_{3\gamma}$&$\frac{1}{im_p\omega_1\omega_3}\left(\frac{2}{\omega_1}-\frac{1}{\omega_3}\right)\rho[123]$\\
\hline
$ \overline{\kappa}_{3\gamma}$ & $\frac{1}{im_p\omega_1\omega_3}\left(\frac{1}{\omega_1}-\frac{2}{\omega_3}\right)\rho[321]$\\
\hline
$ \zeta_\gamma$ & $\frac{\hbar}{(2m_p\omega_1\omega_3)^2}\Big\{2(\omega_3-2\omega_1)\Big(1+\bose_1+\bose_2\Big)\rho[321] -(\omega_3-2\omega_1)(1+2\bose_2)\rho[123] \Bigl.$\\
& $\Bigr.\qquad
+\omega_2(1+2\bose_3)\rho[123]\Big\}$\\
\hline
$\widehat{\kappa}_{3\gamma} $ & $ -3\frac{\hbar}{(2m_p\omega_1\omega_3)^2}
\Big(\Big\{2\omega_1\Big(1+\bose_1+\bose_2\Big)-\omega_3(1+2\bose_2)\Big\}\rho[321] \Bigr.$\\
&$\qquad+\Big\{2\omega_3\Big(1+\bose_2+\bose_3\Big)-\omega_1(1+2\bose_2)\Big\}\rho[123]\Big)$\\
\hline
 $\zeta_\mu$&$\frac{\hbar}{im_p^2\omega_1\omega_3}\Big\{(1+\bose_1+\bose_2)\rho[321]-(1+\bose_2+\bose_3)\rho[123]\Big\}\ $\\
 \hline
  $ \zeta_N$&$\frac{\hbar^2}{4m_p^3\omega_1\omega_3}
(1+2\bose_3)\Big\{(1+2 \bose_2) \rho[1 2 3]-2(1+\bose_1+\bose_2) \rho[3 2 1] \Big\}$\\
  \hline
\end{tabular}
\end{center}
\end{table}

\begin{table}[ht]
\caption{Sum rules for Coupling  (High Temperature limit)}
\label{table:couplings3Thi}
\begin{center}
\begin{tabular}{ |c| c|  }
\hline
$g$ & $\mathcal{I}[g]$\\
  \hline
 \hline
 $\overline{\lambda}_3$ & $\frac{2}{m_p\omega_1\omega_3}\rho [123]$\\
\hline
$\overline{\kappa}_3$ & $\frac{2}{m_p\omega_1\omega_3}\rho [321]$\\
\hline
$\overline{\lambda}_{3\gamma}$&$\frac{1}{im_p\omega_1\omega_3}\left(\frac{2}{\omega_1}-\frac{1}{\omega_3}\right)\rho[123]$\\
\hline
$ \overline{\kappa}_{3\gamma}$ & $\frac{1}{im_p\omega_1\omega_3}\left(\frac{1}{\omega_1}-\frac{2}{\omega_3}\right)\rho[321]$\\
\hline
$ \zeta_\gamma$ & $\frac{v_{th}^2}{m_p} \frac{1}{\omega_1\omega_3}\Big\{\left(\frac{1}{\omega_1^2}+\frac{3}{\omega_1\omega_2}\right)\rho[321] +\left(\frac{3}{\omega_3\omega_2}-\frac{1}{\omega_3^2}\right)\rho[123]\Big\}$\\
\hline
$\widehat{\kappa}_{3\gamma} $ & $ 3\frac{v_{th}^2}{m_p}\frac{1}{\omega_1\omega_2\omega_3}
\Big\{\frac{1}{\omega_1}\rho[321] 
+\frac{1}{\omega_3}\rho[123]\Big\}$\\
\hline
 $\zeta_\mu$&$ \frac{v_{th}^2}{m_p}  \frac{1}{i\omega_2}\Big\{\frac{1}{\omega_3^2}\rho[123]-\frac{1}{\omega_1^2}\rho[321]\Big\}  $\\
 \hline
  $ \zeta_N$&$\frac{v_{th}^4}{m_p} \frac{1}{\omega_1\omega_2\omega_3}
\Big\{\frac{1}{\omega_1}\rho[321] 
+\frac{1}{\omega_3}\rho[123]\Big\} $\\
  \hline
\end{tabular}
\end{center}
\end{table}

\section{Introduction to the qXY model}\label{sec:qXY}
In this section, we will begin by describing a microscopic model of an oscillator coupled to bath oscillator degrees of freedom in a way that results in an
effective non-linear Langevin equation for the original oscillator. Our motivation here is to construct a physical microscopic description in which one can check 
the Markovian assumption and the relation between the effective couplings that emerge therein.

\subsection{Model of the bath}
We will now begin with the Caldeira-Leggett model and then modify it to suit our requirements. As described before, the model is that of a single system oscillator (denoted by a degree of freedom $q$)
coupled  to a bath of  oscillators (denoted by  degrees of freedom $X$). One starts with a distribution of couplings and masses of bath oscillators specified by a characteristic
distribution function, defined by 
\begin{equation}\begin{split}
\Big\langle \Big\langle \frac{g_x^2}{m_x} \Big\rangle\Big\rangle &\equiv \sum_i    \frac{g^2_{x,i}}{m_{x,i}} 2\pi\delta(\mu_x-\mu_i)\ .
\end{split}\end{equation}
This distribution function multiplied by the spectral contribution of  each bath oscillator can be summed to give the Caldeira-Leggett
spectral function :
\begin{equation}\begin{split}
\rho[12]_{CL} 
&\equiv \int_0^\infty \frac{d\mu_x}{2\pi}\  2\pi \delta(\omega_1+\omega_2) \times  (2\pi)\ \text{sgn}(\omega_1) \delta(\omega_1^2-\mu_x^2)\ \Big\langle \Big\langle \frac{g_x^2}{m_x}\Big\rangle\Big\rangle\ .
\end{split}\end{equation}
To obtain a Lorentz-Drude spectral function, we consider a continuum of oscillators whose couplings add  up to give a distribution of the form
\begin{equation}
\begin{split}
\Big\langle \Big\langle \frac{g_x^2}{m_x}\Big\rangle\Big\rangle = m_p  \gamma_x  \frac{4\mu_x^2\Omega^2}{\mu_x^2+\Omega^2}\ ,
\end{split}
\end{equation}
where $\gamma_x$ denotes the contribution of X oscillators to the damping constant $\gamma$ of the particle. The contribution to the noise
is determined by fluctutation-dissipation relation \[ \langle f^2\rangle=2\gamma v_{th}^2. \]

To get a simple nonlinear generalisation, we will double the number of oscillators into two kinds of bath both at same temperature. We imagine them to be\
two different sets of bath oscillators distinguished by the letters $X$ and $Y$. For simplicity, we will assume that the $Y$ type oscillators also have similar coupling distribution 
as  $X$ type oscillators 
\begin{equation}
\begin{split}
\Big\langle \Big\langle \frac{g_y^2}{m_y}\Big\rangle\Big\rangle = m_p  \gamma_y  \frac{4\mu_y^2\Omega^2}{\mu_y^2+\Omega^2}\ ,
\end{split}
\end{equation}
and they add to the damping due to  $X$ type oscillators. Thus, $\gamma= \gamma_x+ \gamma_y$ and the noise contributions from the two sets of oscillators also add up. Till now, this is merely a relabelling of the original model, and the model is hence exactly solvable and yields linear Langevin theory.

We will now introduce non-linearity into this theory by introducing a very small 3-body interaction term of the form $qXY$. More precisely, one considers a system of oscillators
with the Lagrangian
\begin{equation}\begin{split} \label{Lagrangian}
L[q, X, Y]=L_B[X, Y]+\frac{1}{2}m_{p0}(\dot{q}^2-\overline{\mu}_0^2 q^2)+ q \left( \sum_i g_{x,i} X_i+\sum_j g_{y,j} Y_j+\sum_{i,j} g_{xy,ij} X_iY_j \right)
\end{split}\end{equation}
where we denote the oscillator's position by $q$ and  $L_B[X, Y]$ is the free Lagrangian of the harmonic bath. The operator that acts on the Hilbert space of the bath and couples to $q$ is 
\begin{equation}\label{eq:Opq}
\mathcal{O}\equiv\sum_i g_{x,i} X_i+\sum_j g_{y,j} Y_j+\sum_{i,j} g_{xy,ij} X_iY_j.
\end{equation}
When we integrate out the effects of the bath degrees of freedom, the effective couplings are now induced for the oscillator and there are corrections to the linear Langevin theory.
These Langevin effective couplings  have their microscopic origin in the thermal correlators/spectral functions of the operator above.

One can think of the above as a toy model for say an atom coupled to photons whereby, apart from the standard, dominant linear dipole coupling responsible for single photon emission/absorption processes, one also has two photon processes involving two photons of two different frequencies. The physics here is familiar one say from Brillouin scattering of photons 
against phonons or the inelastic Raman scattering of photons against molecules. At the level of linear Langevin couplings, both the noise and the damping constant receive
contributions from the inelastic 3-body scattering : first, there is an effect due to two `photon' emission and absorption into/from the thermal bath which gives a contribution 
proportional to 
\begin{equation}\begin{split}
\hbar (\mu_x+\mu_y) \
 \left[(1+\bose_x) (1+\bose_y)-\bose_x \bose_y\right] 
 \end{split}\end{equation}
in the spectral function. The second effect is due to inelastic scattering whose contribution is proportional to 
\begin{equation}\begin{split}
\hbar(\mu_x-\mu_y)  \left[\bose_x(1+\bose_y) -\bose_y(1+\bose_x)\right]\ .
 \end{split}\end{equation}
Putting these effects together, we get a spectral function\footnote{The correction to the spectral function appearing in the last two lines comes from computing the Fourier transform of the thermal commutator 
\be\begin{split} \langle [X(t_1)Y(t_1),X(t_2)Y(t_2)] \rangle\ . \end{split}\ee} 
  \begin{equation}\begin{split}
 \rho[1 2 ]&= (2 \pi) \delta(\omega_1+\omega_2 ) \int_0^\infty \frac{d\mu_x}{2\pi} \Big\langle \Big\langle \frac{g_x^2}{m_x}\Big\rangle\Big\rangle  (2 \pi)\text{sgn} (\omega_1)\delta(\omega_1 ^2-\mu_x^2)  \\
&\qquad +(2 \pi) \delta(\omega_1+\omega_2 ) \int_0^\infty \frac{d\mu_y}{2\pi}\Big\langle \Big\langle \frac{g_y^2}{m_y}\Big\rangle\Big\rangle   (2 \pi)\text{sgn} (\omega_1)\delta(\omega_1 ^2-\mu_y^2) \\
&\qquad + (2 \pi) \delta(\omega_1+\omega_2 ) \int_0^\infty \frac{d\mu_x}{2\pi} \int_0^\infty \frac{d\mu_y}{2\pi} \Big\langle \Big\langle \frac{g_{xy}^2}{m_xm_y}\Big\rangle\Big\rangle  \\
 &\qquad \frac{1}{(2\mu_x)( 2\mu_y)} \Big\{2\hbar (\mu_x+\mu_y) \
 \left[(1+\bose_x) (1+\bose_y)-\bose_x \bose_y\right] 
(2 \pi)  \text{sgn} (\omega_1)\delta\left(\omega_1 ^2-(\mu_x+\mu_y)^2\right)  \\
 &\qquad + 2\hbar (\mu_x-\mu_y)  \left[\bose_x(1+\bose_y) -\bose_y(1+\bose_x)\right] 
(2 \pi)  \text{sgn} (\omega_2 ) \delta\left(\omega_2 ^2-(\mu_x-\mu_y)^2\right)\Big\}\ .
 \end{split}\end{equation}

Here we have introduced the  cubic coupling distribution 
\begin{equation}\begin{split}
\Big\langle \Big\langle \frac{g_{xy}^2}{m_xm_y}\Big\rangle\Big\rangle &\equiv \sum_{ij}    \frac{g^2_{xy,ij}}{m_{x,i}m_{y,j}} 2\pi\delta(\mu_x-\mu_i) 2\pi\delta(\mu_y-\mu_j)\ .
\end{split}\end{equation}
This   cubic coupling distribution should be judiciously chosen so that its dynamics does not destroy the Markovian approximation. We 
find it convenient to choose a distribution of couplings such that 
\begin{equation}\begin{split}\label{eq3}
\Big\langle \Big\langle \frac{g_{xy}^2}{m_xm_y}\Big\rangle\Big\rangle \sim \Big\langle \Big\langle \frac{g_x^2}{m_x} \Big\rangle\Big\rangle \Big\langle \Big\langle \frac{g_y^2}{m_y}\Big\rangle\Big\rangle \ .
\end{split}\end{equation}
More precisely, we take 
\begin{equation}\begin{split}
\Big\langle \Big\langle \frac{g_{xy}^2}{m_xm_y}\Big\rangle\Big\rangle
=\Gamma_{xy}\   \frac{4\mu_x^2\Omega^2}{\mu_x^2+\Omega^2}\   \frac{4\mu_y^2\Omega^2}{\mu_y^2+\Omega^2}\ .
\end{split}\end{equation}
As we will show later, this distribution function is sufficient to give a fast decay of correlator at timescales larger than $\Omega^{-1}$. In the large temperature limit i.e. $\beta\rightarrow 0$, this distribution function gives the following form of $\rho[12]$ upto $\text{O}(\beta^{0})$ :
\begin{equation}
\begin{split}
\rho[12]&=2m_p\Big[\frac{1}{2}\Gamma_{xy}\Omega\ v_{th}^2\frac{4\Omega^2}{\omega_1^2+4\Omega^2}+(\gamma_{x}+\gamma_{y})\frac{\Omega^2}{\omega_1^2+\Omega^2}\Big]\omega_1(2\pi)\delta(\omega_1+\omega_2)\ .
\end{split}
\end{equation} 
Thus, the damping constant acquires a correction of the order $\Gamma_{xy}\Omega\ v_{th}^2$. This is a small correction provided we have 
\[ \Gamma_{xy}\ll\frac{\gamma}{\Omega v_{th}^2}=\frac{m_p\gamma}{\Omega k_BT},\]
a condition we will assume from now on.

We can now turn to the three point spectral functions obtained by Fourier transforming the nested commutators of the bath operator
$\mathcal{O}$ defined in Eq.\eqref{eq:Opq}. This yields
 \begin{equation}\begin{split}
 \rho[1 2 3]=& (2 \pi) \delta(\omega_1+\omega_2+\omega_3 ) \int_0^\infty \frac{d\mu_x}{2\pi} \int_0^\infty \frac{d\mu_y}{2\pi} \Big\langle \Big\langle \frac{g_x g_y g_{xy}}{m_xm_y}\Big\rangle\Big\rangle\\
 & \Big\{ \text{sgn}( \omega_3)  (2 \pi)\delta (\omega_3^2 -\mu_x^2) \
 \Big[\text{sgn}( \omega_2)  (2 \pi)\delta (\omega_2^2 -\mu_y^2)-\text{sgn}( \omega_1)  (2 \pi)\delta (\omega_1^2 -\mu_y^2)\Big]\\
 &+\text{sgn}( \omega_3)  (2 \pi)\delta (\omega_3^2 -\mu_y^2)\Big[\text{sgn}( \omega_2)  (2 \pi)\delta (\omega_2^2 -\mu_x^2)
 -\text{sgn}( \omega_1)  (2 \pi)\delta (\omega_1^2 -\mu_x^2) \Big]\Big\}.
 \end{split}\end{equation}
and
\begin{equation}\begin{split}
 \rho[3 2 1]=& (2 \pi) \delta(\omega_1+\omega_2+\omega_3 ) \int_0^\infty \frac{d\mu_x}{2\pi} \int_0^\infty \frac{d\mu_y}{2\pi}  \Big\langle \Big\langle \frac{g_x g_y g_{xy}}{m_xm_y}\Big\rangle\Big\rangle\\
 & \Big\{ \text{sgn}( \omega_1)  (2 \pi)\delta (\omega_1^2 -\mu_x^2) \
 \Big[\text{sgn}( \omega_2)  (2 \pi)\delta (\omega_2^2 -\mu_y^2)-\text{sgn}( \omega_3)  (2 \pi)\delta (\omega_3^2 -\mu_y^2)\Big]\\
 &+\text{sgn}( \omega_1)  (2 \pi)\delta (\omega_1^2 -\mu_y^2)\Big[\text{sgn}( \omega_2)  (2 \pi)\delta (\omega_2^2 -\mu_x^2)
 -\text{sgn}( \omega_3)  (2 \pi)\delta (\omega_3^2 -\mu_x^2) \Big]\Big\},
 \end{split}\end{equation}
 where we have  defined the distribution function 
 \begin{equation}\begin{split}
 \Big\langle \Big\langle \frac{g_x g_y g_{xy}}{m_xm_y}\Big\rangle\Big\rangle &\equiv \sum_{ij}    \frac{g_{x,i} g_{y,j}g_{xy,ij}}{m_{x,i}m_{y,j}} 2\pi\delta(\mu_x-\mu_i) 2\pi\delta(\mu_y-\mu_j)\ .
\end{split}\end{equation}
 We will find it convenient to assume
 \begin{equation}\begin{split}
| \Big\langle \Big\langle \frac{g_x g_y g_{xy}}{m_xm_y}\Big\rangle\Big\rangle |^2
\sim \Big\langle \Big\langle \frac{g_{xy}^2}{m_xm_y}\Big\rangle\Big\rangle \Big\langle \Big\langle \frac{g_x^2}{m_x} \Big\rangle\Big\rangle \Big\langle \Big\langle \frac{g_y^2}{m_y}\Big\rangle\Big\rangle\ .
\end{split}\end{equation}
More precisely, we take 
\begin{equation}\begin{split}
 \Big\langle \Big\langle \frac{g_x g_y g_{xy}}{m_xm_y}\Big\rangle\Big\rangle
= \frac{m_p}{4}\Gamma_3 \   \frac{4\mu_x^2\Omega^2}{\mu_x^2+\Omega^2}\   \frac{4\mu_y^2\Omega^2}{\mu_y^2+\Omega^2}.
\end{split}\end{equation}
Here, the parameter $\Gamma_3$ can roughly be thought of as an inverse penetration depth for the three body scattering. This induces a small correction to the usual
Langevin dynamics if 
\[ \Gamma_3\ll\frac{\gamma}{v_{th}}=\frac{m_p\gamma}{k_BT},\]
a condition we will assume from now on. In the next  subsection, we will  examine the correlation functions of the bath and to what extent they  justify a Markovian appproximation.

\subsection{KMS relations and decay of bath correlations }

Consider the model described in the above section of a single oscillator coupled to two harmonic baths. In this subsection, we will be interested in the evolution starting from 
an initial time $t_0$ and whether and how the bath correlators decay with time. 

We will assume that the bath and the oscillator are unentangled at an initial time $t_0$.
Therefore, the initial density matrix of the oscillator and the bath is given by
\begin{equation}
\rho(t_0)= \frac{e^{-\frac{H_B}{k_B T}}}{Z_B}\otimes \rho_\text{p}\ ,
\end{equation}
where $H_B$ is the Hamiltonian of the bath,  $Z_B$ is its partition function, and $\rho_\text{p}$ is the initial density matrix of the oscillator at time $t_0$.

 The effective theory of the Brownian particle is obtained after integrating out the degrees of freedom of the thermal bath. In the process, the bath correlation functions (in general out-of-time-order) imprint themselves on the effective couplings of the Brownian particle.  In this section we are interested in studying the out-of-time-order bath correlators of the operator $\mathcal{O}$. We will use these correlators later to determine the effective couplings of the particle.

 Since the bath is in a thermal state, not all the Wightman correlators of $\mathcal{O}$ are independent. The bath correlators that are related to each other by cyclic permutations of insertions, satisfy the KMS (Kubo-Martin-Schwinger) conditions \cite{Kubo:1957mj, Martin:1959jp, Chou:1984es, Haag:1967sg, Hou:1998yc, Wang:1998wg, Weldon:2005nr, Evans:1991ky,Guerin:1993ik, Tsuji:2016kep, Haehl:2017eob}. 
For a thermal $n$-point function of $\mathcal{O}(t)$, the KMS condition in time domain gives the following  condition on the connected parts (cumulants) of the bath correlators :
\begin{equation}\begin{split}
\langle \mathcal{O}(t_1) \mathcal{O}(t_2)...\mathcal{O}(t_n)\rangle_c=\langle \mathcal{O}(t_n-i\beta) \mathcal{O}(t_1) \mathcal{O}(t_2)...\mathcal{O}(t_{n-1})\rangle_c.
\end{split}\end{equation}
In frequency space the KMS condition simplifies to
\begin{equation}\begin{split}
\langle \mathcal{O}(\omega_1) \mathcal{O}(\omega_2)... \mathcal{O}(\omega_n)\rangle_c=e^{-\beta \omega_n}\langle  \mathcal{O}(\omega_n)\mathcal{O}(\omega_1)...\mathcal{O}(\omega_{n-1})\rangle_c.
\end{split}\end{equation}
This follows straightforwardly from the fact that the frequency domain analogue of $\mathcal{O}(t_n-i\beta)$ is $e^{-\beta \omega_n} \mathcal{O}(\omega_n)$.

 At the level of the two point function, the statement implies that there is only one independent two-point correlator of $\mathcal{O}$. We choose that to be $\rho[12]$. 
Then, using the KMS relations, the expectation value of the anticommutator is given by
 \begin{equation}\begin{split} \label{KMS1}
\rho [12_+]=-\hbar\left( 1+2\bose_2 \right)\rho[12]\  ,
\end{split}\end{equation}
where  $\bose_1=\frac{1}{e^{\beta \omega_1}-1}$ and $\bose_2=\frac{1}{e^{\beta \omega_2}-1}$ are the Bose-Einstein distribution functions. 

Similarly all the three point functions of $\mathcal{O}$ are determined by $\rho[1 2 3]$ and $\rho[3 2 1]$.
The other 3-point correlators are related to the two by the following KMS relations \cite{Haehl:2017eob}
 \begin{equation}\begin{split}
 \rho[1 2 3_+]=& -\hbar(1+2\bose_3) \rho[1 2 3]\ ,\\
  \rho[3 2 1_+]=& -\hbar(1+2\bose_1) \rho[3 2 1]\ ,\\
  \rho[1 2_+ 3]=& -\hbar(1+2\bose_2) \rho[1 2 3]+2(1+\bose_1+\bose_2) \rho[3 2 1]\rangle\ ,\\
   \rho[1 2_+ 3_+]=& \hbar^2(1+2\bose_3)\Big[(1+2 \bose_2) \rho[1 2 3]-2(1+\bose_1+\bose_2) \rho[3 2 1] \Big].
 \end{split}\end{equation}
 Hence the spectral functions $\rho[12]$, $ \rho[1 2 3]$ and  $\rho[3 2 1]$ are sufficient to determine all two-point and three-point bath correlators.

For the two-point function the correlator of the anticommutator in equation \eqref{KMS1} provides a measure of the thermal noise arising from the thermal fluctuations in the bath whereas the correlation function of the commutator in that equation gives a measure of the dissipation/damping in the bath due to the motion of the Brownian particle \cite{Feynman:1963fq}. We denote the connected part of nested commutators of operators in time domain by a tilde in the following :
\begin{equation}\begin{split}
\widetilde {\langle [123] \rangle} &\equiv\langle [[O(t_{1}),O(t_{2})],O(t_{3})] \rangle_c\ . \\
\end{split}\end{equation}
For the cumulants of the nested anticommutators we use a similar notation with an extra `$+$' sign inside the square bracket indicating the position of the anticommutator as follows 
\begin{equation}\begin{split}
\widetilde {\langle [12_+3] \rangle} &\equiv\langle [\{O(t_{1}),O(t_{2})\},O(t_{3})] \rangle_c\ ,\\
\widetilde {\langle [321_+] \rangle} &\equiv\langle \{[O(t_{3}),O(t_{2})],O(t_{1})\} \rangle_c\ .
\end{split}\end{equation}

We can use the forms of the spectral functions to get the bath correlators in time domain. The bath correlators decay with increase in separation between any two insertions.
In the following, we provide the two-point and three-point cumulants for the slowest decaying modes with frequency $\Omega$. 

For our model, the two point  cumulants decay as 
\begin{equation}
\begin{split}
\frac{i}{\hbar}\langle\widetilde{[12]}\rangle&=  \Omega^2\Big[m_p  (\gamma_x+\gamma_y) \exp(-\Omega t_{12}) +\hbar\Gamma_{xy}\Omega^2 \exp(-2 \Omega t_{12})\cot\left(\frac{\beta\Omega}{2}\right)\Big]\ ,\\ 
\langle\widetilde{[12_+]}\rangle&= \frac{\Omega^2}{2}\Big[ (\gamma_x+\gamma_y) \hbar\csc\left(\frac{\beta\Omega}{2}\right) \exp(-\Omega t_{12})  
 + \hbar^2\Gamma_{xy}\Omega^2 \left(\cot^2\left(\frac{\beta\Omega}{2}\right) -1\right) \exp(-2 \Omega t_{12})\Big].
 \end{split}
\end{equation} 
In the high temperature limit, this yields
\begin{equation}
\begin{split}
\frac{i}{\hbar}\langle\widetilde{[12]}\rangle&=  m_p\Omega^2\Big[  (\gamma_x+\gamma_y) \exp(-\Omega t_{12}) +2\Gamma_{xy}\Omega v_{th}^2 \exp(-2 \Omega t_{12})\Big]\ ,\\ 
\langle\widetilde{[12_+]}\rangle&= m_p^2v_{th}^2\Omega\Big[ (\gamma_x+\gamma_y) \exp(-\Omega t_{12})  
 +2\Gamma_{xy}\Omega v_{th}^2 \exp(-2 \Omega t_{12})\Big].
 \end{split}
\end{equation} 
The three point nested  cumulants are given by 
\begin{equation}
\begin{split}
\frac{i^2}{\hbar^2}\langle[\widetilde{321}]\rangle = & m_p \frac{\Gamma_3\Omega^4 }{2} \exp(-\Omega t_{13}) \bigg(1+
 \exp(-\Omega t_{23})\bigg)\ , \\
\frac{i^2}{\hbar^2}\langle[\widetilde{123}]\rangle=& m_p \frac{\Gamma_3\Omega^4 }{2} \exp(-\Omega t_{13}) \bigg(1+
 \exp(-\Omega t_{12})\bigg)\ .
 \end{split}
\end{equation}
 The other cumulants can also be computed to yield
 \begin{equation}
\begin{split}
\frac{i}{\hbar} \langle[\widetilde{{321}_+}]\rangle =&- m_p\Gamma_3\Omega^4 \frac{\hbar}{2}\cot\left(\frac{\beta\Omega}{2}\right) \exp(-\Omega t_{13})\bigg(1+
 \exp(-\Omega t_{23})\bigg)\ , \\
\frac{i}{\hbar} \langle[\widetilde{{123}_+}]\rangle   =& m_p\Gamma_3\Omega^4 \frac{\hbar}{2}\cot\left(\frac{\beta\Omega}{2}\right) \exp(-\Omega t_{13}) \bigg(1+
 \exp(-\Omega t_{13})\bigg)\ , \\
\frac{i}{\hbar} \langle[\widetilde{{12_+3}}]\rangle = & m_p\Gamma_3\Omega^4\frac{\hbar}{2}\cot\left(\frac{\beta\Omega}{2}\right) \exp(-\Omega t_{13})\bigg( 1+2 \exp(-\Omega t_{23})
+\exp(-\Omega t_{12})\bigg)\ 
 \end{split}
\end{equation}
and
   \begin{equation}
\begin{split}
 \langle[\widetilde{{12_+3_+}}]\rangle 
= &\hbar^2 m_p \frac{\Gamma_3\Omega^4}{2} \exp(-\Omega t_{13})\Big[\cot^2\left(\frac{\beta\Omega}{2} \right)\left\{ 1+\exp(-\Omega t_{12})+\exp(-\Omega t_{23}) \right\}-\exp(-\Omega t_{23}) \Big] \ .
\end{split}
\end{equation}
In the high temperature limit, this yields
 \begin{equation}
\begin{split}
\frac{i}{\hbar} \langle[\widetilde{{321}_+}]\rangle =&- \Gamma_3\ m_p^2 v_{th}^2\Omega^3  \exp(-\Omega t_{13})\bigg(1+
 \exp(-\Omega t_{23})\bigg)\ , \\
\frac{i}{\hbar} \langle[\widetilde{{123}_+}]\rangle   =& \Gamma_3\ m_p^2 v_{th}^2\Omega^3 \exp(-\Omega t_{13}) \bigg(1+
 \exp(-\Omega t_{13})\bigg)\ , \\
\frac{i}{\hbar} \langle[\widetilde{{12_+3}}]\rangle = & \Gamma_3\ m_p^2 v_{th}^2\Omega^3 \exp(-\Omega t_{13})\bigg( 1+2 \exp(-\Omega t_{23})
+\exp(-\Omega t_{12})\bigg)\ 
 \end{split}
\end{equation}
and
   \begin{equation}
\begin{split}
 \langle[\widetilde{{12_+3_+}}]\rangle 
= &2\Gamma_3\ m_p^3v_{th}^4\Omega^4 \exp(-\Omega t_{13})\left( 1+\exp(-\Omega t_{12})+\exp(-\Omega t_{23}) \right)\ .
\end{split}
\end{equation}

Thus, given the decay of the memory in the bath at time-scales much larger than $(\frac{1}{\Omega})$ we expect to obtain  a local effective theory for the particle at long time-scales.
In the next section, we will describe how such an effective theory be obtained starting from the microscopic description.

\section{Effective theory of the Brownian particle}\label{sec:qXYtoLangevin}

\subsection{OTO Influence Phase}
We will begin by describing the procedure we employ to derive the effective theory for the particle at long time-scales, closely following \cite{Chaudhuri:2018ihk}. Our aim is to obtain the couplings in the effective action of the Brownian particle (after systematically integrating out the degrees of freedom of the bath) in a way that keeps track of  out-of-time order correlations. Thus, we are interested in an effective theory which has sufficient number of effective couplings that can compute along with the time-ordered correlators, also the out-of-time order correlators (OTOCs) of the particle. 

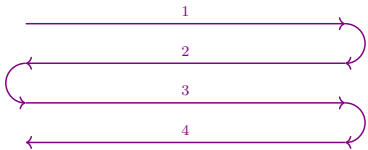
\begin{figure}[h!]
\centering
\begin{center}
\caption{A contour with 2 time-folds}
\scalebox{0.7}{\begin{tikzpicture}[scale = 1.5]
\draw[thick,color=violet,->] (-2,0 cm) -- (2,0 cm) node[midway,above] {\scriptsize{1}} ;
\draw[thick,color=violet,->] (2, 0 cm - 0.5 cm) -- (-2, 0 cm - 0.5 cm)node[midway,above] {\scriptsize{2}}  ;
\draw[thick,color=violet,->] (2, 0 cm) arc (90:-90:0.25);
\draw[thick,color=violet,->] (-2,-1.0 cm) -- (2,-1.0 cm) node[midway,above] {\scriptsize{3}} ;
\draw[thick,color=violet,->] (2, -1.0 cm - 0.5 cm) -- (-2, -1.0 cm - 0.5 cm) node[midway,above] {\scriptsize{4}} ;
\draw[thick,color=violet,->] (2, -1.0 cm) arc (90:-90:0.25);
\draw[thick,color=violet,->] (-2,-1.0 cm + 0.5 cm) arc (90:270:0.25);
\end{tikzpicture}}
 \label{fig:contour}
 \end{center}
\end{figure}

We use the generalised Schwinger-Keldysh path integral formalism to arrive at an effective action/generalised influence phase for the Brownian particle. The path integral representation of $2$-OTO correlators (correlators with two insertions whose immediate neighbours lie to their pasts)
\footnote{E.g. for $t_1> t_2> t_3>t_0, \langle \mathcal{O}(t_1) \mathcal{O}(t_3)\mathcal{O}(t_2)\rangle=Tr\left(\rho(t_0) \mathcal{O}(t_1) \mathcal{O}(t_3)\mathcal{O}(t_2) \right)$ has $\mathcal{O}(t_1)$ and $\mathcal{O}(t_2)$ whose immediate neighbours lie to their pasts. Hence this is an example of a $2$-OTO correlator.} is then defined on a contour with two time-folds as shown in figure \ref{fig:contour}. 
There are four copies of the degrees of freedom of the particle and bath 
$\{q_1,X_{i1}, Y_{j1}\},\{q_2,X_{i2}, Y_{j2}\},\{q_3,X_{i3}, Y_{j3}\}$ and $\{q_4,X_{i4}, Y_{j4}\}$ living on the four legs of the contour. 
The action that enters in the path integral is
\begin{equation}
\begin{split}
S_{\text{2-fold}}=\int_{t_0}^T dt &\Big\{L[q_1,X_{i1}, Y_{j1}]-L[-q_2,X_{i2}, Y_{j2}]
+L[q_3,X_{i3}, Y_{j3}]-L[-q_4,X_{i4}, Y_{j4}]\Big\}.
\end{split}
\end{equation}
The degrees of freedom of the particle are identified at the turning points of the contour while performing the path integral.
After integrating out the degrees of freedom of the bath, one obtains an out-of-time-order generalised influence phase $W$ for the particle which can be expanded in powers of  the particle-bath coupling:
\begin{equation}\label{eq:genInf}
W= W_1+W_2+ W_3+\ldots\ \ \ \ .
\ee
The $n$-th term in this perturbative expansion is given by
\be
\begin{split}
W_n=\frac{i^{n-1}}{n!\hbar^{n-1}}   \int_{t_0}^T dt_1\cdots \int_{t_0}^{T} dt_n
\sum_{i_1,\cdots,i_n=1}^4 \langle \mathcal{T}_C \mathcal{O}_{i_1}(t_1)\cdots \mathcal{O}_{i_n}(t_n)\rangle_c\ q_{i_1}(t_1) \cdots q_{i_n}(t_n)
\end{split}
\label{infphase}
\end{equation}
Here the subscripts denote the contour legs and the expectation values are contour ordered cumulants computed in the initial state of the bath.

\subsection{Markovian approximation and effective action}
The correlators of the Brownian particle calculated from the generalised influence phase can be obtained from a 1-PI  effective action. The 1-PI effective action is generally non-unitary and non-local.
However since in our model the cumulants of $\mathcal{O}(t)$ decay sufficiently fast, with an increase in separation between the insertions compared to the natural time scale of the particle, we can work in a Markovian limit \cite{BRE02}.
In this limit we get a local, non-unitary 1-PI effective action.
We assume that in the action the terms  with two or more time derivatives on q are negligible.
Such a  local 1-PI effective Lagrangian for the Brownian particle has the following form \cite{Avinash:2017asn,Chaudhuri:2018ihk}
\begin{equation}
L_{\text{1PI}}=L_{\text{1PI}}^{(1)}+L_{\text{1PI}}^{(2)}+L_{\text{1PI}}^{(3)}+\cdots
\end{equation}
where the $L_{\text{1PI}}^{(1)}$, $L_{\text{1PI}}^{(2)}$ and $L_{\text{1PI}}^{(3)}$ correspond to terms that are linear, quadratic and cubic in q's respectively. 
The linear term is given by
\begin{equation}
L_{\text{1PI}}^{(1)}= \widehat{F}(q_1+q_2+q_3+q_4),
\label{1PIeff1234:linear}
\end{equation}
The quadratic term is given by
\begin{equation}
\begin{split}
L_{\text{1PI}}^{(2)}&=\frac{1}{2} Z (\dot{q}_1^2+\dot{q}_3^2)-\frac{1}{2} Z^* (\dot{q}_2^2+\dot{q}_4^2)+i\ Z_\Delta\sum_{i<j}\dot{q}_i \dot{q}_j\\
&\quad-\frac{m^2}{2}(q_1^2+q_3^2)+\frac{(m^2)^*}{2}(q_2^2+q_4^2)\\
&\quad-i m_{\Delta}^2 \sum_{i<j}q_i q_j+\frac{\widehat{\gamma}}{2} \sum_{i<j} ({q_i} \dot{q}_j-\dot{q_i} {q_j}),
\end{split}
\label{1PIeff1234:quadratic}
\end{equation}
The cubic term $L_{\text{1PI}}^{(3)}$ can be split into 2 parts:
One part, which  reduces to the cubic terms in the Schwinger-Keldysh effective theory under identification of the degrees of freedom on any two successive legs, is given by 
\begin{equation}
\begin{split}
L_{\text{1PI,SK}}^{(3)}&=-\frac{\lambda_3}{3!} (q_1^3+q_3^3)-\frac{\lambda_3^*}{3!} (q_2^3+q_4^3)\\
&\quad+\frac{\sigma_3}{2!}\Big[q_1^2(q_2+q_3+ q_4)- q_2^2 (q_3+q_4)+q_3^2  q_4\Big]\\
& \quad+\frac{\sigma_3^*}{2!}\Big[q_1(q_2^2-q_3 ^2+q_4^2)-q_2(q_3^2-q_4^2)+q_3 q_4^2\Big]\\
&\quad+\frac{\sigma_{3 \gamma}}{2!} \Big[q_1^2 (\dot{q}_2+\dot{q}_3+\dot{q}_4)-q_2^2(\dot{q}_3+\dot{q}_4)+q_3^2 \dot{q}_4-(q_2^2\dot{q}_2+q_4^2\dot{q}_4)\Big]\\
&\quad+\frac{\sigma_{3 \gamma}^*}{2!} \Big[\dot{q}_1 ({q}_2^2-{q}_3^2+{q}_4^2)-\dot{q}_2({q}_3^2-{q}_4^2)+\dot{q}_3 {q}_4^2-(q_1^2\dot{q}_1+q_3^2\dot{q}_3)\Big],
\end{split}
\label{1PIeff1234:cubicSK}
\end{equation}
The other part, which  vanishes under such identifications, is given by
\begin{equation}
\begin{split}
L_{\text{1PI,2-OTO}}^{(3)}
&=-\Big(\kappa_3+\frac{1}{2} \text{Re}[\lambda_3-\sigma_3]\Big) (q_1+q_2)(q_2+q_3)(q_3+q_4) \\
&\quad-({q}_2+{q}_3) \Big[\Big(\kappa_{3\gamma}- \text{Re}[\sigma_{3 \gamma}]\Big) (\dot{q}_1+\dot{q}_2) ({q}_3+{q}_4) \\
&\qquad\qquad\qquad+\Big(\kappa_{3\gamma}^*- \text{Re}[\sigma_{3 \gamma}]\Big) ({q}_1+{q}_2) (\dot{q}_3+\dot{q}_4)\Big].
\end{split}
\label{1PIeff1234:cubicOTO}
\end{equation}
The effective action constructed from this Lagrangian has to satisfy certain conditions as elaborated in the appendix \ref{a1}. The conditions arise due to the fact that the Brownian particle and the bath together comprise a closed unitary quantum system. 

\subsection{Comparison with the Nonlinear Langevin system}
We will now relate the description in terms of generalised influence phase and generalised Schwinger Keldysh effective action to the classical stochastic description in
terms of non-linear Langevin equation. This relation, familiar in the stochastic quantisation literature, can be studied at various levels : we can for example, study the correlators predicted by the two descriptions and match them against each other. We will instead derive the dictionary between the quantum and stochastic descriptions by deriving 
a path integral which generates the non-linear Langevin correlators and then matching its terms against the  influence phase obtained by ignoring out of time ordered contributions.

We are interested in the non-linear Langevin theory described by the  stochastic equation :
\begin{equation}\begin{split}
\mathcal{E}[q]\equiv \frac{d^2q}{dt^2}+(\gamma+\zeta_\gamma \mathcal{N})  \frac{dq}{dt} + (\bar{\mu}^2+\zeta_{\mu} \mathcal{N})\ q +\left(\overline{\lambda}_3-\overline{\lambda}_{3\gamma}\ \frac{d}{dt}\right)\frac{q^2}{2!} -F =\langle f^2\rangle \mathcal{N}\ \ . 
  \end{split}
\end{equation} 
Here, we will take $\mathcal{N}$ to be a random noise drawn from the non-Gaussian probability distribution
\begin{equation}
\begin{split} 
P[\mathcal{N}] \propto \exp\left\{-\frac{1}{2 \langle f^2\rangle}\int dt\ \Bigl(\langle f^2\rangle \mathcal{N}-\zeta_N \mathcal{N}^2\Bigr)^2-\frac{1}{2}Z_I\int dt\ \dot{\mathcal{N}}^2\right\}\ . 
  \end{split}
\end{equation} 
We will assume that the corrections to the Langevin equation are small : this is equivalent to assuming the  parameters
$\{\zeta_\gamma,\zeta_{\mu} ,\overline{\lambda}_3,\overline{\lambda}_{3\gamma},\zeta_N,Z_I\}$ are small. 

The equation above is a non-linear stochastic ODE with \emph{multiplicative noise}, i.e., the noise variable $\mathcal{N}$ appears
in the equation multiplied by the functions of the fundamental stochastic variable $q$ of the differential equation. In the theory of stochastic ODEs,
such ODEs need a definite prescription for equal time stochastic products to be well-defined. In this work, we will adopt 
a time-symmetric (or Stratonovich) prescription for equal time stochastic products. But we will only need leading order 
corrections due to the multiplicative noise terms, where the subtleties regarding various prescriptions for stochastic products will not matter.

 To study this non-linear Langevin  theory in the context of path integrals, one can employ the following method (often attributed to Martin-Siggia-Rose\cite{1973PhRvA...8..423M}-De Dominicis-Peliti\cite{1978PhRvB..18..353D}-Janssen\cite{1976ZPhyB..23..377J}) :
We start by thinking about the functional integral\footnote{In this integral, we ignore the Jacobian $\det\Big[\frac{\delta \mathcal{E}[q(t)]}{\delta q(t^\prime)} \Big]$ as it does not correct the coefficients of the terms obtained in \eqref{SK-NLD duality} up to leading order in the particle-bath coupling.} over noise realisations along with the imposition of the non-linear Langevin equation on a
variable $q_a(t)$:
\begin{equation}
\begin{split}
 & \int [dq_a][d\mathcal{N}]\   P[\mathcal{N}]\ \delta \Bigl[ \langle f^2\rangle \mathcal{N} -\mathcal{E}[q_a]\Bigr]\\
 &=\int  [dq_a][dq_d]  [d\mathcal{N}] \ P[\mathcal{N}]\ \exp\left\{i \frac{m_p}{\hbar}\int dt\ q_d \Bigl[ \langle f^2\rangle \mathcal{N} -\mathcal{E}[q_a]\Bigr]\right\}\ ,
 \end{split}
 \end{equation} 
where we have given the standard  functional integral representation of the delta function.
We can now discretise the noise integral, add appropriate counterterms and perform the path integral perturbatively in the small parameters $\{\zeta_\gamma,\zeta_{\mu} ,\overline{\lambda}_3,\overline{\lambda}_{3\gamma},\zeta_N,Z_I\}$.  This exercise yields
\begin{equation}
\begin{split}
&\lim_{\delta t\to 0}  \int [d\mathcal{N}]  e^{-i\frac{m_p}{\hbar}\frac{3\zeta_N}{\langle f^2\rangle \delta t}\int dt\ q_d}\ P[\mathcal{N}] \exp\left\{i \frac{m_p}{\hbar}\int dt\ q_d \Bigl[ \langle f^2\rangle \mathcal{N} -\mathcal{E}[q_a]\Bigr]\right\}\\
&\approx  \exp\left\{\frac{i}{\hbar}\int dt\ \Bigl[ \frac{i}{2} \frac{m_p^2}{\hbar}\langle f^2\rangle q_d^2 - \frac{i}{2} \frac{m_p^2}{\hbar}Z_I \dot{q}_d^2 -\frac{m_p^3}{\hbar^2}\zeta_N q_d^3 -m_p q_d\mathcal{E}[q_a]_{\mathcal{N}=0}-i\frac{m_p^2}{\hbar}q_d^2 \frac{\partial \mathcal{E}[q_a]}{\partial \mathcal{N}}\Bigr]\right\}\ .
 \end{split}
 \label{SK-NLD duality}
 \end{equation} 
This MSRDPJ effective action can be connected to Schwinger Keldysh effective action by identifying  $q_d=q_1+q_2$, $q_a= \frac{1}{2}(q_1-q_2)$. We will 
refer the reader to \cite{kamenev_2011} for a textbook level discussion of why this is the correct identification that maps the Schwinger Keldysh boundary conditions
on the quantum side to the causal boundary conditions of the stochastic path integral. Using this map, we can write the above effective action in the form :
\begin{equation}
\begin{split}
L_{\text{1PI,SK}}^{(1)}&= \widehat{F}(q_1+q_2),\\
L_{\text{1PI,SK}}^{(2)}&=\frac{1}{2} Z \dot{q}_1^2-\frac{1}{2} Z^* \dot{q}_2^2+i\ Z_\Delta\dot{q}_1\dot{q}_2-\frac{m^2}{2}q_1^2+\frac{(m^2)^*}{2}q_2^2-i m_{\Delta}^2 q_1 q_2+\frac{\widehat{\gamma}}{2} ({q_1} \dot{q}_2-\dot{q_1} {q_2}),\\
L_{\text{1PI,SK}}^{(3)}&=-\frac{\lambda_3}{3!} q_1^3-\frac{\lambda_3^*}{3!} q_2^3+\frac{\sigma_3}{2!}q_1^2q_2+\frac{\sigma_3^*}{2!}q_1q_2^2+\frac{\sigma_{3 \gamma}}{2!} q_1^2 \dot{q}_2+\frac{\sigma_{3 \gamma}^*}{2!} \dot{q}_1 {q}_2^2\ ,
\end{split}
\end{equation}
where the  Schwinger Keldysh effective couplings are given in terms of Langevin couplings via
  \begin{equation}
\begin{split}
 \widehat{F} &\equiv m_p F\ ,\qquad
\widehat{\gamma} \equiv m_p \gamma\ ,\\
 Z &\equiv m_p-i\frac{m_p^2}{\hbar}Z_I\ ,\qquad
 Z_\Delta \equiv -\frac{m_p^2}{\hbar}Z_I\ ,\\
 m^2 &\equiv m_p\bar{\mu}^2-i\frac{m_p^2}{\hbar}\langle f^2\rangle\ ,\qquad
 m_\Delta^2\equiv -\frac{m_p^2}{\hbar}\langle f^2\rangle\ ,\\
 \lambda_3 &\equiv \frac{3}{4}m_p\overline{\lambda}_3 +6\frac{m_p^3}{\hbar^2} \zeta_N+3i \frac{m_p^2}{\hbar} \zeta_{\mu}\ , \\
\sigma_3 &\equiv \frac{1}{4}m_p\overline{\lambda}_3 -6\frac{m_p^3}{\hbar^2}\zeta_N-i \frac{m_p^2}{\hbar}\zeta_{\mu}\ , \\
\sigma_{3\gamma} &\equiv -\frac{1}{2}m_p\overline{\lambda}_{3\gamma}+2i \frac{m_p^2}{\hbar} \zeta_{\gamma}\ .
\end{split}
\end{equation}
We recognise the above form as the most general Schwinger Keldysh effective action obtained by collapsing two successive time contours
as mentioned in the previous subsection. As will be described elsewhere\cite{SC}, one can extend the nonlinear Langevin theory
to an `out of time order' stochastic theory which can capture all the couplings of the generalised Schwinger Keldysh effective action. For now, we 
will content ourselves with matching the out of time-ordered couplings by looking at the OTOCs of the system. This 
yields a map
\begin{equation}
\begin{split}
\kappa_{3} &\equiv -\frac{1}{2}m_p\overline{\kappa}_{3}\ ,\\
\kappa_{3\gamma} &\equiv -\frac{1}{2}m_p\overline{\kappa}_{3\gamma}+\frac{2}{3}i \frac{m_p^2}{\hbar}\widehat{\kappa}_{3\gamma}\ .
\end{split}
\end{equation}

The couplings in the 2-OTO effective theory, and hence the non-linear Langevin couplings are then determined by the bath correlators which enter into the generalised
influence phase Eq.\eqref{eq:genInf}.  Such relations between the couplings and the correlators can be obtained by computing the particle's correlators 
with generalised influence phase given in Eq.\eqref{eq:genInf} and then comparing them with the results obtained from the effective theory. The expressions of the
influence phase couplings in terms of correlators in the time domain were given in \cite{Chaudhuri:2018ihk} and in frequency domain, they take the forms summarised earlier 
in section\S\ref{sec:nonLinearLangevin}. In the following subsection, we will use the expressions quoted in  section\S\ref{sec:nonLinearLangevin} to compute explicitly the effective couplings for the qXY model.

\subsection{Influence couplings in the qXY model }
We will now describe the computation of non-linear Langevin couplings starting from the spectral functions of the qXY model.
In our model, quadratic spectral function is given at high temperatures by the expression
\begin{equation}
\begin{split}
\rho[12]&=2m_p\Big[(\gamma_{x}+\gamma_{y})\frac{\Omega^2}{\omega_1^2+\Omega^2}+\frac{1}{2}\Gamma_{xy}\Omega\ v_{th}^2\frac{4\Omega^2}{\omega_1^2+4\Omega^2}\Big]\omega_1(2\pi)\delta(\omega_1+\omega_2)\ .
\end{split}
\end{equation} 
This two point spectral function obeys the following integral identities
\be
\begin{split}
\int_{\mathcal{C}_2}\frac{\rho[12]}{\omega_1}&=m_p\Omega(\gamma_x+\gamma_y+\Gamma_{xy}\Omega v_{\text{th}}^2)\ ,\\
\int_{\mathcal{C}_2}\frac{\rho[12]}{i\omega_1^2}&=m_p(\gamma_x+\gamma_y+\Gamma_{xy}\Omega v_{\text{th}}^2)\ ,\\
\int_{\mathcal{C}_2}\frac{\rho[12]}{\omega_1^3}&=-\frac{m_p}{\Omega}(\gamma_x+\gamma_y+\frac{1}{4}\Gamma_{xy}\Omega v_{\text{th}}^2)\ ,\\
-\int_{\mathcal{C}_2}\frac{\rho[12]}{i\omega_1^4}&=\frac{m_p}{\Omega^2}(\gamma_x+\gamma_y+\frac{1}{8}\Gamma_{xy}\Omega v_{\text{th}}^2)\ ,
\end{split}
\ee
which yield the following quadratic couplings at high temperature :
\begin{equation}
\begin{split}
\Delta m_p &\equiv m_p-m_{p0}=-\frac{m_p}{\Omega} \left(\gamma_x+\gamma_y+\frac{1}{4}\Gamma_{xy} \Omega \ v^2_{th}\right)\ ,\\
Z_I&=\frac{v^2_{th}}{\Omega^2}\left(2\gamma_x+2\gamma_y+\frac{1}{4}\Gamma_{xy} \Omega \ v^2_{th}\right)\ ,\\
\Delta \overline{\mu}^2 &\equiv \overline{\mu}^2-\overline{\mu}_0^2=-\Omega\left(\gamma_x+\gamma_y+ \Gamma_{xy} \Omega \ v^2_{th}\right)\ ,\\
\langle f^2\rangle &=2v^2_{th}\left(\gamma_x+\gamma_y+\frac{1}{2}\Gamma_{xy} \Omega \ v^2_{th}\right)\ ,\\
 \gamma &=\gamma_x+\gamma_y+\frac{1}{2}\Gamma_{xy} \Omega \ v^2_{th}\ .
 \label{quadcoup}
\end{split}
\end{equation}

Similarly, the cubic spectral functions are given by 
\begin{equation}\begin{split} \label{spectral3}
\rho[123]&=2\pi\delta(\omega_1+\omega_2+\omega_3)\times 2m_p\Gamma_3 (\omega_1^2-\omega_2^2) \left(1-\frac{\omega_1\omega_2}{\Omega^2}\right)\times \prod_{k=1}^3\frac{ \Omega^2}{\omega_k^2+\Omega^2}\ , \\
\rho[321]&=2\pi\delta(\omega_1+\omega_2+\omega_3)\times 2m_p\Gamma_3 (\omega_3^2-\omega_2^2)\left(1-\frac{\omega_3\omega_2}{\Omega^2}\right)\times \prod_{k=1}^3\frac{ \Omega^2}{\omega_k^2+\Omega^2}\ .
\end{split}\end{equation}
These three point  spectral functions obeys the following integral identities at high temperature :
\be
\begin{split}
 \int_{\mathcal{C}_3} \frac{\rho[123]}{\omega_1\omega_3}=  \int_{\mathcal{C}_3} \frac{\rho[321]}{\omega_1\omega_3}= -\frac{3}{4}m_p \Gamma_3 \Omega^2\ ,
\end{split}
\ee
\be
\begin{split}
\frac{2}{3} \int_{\mathcal{C}_3} \frac{\rho[123]}{i\omega_1\omega_3^2} 
&= \frac{4}{5} \int_{\mathcal{C}_3} \frac{\rho[321]}{i\omega_1\omega_3^2}
= - \frac{4}{5} \int_{\mathcal{C}_3} \frac{\rho[123]}{i\omega_1^2\omega_3}
=-\frac{2}{3} \int_{\mathcal{C}_3} \frac{\rho[321]}{i\omega_1^2\omega_3} 
= \int_{\mathcal{C}_3} \frac{\rho[132]}{i\omega_1\omega_2\omega_3}\\
&=-\frac{1}{4}\int_{\mathcal{C}_3} \frac{1}{i\omega_2}\Big\{\frac{\rho[123]}{\omega_3^2}-\frac{\rho[321]}{\omega_1^2}\Big\} 
=\frac{1}{2}m_p\Gamma_3\Omega\ ,
\end{split}
\end{equation}
\be
\begin{split}
 \frac{4}{3} \int_{\mathcal{C}_3}\frac{1}{\omega_1^3\omega_3}\rho[321] =\frac{4}{3} \int_{\mathcal{C}_3}\frac{1}{\omega_1\omega_3^3}\rho[123] =\int_{\mathcal{C}_3} \frac{1}{\omega_1\omega_2\omega_3} \Big\{\frac{1}{\omega_1}\rho[321] 
+\frac{1}{\omega_3}\rho[123]\Big\} &= m_p\Gamma_3 \ ,
\end{split}
\ee
which yields the following values for cubic couplings 
\begin{equation}\label{cubiccouplings}
\begin{split}
\overline{\lambda}_3 &=\overline{\kappa}_3= -\frac{3}{2} \Gamma_3 \Omega^2\ ,\\
 \overline{\lambda}_{3\gamma} &=\overline{\kappa}_{3\gamma} =-2\Gamma_3 \Omega\ ,\quad \zeta_{\mu} = -2 \Gamma_3\Omega\ v_{th}^2\ , \\
\zeta_{\gamma} &= \frac{1}{2}\widehat{\kappa}_{3\gamma} =\frac{3}{2} \Gamma_3 v_{th}^2\ , \quad
 \zeta_N =  \Gamma_3 v_{th}^4\  .
\end{split}
\end{equation}
We provide more details about these integrals and the poles which contribute via their residues in the appendix~\ref{app:poles} of this work.

It is evident from the expressions above that many of the couplings in the effective theory are related to each other by a series of relations.
As we will elaborate in the next section, quite a few of these relations can be explained on general grounds using the fact that the bath correlators exhibit
microscopic time-reversal invariance and obey KMS conditions.

\section{Relations between effective couplings}\label{sec:Relations}

In this section we discuss the origin of the relations between the cubic effective couplings given in Eq.\eqref{cubiccouplings}. As we show in the following two subsections, 
most of these relations are based on the following two general features of our model:
\begin{enumerate}
\item Microscopic time-reversal invariance in the bath,
\item KMS relations between bath correlators.
\label{relationcouplingconditon}
\end{enumerate}
While discussing the consequences of these features, we will first give a general proof of the relations between the couplings, and then describe why our particular model satisfies the conditions that go into the proof. The arguments in the following two subsections will show that most of the relations between the effective couplings  in our model are not just particular features  of our model alone. Rather, they are generally valid for a broad class of systems, whenever the two conditions mentioned above are satisfied.

\subsection{Consequences of time-reversal invariance}

First, let us discuss the consequence of microscopic time-reversal invariance in the bath. As we mentioned in the introduction, the implications of such microscopic time-reversal invariance for systems with multiple degrees of freedom were analysed by Onsager in \cite{PhysRev.37.405,PhysRev.38.2265} where he showed that the quadratic effective couplings such as $\gamma_{_{AB}}$ and $\langle f^2_{_{AB}}\rangle$  are symmetric under the exchange of the indices. The derivation of such reciprocal relations relied on the operators $\{\mathcal{O}_A\}$ being invariant under  time-reversal. These relations were later generalised by Casimir \cite{RevModPhys.17.343} to the case  where the operators $\{\mathcal{O}_A\}$ have the parities $\{\eta_{_A}\}$ under time-reversal. The corresponding generalisation of the Onsager relations is as follows:
\be
\begin{split}
\gamma_{_{AB}}&=\eta_{_A}\eta_{_B}\ \gamma_{_{BA}},\\
\langle f^2_{_{AB}}\rangle&=\eta_{_A}\eta_{_B}\ \langle f^2_{_{BA}}\rangle.
\end{split}
\ee

Here, we generalise the Onsager-Casimir reciprocal relations to cubic couplings in the OTO effective theory. We find that  microscopic time-reversal invariance in the bath leads to certain relations between the 2-OTO couplings and the 1-OTO couplings which are derived below. We would like to point out that, unlike the scenario considered by Onsager and Casimir, our system (the Brownian particle) has a single degree of freedom. Nevertheless, the relations that we obtain between the couplings are based on principles similar to those for the reciprocal relations.

To keep the discussion precise, let us note that the operator $\mathcal{O}(t)$ is defined with respect to some particular reference point in time when it coincides with $\mathcal{O}$ (the Schrodinger-picture operator). While calculating the contribution of the correlators of this operator to the particle's dynamics, this reference point must be the instant $t_0$ at which the particle starts interacting with the bath. However, if the bath's initial state (described by the density matrix $\rho_B$) is time-translation invariant i.e 
\be
[\rho_B,H_B]=0,
\ee
then such correlators are independent of the choice of the reference point and depend only on the intervals between the insertions. This is true, for instance, in our model where the bath is assumed to be in a thermal state.
 
For such an initial state of the bath, one can shift the reference point to $t=0$ which can be chosen well into the domain of validity of the particle's effective theory. With respect to this new reference point, the bath's correlators with insertions at both positive and negative values of time are relevant for the particle's dynamics. In the following discussion we are going to assume time-translation invariance of the initial state of the bath and choose the reference point for the operators to be at the origin of time  $t=0$.

Let us now assume that the bath's dynamics has time-reversal invariance and the initial state of the bath respects this symmetry.
Then, there exists an anti-linear and anti-unitary time-reversal operator $\textbf{T}$  such that\footnote{See \cite{weinberg1995quantum} for a proof of the existence of such an operator.}
\be
\begin{split}
[\textbf{T},H_B]&=0,\quad
\textbf{T}\rho_B\textbf{T}^\dag=\rho_B\ .
\end{split}
\ee
At the level of correlators, this symmetry implies
\be
\text{Tr}\Big[\textbf{T} \rho_B \textbf{T}^\dag \textbf{T}\mathcal{O}(t_1)\textbf{T}^\dag \cdots \textbf{T}\mathcal{O}(t_n)\textbf{T}^\dag\Big]=
\text{Tr}\Big[ \rho_B \mathcal{O}(t_1) \cdots \mathcal{O}(t_n)\Big]^*.
\label{timereversal_timedomain1}
\ee

Now, say the operator $\mathcal{O}$  has a definite time parity   i.e.,
\be
\textbf{T}\mathcal{O}\textbf{T}^\dag=\eta_{_\mathcal{O}} \mathcal{O}\ ,
\ee
where $\eta_{_\mathcal{O}}=\pm 1$. The fact that $\textbf{T}$ is an anti-linear operator which commutes with $H_B$,  implies
\be
\textbf{T}\mathcal{O}(t)\textbf{T}^\dag
=\eta_{_\mathcal{O}}\mathcal{O}(-t)\ .
\ee

Inserting this transformation of $\mathcal{O}(t)$  into the equation \eqref{timereversal_timedomain1} and imposing the time-reversal invariance of the initial state, we get
\be
\langle \mathcal{O}(-t_1)\cdots \mathcal{O}(-t_n)\rangle=\eta_{_\mathcal{O}}^n
\langle \mathcal{O}(t_1)\cdots \mathcal{O}(t_n)\rangle^*.
\label{timereversal_timedomain2}
\ee

As the operator $\mathcal{O}$ is Hermitian, the complex conjugation in the RHS of the above equation implies reversing the order of the insertions. So we have
\be
\langle \mathcal{O}(-t_1)\cdots \mathcal{O}(-t_n)\rangle=\eta_{_\mathcal{O}}^n
\langle \mathcal{O}(t_n)\cdots \mathcal{O}(t_1)\rangle.
\label{timereversal_timedomain3}
\ee

Such relations, in general, imply that correlators with different OTO numbers (i.e., the number of minimum time-folds required to compute the correlators\cite{2017arXiv170102820H})  get related to each other. In case of 3-point functions, as was mentioned earlier, we have at most 2-OTO correlators. For three time instants $t_1>t_2>t_3$, the 2-OTO correlators are: 
\be 
\langle \mathcal{O}(t_1)\mathcal{O}(t_3)\mathcal{O}(t_2)\rangle \text{ and } \langle \mathcal{O}(t_2)\mathcal{O}(t_3)\mathcal{O}(t_1)\rangle.
\ee
In both these correlators, we have 2 future turning point insertions: $\mathcal{O}(t_1)$ and $\mathcal{O}(t_2)$, and hence they are 2-OTO correlators.
From \eqref{timereversal_timedomain3} we can see that these correlators are related to their time-reversed counterparts as follows:
\be
\begin{split}
&\langle \mathcal{O}(t_1) \mathcal{O}(t_3)\mathcal{O}(t_2)\rangle=\eta_{_\mathcal{O}}
\langle \mathcal{O}(-t_2) \mathcal{O}(-t_3)\mathcal{O}(-t_1)\rangle,\\
&\langle \mathcal{O}(t_2) \mathcal{O}(t_3)\mathcal{O}(t_1)\rangle=\eta_{_\mathcal{O}}
\langle \mathcal{O}(-t_1) \mathcal{O}(-t_3)\mathcal{O}(-t_2)\rangle.\\
\end{split}
\label{timereversal_timedomain4}
\ee
The correlators in the RHS of the above equations are 1-OTO correlators. So we see that all 2-OTO 3-point correlators of the bath get related to 1-OTO correlators.

It is natural to ask whether such relations between the bath's correlators lead to any relation between the 2-OTO couplings and the 1-OTO couplings in the effective theory of the particle. To answer this question, first notice that the relation \eqref{timereversal_timedomain2} implies that, when all the frequencies are real, correlators of $\mathcal{O}$ in frequency space are either purely real or purely imaginary i.e.
\begin{equation}
\langle \mathcal{O}(\omega_1)\cdots\mathcal{O}(\omega_n)\rangle=\eta_{_\mathcal{O}}^n
\langle \mathcal{O}(\omega_1)\cdots\mathcal{O}(\omega_n)\rangle^*.
\end{equation}
This reality property gets carried over to the connected parts i.e. the cumulants of these correlators. 
Equivalently, the spectral functions satisfy relations of the form
\begin{equation}
\rho[1\ldots n]=\eta_{_\mathcal{O}}^n
\rho[1\ldots n]^* .
\end{equation}
We use this property of the cumulants in frequency space to derive relations between the 2-OTO and 1-OTO couplings.

From the expressions of the cubic couplings, we see that the couplings can be divided into doublets and singlets as shown in the table \ref{table:couplingsTimeReve} . We will show that the pair of couplings in each doublet are related to each other due to time-reversal invariance. On the other hand, time-reversal maps the singlets to themselves upto a  factor  of $\eta_{_\mathcal{O}}$. So when $\eta_{_\mathcal{O}}=1$, these relations are trivial. But when $\eta_{_\mathcal{O}}=-1$, these relations imply that these singlets vanish.

\begin{table}[ht]
\caption{Coupling doublets/singlets under microscopic time-reversal (general environment)}
\label{table:couplingsTimeReve}
\begin{center}
\begin{tabular}{ |c| c|  }
\hline
$g$ & $\mathcal{I}[g]$\\
  \hline
 \hline
 $\overline{\lambda}_3$ & $\frac{2}{m_p\omega_1\omega_3}\rho [123]$\\
\hline
$\overline{\kappa}_3$ & $\frac{2}{m_p\omega_1\omega_3}\rho [321]$\\
\hline
\hline
$\overline{\lambda}_{3\gamma}$&$\frac{1}{im_p\omega_1\omega_3}\left(\frac{2}{\omega_1}-\frac{1}{\omega_3}\right)\rho[123]$\\
\hline
$ \overline{\kappa}_{3\gamma}$ & $\frac{1}{im_p\omega_1\omega_3}\left(\frac{1}{\omega_1}-\frac{2}{\omega_3}\right)\rho[321]$\\
\hline
\hline
$ \zeta_\gamma$ & $\frac{1}{(2m_p\omega_1\omega_3)^2}\Bigl((2\omega_1-\omega_3)\ \rho[321_+]+(\omega_3-2\omega_1)\ \rho[132_+]+(\omega_1+\omega_3)\ \rho[123_+] \Bigr)$\\
\hline
$\widehat{\kappa}_{3\gamma} - \zeta_\gamma$ & $\frac{1}{(2m_p\omega_1\omega_3)^2}\Bigl((\omega_1+\omega_3)\ \rho[321_+]+(2\omega_3-\omega_1)\ \rho[132_+]+(2\omega_3-\omega_1)\ \rho[123_+] \Bigr)$\\
\hline
\hline
$\widehat{\kappa}_{3\gamma} $ & $\frac{3}{(2m_p\omega_1\omega_3)^2}\Bigl(\omega_1\ \rho[321_+]+(\omega_3-\omega_1)\ \rho[132_+]+\omega_3\ \rho[123_+] \Bigr)$\\
\hline
\hline
 $\zeta_\mu$&$ \frac{1}{2im_p^2\omega_1\omega_3}\Big(\rho[123_+]-\rho [321_+]+ \rho [132_+] \Big)$\\
 \hline
 \hline 
  $ 2\zeta_N+\frac{\hbar^2}{8m_p^2}\left(\overline{\lambda}_3-\overline{\kappa}_3\right)$&$\frac{1}{8m_p^3\omega_1\omega_3}\Bigl(\rho [12_+3_+]+\rho [32_+1_+]\Bigr)$\\
  \hline
\end{tabular}
\end{center}
\end{table}

As the couplings mentioned in table \ref{table:couplingsTimeReve} are all real, their complex conjugates are equal to them. This can be used to obtain alternative expressions for these couplings by complex conjugating the integrals. Such a complex conjugation in the frequency space maps the contour of integration from $\mathcal{C}_3$ :
\be
\begin{split}
\int_{\mathcal{C}_3}\equiv \int_{-\infty-i\epsilon_1}^{ \infty-i\epsilon_1} \frac{d\omega_1}{2\pi}\int_{-\infty+i\epsilon_1-i\epsilon_3}^{ \infty+i\epsilon_1-i\epsilon_3} \frac{d\omega_2}{2\pi}
\int_{-\infty+i\epsilon_3}^{ \infty+i\epsilon_3} \frac{d\omega_3}{2\pi}
\end{split}
\ee
to $\mathcal{C}_3^\ast$ where the frequencies run over the following values:
\be
\begin{split}
\int_{\mathcal{C}_3^\ast}\equiv \int_{-\infty+i\epsilon_1}^{ \infty+i\epsilon_1} \frac{d\omega_1^c}{2\pi}\int_{-\infty-i\epsilon_1+i\epsilon_3}^{ \infty-i\epsilon_1+i\epsilon_3} \frac{d\omega_2^c}{2\pi}
\int_{-\infty-i\epsilon_3}^{ \infty-i\epsilon_3} \frac{d\omega_3^c}{2\pi}\ .
\end{split}
\ee
Note that the integration over $\omega_1^c$  in the $\mathcal{C}_3^\ast$ contour runs just above the real axis, exactly like the integration over $\omega_3$  in the $\mathcal{C}_3$ contour. Similarly,  integration over $\omega_3^c$  in the $\mathcal{C}_3^\ast$ contour runs just below the real axis exactly like the integration over $\omega_1$  in the $\mathcal{C}_3$ contour.
Therefore, under the following redefinitions:
\be
\begin{split}
\omega_1\equiv \omega_3^c,\quad
\omega_2\equiv \omega_2^c,\quad
\omega_3\equiv \omega_1^c,\ 
\label{freq_redef}
\end{split}
\ee
the contour of integration gets mapped back to $\mathcal{C}_3$ with $\epsilon_1$ and $\epsilon_3$ exchanged (and  the imaginary part of $\omega_2$ reversed). This exchange of $\epsilon_1$ and $\epsilon_3$ and the concomitant reversal of the imaginary part of $\omega_2$ can be undone by a contour deformation, provided our integrands have 
no $\omega_2$  discontinuities near real axis (as required for the validity of Markovian approximation). To conclude, assuming appropriate analyticity in 
 $\omega_2$, the complex conjugation and the above redefinition leave $\mathcal{C}_3$ contour invariant.

Now, let us turn to how the integrands are modified under the above operation. Notice that each term in the integrand has the following form : it is a product of a rational function of the frequencies and the bath cumulants. The modification of the rational functions is simple : the rational functions are modified by complex conjugating them and then performing the above frequency redefinition.
This has the effect of replacing any explicit $i$ by $(-i)$ and exchanging $\omega_1$ and $\omega_3$.

Turning to the cumulants, time-reversal invariance implies that, when the frequencies are real, these cumulants  are either purely real or purely imaginary  depending on their time-reversal parity $\eta_{_\mathcal{O}}$. Thus, the cumulants can be complex-conjugated by conjugating the frequencies in the argument of the cumulants followed by a multiplication by $\eta_{_\mathcal{O}}$. The  frequency redefinition above then results in the exchange of $\omega_1$ and $\omega_3$ in the arguments. To summarise,  the modified integrands are obtained from the original ones by the following rules:
\begin{itemize}
\item Replace $i$ by $-i$ in the coefficients,
\item Exchange  $\omega_1$ and $\omega_3$ in the rational functions and the cumulants,
\item Multiply by $\eta_{_\mathcal{O}}$.
\end{itemize}

After re-expressing the couplings in terms of these modified integrals over the same contour $\mathcal{C}_3$, one can compare them with the expression given in table \ref{table:couplingsTimeReve} and  find the following relations :
\be
\begin{split}
\overline{\kappa}_3=\eta_{_\mathcal{O}}\overline{\lambda}_3\ ,\quad
\overline{\kappa}_{3\gamma} =\eta_{_\mathcal{O}}\overline{\lambda}_{3\gamma}\ ,\quad
&\widehat{\kappa}_{3\gamma}  -\zeta_\gamma 
=\eta_{_\mathcal{O}}\zeta_\gamma\ , \quad
\zeta_\mu=\eta_{_\mathcal{O}}\zeta_\mu\ , \\
2\zeta_N+\frac{\hbar^2}{8m_p}(\overline{\lambda}_3-\overline{\kappa}_3)&=\eta_{_\mathcal{O}}\Big(2\zeta_N+\frac{\hbar^2}{8m_p}(\overline{\lambda}_3-\overline{\kappa}_3)\Big)\ .
\end{split}
\label{2OTOfrom1OTO}
\ee
This, in turn,  implies
\be
\begin{split}
\overline{\kappa}_3&=\eta_{_\mathcal{O}}\overline{\lambda}_3\ ,\quad
\overline{\kappa}_{3\gamma} =\eta_{_\mathcal{O}}\overline{\lambda}_{3\gamma}\ ,\quad
\widehat{\kappa}_{3\gamma} 
=(1+\eta_{_\mathcal{O}})\zeta_\gamma \ , \quad
\widehat{\kappa}_{3\gamma} 
=\eta_{_\mathcal{O}}\widehat{\kappa}_{3\gamma} \ , \\
\zeta_\mu&=\eta_{_\mathcal{O}}\zeta_\mu\ , \quad
2(1-\eta_{_\mathcal{O}})\zeta_N=(\eta_{_\mathcal{O}}-1)\frac{\hbar^2}{8m_p}(\overline{\lambda}_3-\overline{\kappa}_3)\ .
\end{split}
\label{2OTOfrom1OTOp}
\ee
When the operator $\mathcal{O}$ is even under time-reversal i.e. when $\eta_{_\mathcal{O}}=1$, the last two relations are trivial. The other relations reduce to
\be
\begin{split}
&\overline{\kappa}_3=\overline{\lambda}_3\ ,\quad
\overline{\kappa}_{3\gamma} =\overline{\lambda}_{3\gamma}\ ,\quad
\widehat{\kappa}_{3\gamma}  =2\zeta_\gamma\ .
\end{split}
\label{2OTOfrom1OTOeven}
\ee
On the other hand, when the operator $\mathcal{O}$ is odd under time-reversal i.e. when $\eta_{_\mathcal{O}}=-1$, then the relations in \eqref{2OTOfrom1OTO} reduce to
\be
\begin{split}
\overline{\kappa}_3&=-\overline{\lambda}_3\ ,\quad
\overline{\kappa}_{3\gamma} =-\overline{\lambda}_{3\gamma}\ ,\quad
\widehat{\kappa}_{3\gamma}   =0\ ,\quad
\zeta_\mu=0\ ,\quad
\zeta_N =-\frac{\hbar^2}{16m_p}(\overline{\lambda}_3-\overline{\kappa}_3)\ .
\end{split}
\label{2OTOfrom1OTOodd}
\ee

\subsection{Time-reversal invariance of the bath in qXY model } 

We found in the preceding discussion that the bath needs to satisfy the following conditions for the relations given in equation \eqref{2OTOfrom1OTOeven} to hold true:
\begin{enumerate}
\item Time-translation invariance of the initial state,
\item Time-reversal invariance in the dynamics,
\item Time-reversal invariance of the initial state,
\item Time-reversal invariance of the operator that couples to the particle.
\end{enumerate}
Let us check whether these conditions are satisfied in our model one by one. 

As we have already mentioned, the initial state of the bath is a thermal state and hence it is invariant under time-translations.

To see the time-reversal invariance in the bath's dynamics, first note that the bath consists of two sets of harmonic oscillators. 
We can denote the lowering and raising operators of these oscillators by $a_i$ and $a_i^\dag$ for the X-type oscillators and 
by $b_j$ and $b_j^\dag$ for the Y-type oscillators. Therefore, the Hamiltonian of the bath is given by
\be
H_B=\sum_i \hbar \mu_{x,i}\Big(a_i^\dag a_i+\frac{1}{2}\Big)+\sum_j \hbar \mu_{y,j}\Big(b_j^\dag b_j+\frac{1}{2}\Big)\ .
\label{hamiltonian_model}
\ee
Now, the action of the time-reversal operator on the raising and lowering operators is as follows:
\be
\begin{split}
\textbf{T}a_i\textbf{T}^\dag=a_i,\ \textbf{T}b_j\textbf{T}^\dag=b_j,\\
\textbf{T}a_i^\dag\textbf{T}^\dag=a_i^\dag,\ \textbf{T}b_j^\dag\textbf{T}^\dag=b_j^\dag\ .
\end{split}
\label{raising_lowering_timereversal}
\ee
Using these transformations of the raising and lowering operators under time-reversal and the form of $H_B$ given in \eqref{hamiltonian_model} we see that
\be
\textbf{T}H_B\textbf{T}^\dag=H_B
\implies [\textbf{T},H_B]=0,
\ee
which means that time-reversal is a symmetry of the dynamics.

Now, the initial state is a thermal state i.e. 
\be
\rho_B=\frac{1}{Z_B}e^{-\frac{ H_B}{k_BT}} \ .
\ee 
Therefore, the commutation of the time-reversal operator with the Hamiltonian also implies
\be
\textbf{T}\rho_B\textbf{T}^\dag=\rho_B
\ee
i.e. the initial state is invariant under time-reversal.

Finally, the operator that couples to the particle is
\be
\mathcal{O}=\sum_i g_{x,i}X_i+\sum_j g_{y,j}Y_j+\sum_{i,j} g_{xy,ij}X_i Y_j\ .
\ee
Here the positions of the oscillators are give by
\be
\begin{split}
X_i=\sqrt{\frac{\hbar}{2 m_{x,i}\mu_{x,i}}}(a_i+a_i^\dag),\\
Y_j=\sqrt{\frac{\hbar}{2 m_{y,j}\mu_{y,j}}}(b_j+b_j^\dag).
\end{split}
\ee
Therefore, using the transformations in \eqref{raising_lowering_timereversal} we have 
\be
\begin{split}
&\textbf{T}X_i\textbf{T}^\dag=X_i ,\\
&\textbf{T}Y_j\textbf{T}^\dag=Y_j.\\
\end{split}
\ee
This implies that the operator $\mathcal{O}$ is invariant under time-reversal i.e.
\be
\textbf{T}\mathcal{O}\textbf{T}^\dag=\mathcal{O}\ .
\ee
So, all the conditions necessary for the relations \eqref{2OTOfrom1OTOeven}  between the effective couplings are satisfied in our model.

\paragraph{Example of a time-reversal odd operator coupling to the particle:}
\label{time-reversal odd} 
We can slightly modify the qXY model to introduce a piece in the operator $\mathcal{O}$ that is odd under time-reversal. For this, consider the following particle-bath interaction:
\be
L_{\text{int}}=\Big(\sum_i g_{x,i}X_i+\sum_i g_{y,j}Y_j\Big) q-\sum_i \widetilde{g}_{xy,ij}X_i Y_j \dot{q}\ .
\ee
Integrating by parts, we see that the operator that couples to the particle's position is
\be
\mathcal{O}=\sum_i g_{x,i}X_i+\sum_i g_{y,j}Y_j+\sum_i \widetilde{g}_{xy,ij}\dot{X_i} Y_j +\sum_i \widetilde{g}_{xy,ij}X_i\dot{Y_j}\ .
\ee
Here, the operators $\dot{X_i} $ and $\dot{Y_j}$ are odd under time-reversal i.e.
\be
\begin{split}
\textbf{T} \dot{X}_i\textbf{T}^\dag&=- \dot{X}_i\ , \quad
\textbf{T} \dot{Y}_j\textbf{T}^\dag=- \dot{Y}_j\ .
\end{split}
\ee
The thermal correlators of 3 point functions $\mathcal{O}$ receive contributions only from terms of the form:
\be
\begin{split}
\langle \mathcal{O}(t_1)\mathcal{O}(t_2)\mathcal{O}(t_3)\rangle=&\sum_{i,j} g_{x,i} g_{y,j} \widetilde{g}_{xy,ij}\langle X_i(t_1)Y_j(t_2)\dot{X}_i(t_3)Y_j(t_3)\rangle\\
&+\sum_{i,j} g_{x,i} g_{y,j} \widetilde{g}_{xy,ij}\langle X_i(t_1)Y_j(t_2)X_i(t_3)\dot{Y}_j(t_3)\rangle+\cdots
\end{split}
\ee
All such terms are correlators with 3 position operators (which are time-reversal even) and one velocity operator (which is time-reversal odd). Therefore, the overall correlator satisfies the following relation:
\be
\begin{split}
\langle \mathcal{O}(t_1)\mathcal{O}(t_2)\mathcal{O}(t_3)\rangle=&-\langle \mathcal{O}(-t_3)\mathcal{O}(-t_2)\mathcal{O}(-t_1)\rangle\ .
\end{split}
\ee
This, in turn, means that the relations in equation \eqref{2OTOfrom1OTOodd} are satisfied in this model.

\paragraph{A brief comment on couplings of quartic and higher degree terms:}
The fact that all the 2-OTO couplings get related to 1-OTO couplings is not generally true for higher degree terms in the particle's effective action. To see this consider the quartic couplings which receive contributions from 4-point cumulants of the operator $\mathcal{O}$. Such 4-point functions can again be at most 2-OTO. But not all of these 2-OTO correlators are related to 1-OTO correlators by time-reversal. For instance, consider the following correlator:
\be
\langle\mathcal{O}(t_1)\mathcal{O}(t_3)\mathcal{O}(t_2)\mathcal{O}(t_4)\rangle
\ee
where $t_1>t_2>t_3>t_4$. This is a 2-OTO correlator as there are 2 future-turning point insertions in it, viz.,   $\mathcal{O}(t_1)$ and $\mathcal{O}(t_2)$. 

Under time-reversal, this gets related to the another correlator as follows:
 \be
\langle\mathcal{O}(t_1)\mathcal{O}(t_3)\mathcal{O}(t_2)\mathcal{O}(t_4)\rangle=\langle\mathcal{O}(-t_4)\mathcal{O}(-t_2)\mathcal{O}(-t_3)\mathcal{O}(-t_1)\rangle\ .
\ee
Now the correlator on RHS of the above equation also has two future-turning point insertions, viz., $\mathcal{O}(-t_4)$ and $\mathcal{O}(-t_3)$. Thus, this is a genuine 2-OTO correlator which is not related to a 1-OTO correlator under time-reversal. Such 2-OTO correlators  may lead to genuinely new 2-OTO quartic couplings in the effective dynamics of the particle, unrelated to the 1-OTO coupling \cite{Chakrabarty:2019qcp}.

\subsection{Consequence of KMS relations: Generalised Fluctuation-Dissipation Relations}

We have already seen that, at the high temperature limit, the KMS relations between thermal 2-point functions of the operator $\mathcal{O}$ leads to a relation between the damping coefficient $\gamma$ of the particle and the strength of the additive noise $\langle f^2\rangle$ that it experiences:
\be
\langle f^2\rangle=2\gamma v_{th}^2\ .
\label{FDquad}
\ee
This is the fluctuation-dissipation relation that  was originally discovered through the studies of Brownian motion by Einstein, Smoluchowski and Sutherland. An analogous relation in electrical circuits was discovered by Johnson\cite{PhysRev.32.97} and was theoretically derived by Nyquist\cite{PhysRev.32.110}. A general proof of such relations was worked out by Callen and Welton in \cite{PhysRev.83.34} which was further generalised by Stratonovich \cite{stratonovich2012nonlinear}. 

All these fluctuation-dissipation relations are relations between what we now understand to be 1-OTO couplings and are rooted in the KMS relations between the 1-OTO correlators of the bath. However, in \cite{Haehl:2017eob} it was pointed out that the KMS relations can also relate  2-OTO correlators to 1-OTO correlators. For example, consider the following 2-OTO correlator:
\be
\langle \mathcal{O}(t_1)\mathcal{O}(t_3)\mathcal{O}(t_2)\rangle
\ee
where $t_1>t_2>t_3.$
By KMS relations, this 2-OTO correlator is related via analytic continuation to a 1-OTO correlator as follows:
\be
\langle \mathcal{O}(t_1)\mathcal{O}(t_3)\mathcal{O}(t_2)\rangle=\langle \mathcal{O}(t_2-i\beta)\mathcal{O}(t_1)\mathcal{O}(t_3)\rangle.
\ee
It is natural to wonder what imprint do such relations have on the effective dynamics of the particle. Do they lead to generalisations of the relation in \eqref{FDquad}? If so, then do such relations connect 2-OTO couplings with the 1-OTO ones?

In section \S\ref{sec:nonLinearLangevin}, we indeed saw such a relation between the 2-OTO coupling $\widehat{\kappa}_{3\gamma}$ and the 1-OTO coupling $\zeta_N$ at high temperature: 
\be
\begin{split}
\zeta_N +\frac{\hbar^2}{16m_p^2}(\overline{\lambda}_3-\overline{\kappa}_3)& = \frac{1}{3} \widehat{\kappa}_{3\gamma}v_{th}^2.
\end{split}
\label{FD_1}
\ee
As we mentioned there, this followed from using the KMS relations between the 3-point thermal correlators of the bath to express the couplings in terms of the spectral functions $\rho[123]$ and $\rho[321]$, and keeping only the leading order term in $\beta$-expansion. Thus, this relation can be considered to be a generalisation of the fluctuation-dissipation relation between the quadratic couplings in \eqref{FDquad}.

Microscopic time-reversal invariance with $\eta_{_{\mathcal{O}}}=1$ implies $\overline{\lambda}_3=\overline{\kappa}_3$ and $\widehat{\kappa}_{3\gamma} =2\zeta_{\gamma}$.
In this case, the above relation reduces to
\be
\zeta_N  = \frac{1}{3} \widehat{\kappa}_{3\gamma}v_{th}^2=\frac{2}{3} \zeta_{\gamma}v_{th}^2.
\label{FD_classical}
\ee
This is  hence a relation between the coefficient $\zeta_N$ of the cubic non-gaussian term in the probability distribution of the noise and 
the jitter $\zeta_{\gamma}$ in the damping coefficient of the particle.

On the other hand, if $\eta_{_{\mathcal{O}}}=-1$, then $\widehat{\kappa}_{3\gamma} =0$ along with
\be
\begin{split}
\zeta_N = -\frac{\hbar^2}{16m_p^2}(\overline{\lambda}_3-\overline{\kappa}_3)\ .
\end{split}
\ee
This implies that the relation in \eqref{FD_1} is trivially satisfied.

Apart from these relations, we also see another analogous fluctuation-dissipation relation in our model which is given by
\begin{equation}
\begin{split}
\zeta_\mu=v_{\text{th}}^2\overline{\kappa}_{3\gamma}=v_{\text{th}}^2\overline{\lambda}_{3\gamma}\ .
\end{split}
\label{prove}
\end{equation}
As we discussed in the last subsection, the relation between $\overline{\kappa}_{3\gamma}$ and $\overline{\lambda}_{3\gamma}$ in the second half of the above equation is a consequence of time-reversal invariance in the bath. This equation relates the cubic 1-derivative anharmonicity in the particle's motion to the jitter in its frequency. 

As yet, we do not know how generic this relation is. However, we suspect that, with some assumptions about the properties of the spectral functions, a general relation of the following form can be proven:
\be
\zeta_\mu=\frac{v_{\text{th}}^2}{2}\ (\overline{\kappa}_{3\gamma}+\overline{\lambda}_{3\gamma})\ .
\label{FD2}
\ee
Such a relation is consistent with time-reversal invariance, since for the operator $\mathcal{O}$ having a definite parity $\eta_{_{\mathcal{O}}}$, the two sides of the above equation transform similarly under time-reversal (see the relations in \eqref{2OTOfrom1OTO}) .  When $\eta_{_{\mathcal{O}}}=-1$ both sides of the equation are equal to zero and hence the relation \eqref{FD2} is trivially true. Moreover, we find it to be satisfied in the qXY model and a variety of related models. Hence, we expect it to be true for a broad class of models than the one studied in this work. It will be interesting to check this expectation and determine the exact conditions which are required for this relation to hold.

\section{Conclusion and discussion}\label{sec:discussion}

In this paper, we have constructed an effective theory of a Brownian particle which goes beyond the standard Langevin dynamics. We remind the reader that the standard Langevin theory describes a Brownian particle subject to linear damping and a Gaussian thermal noise. The effective theory described in this paper includes in addition anharmonic couplings $\overline{\lambda}_3$ and  $\overline{\lambda}_{3\gamma}$  along with a thermal jitter  $\zeta_\mu$ in the frequency and a jitter $\zeta_\gamma$ in the damping constant.  Apart from these parameters and the usual Langevin couplings, this theory also contains a parameter $\zeta_N$ which is the strength of the 
non-Gaussianity in the thermal noise experienced by the particle. 

When out of time ordered correlations (or more specifically, 2-OTO correlators) transmitted from the bath are kept track of, one has to add three more OTO couplings $\overline{\kappa}_3,\overline{\kappa}_{3\gamma}$ and $\widehat{\kappa}_{3\gamma}$ which are related by time-reversal to the standard (1-OTO) couplings $\overline{\lambda}_3,\overline{\lambda}_{3\gamma}$ and  $\zeta_\gamma$ respectively. To demonstrate how these couplings affect the dynamics of the particle, we have expressed the classical limits of the particle's correlators in terms of these couplings. We find that the OTO couplings show up in the out of time ordered nested Poisson brackets of the particle.

We explore the constraints imposed by microscopic time-reversal invariance of the (particle+bath) combined system on the effective theory of the particle. Such an invariance of the overall dynamics of the combined system under time-reversal holds when the interaction term between the particle and the bath is even under this transformation. This fixes  the OTO   cubic effective couplings in terms of the parameters in the nonlinear Langevin theory via the relations
\[ \overline{\kappa}_3=\overline{\lambda}_3\ ,\quad \overline{\kappa}_{3\gamma}=\overline{\lambda}_{3\gamma}\ , \quad  \widehat{\kappa}_{3\gamma}=2\zeta_\gamma \ .\]
These relations between the cubic couplings are the generalisations of the well known Onsager-Casimir reciprocal relations that originate from the microscopic time-reversal invariance of the combined system. 

Since the bath is in a thermal state, the bath correlators satisfy the KMS relations. In the high temperature limit, these further give rise to a generalised fluctuation-dissipation relation between the 2-OTO cubic derivative coupling $\widehat{\kappa}_{3\gamma}$ and the 1-OTO cubic non-derivative coupling $\zeta_N$. Combining time-reversal invariance and the generalised fluctuation-dissipation relation, the coefficient of the thermal jitter $\zeta_{\gamma}$ in the damping term of the non-linear Langevin equation  gets related to the coefficient of non-Gaussianity  $\zeta_{N}$ of thermal noise. 

To provide a concrete model where these general results are justified, we have constructed an OTO-effective theory of a Brownian particle interacting with a dissipative thermal bath composed of two sets of harmonic oscillators. To this, we 
 add a small 3-body interaction coupling the particle to two other oscillators, one from each set. For this model, we show that all the bath correlators decay exponentially at late times, leading to a local non-unitary effective theory for the particle.

Working out the effective couplings of the particle in this model, we find that the above mentioned relations between these couplings are indeed satisfied. Furthermore, in this model, we find another fluctuation-dissipation type relation between the strength ($\zeta_\mu$) of the thermal jitter in the frequency  and the coefficient ($\overline{\lambda}_{3\gamma}$) of the anharmonic term with a single time derivative.

An immediate future direction would be to check whether the relations between the effective couplings (arising from time-reversal invariance and thermality of the bath) are valid at higher orders in the  system-bath coupling. Moreover, it will be interesting to explore quartic and higher order terms in the effective action of the particle. At the quartic order, there will be genuinely new 2-OTO effective couplings that are not fixed in terms  of the Schwinger-Keldysh 1-OTO effective couplings by time-reversal invariance. 
 
 It will be useful to understand the generalisation of fluctuation-dissipation relations and the consequence of time reversal invariance in the quartic case. Furthermore, such an extension to quartic terms in the effective theory may lead to observing chaotic motion of the particle\footnote{Such chaotic motion is seen, for example, in periodically driven anharmonic oscillators such as the Van der Pol oscillator and the Duffing oscillator.}. If that is indeed the case, then it may be insightful to connect the already existing studies on quantum chaos based on OTOCs with this example.

Our techniques can potentially be extended to the context of quantum optics\cite{klyshko1988photons,kippenberg2007cavity}. As a toy model one can consider an atom  that is interacting electromagnetically with a gas of photons\cite{PhysRevA.45.5056, PhysRevA.45.6816, PhysRevA.52.4214,Liu2018}. 
One can write down an effective theory for the atom and calculate its OTOCs. Such predictions for the behaviour of OTOCs  may be verified with development in experimental techniques to measure them \cite{2017NatPh..13..781G,PhysRevX.7.031011,Zhu:2016uws,Swingle:2016var,2016arXiv160701801Y,2017PhRvL.119d0501G}.
 Moreover, our prediction of a generalised fluctuation-dissipation relation between the thermal jitter in the damping and the non-gaussianity in the noise may be testable in such a setup.

Another possible extension is  to 1-D spin chains  in a thermal environment \cite{2015NatSR...514873B, PhysRevA.94.012341, Ilievski:2014gra}.
It will be interesting to explore the possibility of writing a similar effective theory of the chain (or a part of it) which would aid in studying the thermalisation of its OTOCs and comparing them with those of time-ordered correlators. Such comparisons may be useful in classifying systems according to the behaviour of their OTOCs.

In this work, we saw that the OTO couplings are not just a feature of the quantum mechanical theory of the particle, but they show up in out of time ordered Poisson brackets in the classical limit as well (Similar classical limits of OTOCs have been discussed in \cite{PhysRevLett.121.024101, Scaffidi:2017ghs, Khemani:2018sdn, PhysRevLett.118.086801}). A stochastic interpretation of this classical OTO behaviour would be useful in understanding the significance of such OTOCs in the quantum mechanical framework (where OTO  dynamics has  been mostly studied up till now) as well as devising experiments to measure them. In this context, it will also be interesting to extend  the idea of decoherence of quantum systems to their out of time ordered dynamics. 

One potentially useful way to study the OTO dynamics of open systems connected to holographic baths would be to use the AdS/CFT correspondence for mapping this to a problem in gravity. It would be nice to systematically derive the spectral functions from gravity following \cite{Herzog:2002pc, Skenderis:2008dh, deBoer:2017xdk} and calculate OTOCs in AdS gravity theories.

Another interesting direction to pursue would be to systematically develop diagrammatic methods  for OTO perturbation theory\cite{SC}. 
This would provide a simpler way of computing OTOCs of the system. It would also be nice to derive cutting rules for
OTOCs along the lines of \cite{Kobes:1986za, Kobes:1990ua, Guerin:1993ik}.

\acknowledgments

We would like to thank Subhro Bhattacharjee, Ahana Chakraborty, Chandramouli Chowdhury, Arghya Das, Manas Kulkarni, Gautam Mandal, 
Shiraz Minwalla, Archak Purakaystha, Suvrat Raju, Mukund Rangamani, Rajdeep Sensarma, Shivaji Sondhi, Douglas Stanford, Spenta Wadia  for useful discussions. 
Preliminary versions of these results were reported by SC in a seminar at Institute of Mathematical Sciences (IMSc.), Chennai.

 BC and SC are grateful to the organisers of Strings, 2018 for hospitality while the project was ongoing. BC would like to thank Chennai Mathematical Institute (CMI), Chennai and 
National Institute of Science Education and Research (NISER), Bhubaneswar (during `$ST^4\ 2018$ : Student Talks on Trending Topics in Theory ' workshop) 
for hospitality while the project was in progress. SC thanks Tata Institute of  Fundamental Research (TIFR), Mumbai and Institute of Mathematical Sciences (IMSc.), for warm hospitality towards the final stages of this work. We would also like to thank the 
participants of an OTOCs discussion meeting at ICTS-TIFR, Bangalore for their comments and questions on the subject of OTOCs in general.

 BC, SC and RL are grateful 
for the support from International Centre for Theoretical Sciences (ICTS-TIFR),  Bangalore.
BC and SC thank Infosys International Exchange Program for supporting the travel to Strings 2018. BC also acknowledges Infosys Program for supporting the travel to $ST^4$ 2018. 
We acknowledge our debt to the people of India for their steady and generous support to research in the basic sciences.

\appendix
\begin{appendix}

\section{Dimensional analysis}\label{app:DimensionalAnalysis}
In this appendix, we list the dimensions of various couplings etc. that appear in this work. Though we do not do so in this work, many of the computations described in this work  are simplified by  setting $\hbar=m_p=1$, which is equivalent to 
counting dimensions by setting $M=T^0$ and  $L^2=T$. 
\begin{itemize}
\item Dimensions of position and noise variables :

\be\begin{split} [q] = L\ , \quad [\mathcal{N}] = L^{-1} T\ . \end{split}\ee
\item Dimensions of couplings :
\begin{equation}
\begin{split}
[m_p] &=M\ ,\quad [F] =LT^{-2}\ ,\quad [\gamma]=T^{-1}\ ,\quad [\bar{\mu}^2]=T^{-2}\ ,\\
  [\zeta_\gamma] &= [\widehat{\kappa}_{3\gamma}]=LT^{-2} \ ,\quad [\zeta_\mu]=LT^{-3} \ ,\\
\quad [\overline{\lambda}_3]&= [\overline{\kappa}_3]=L^{-1}T^{-2}\ ,\quad [\overline{\lambda}_{3\gamma}]= [\overline{\kappa}_{3\gamma}]=L^{-1}T^{-1}\ . 
\end{split}
\end{equation}
\item Dimensions of Noise parameters :
\begin{equation}
\begin{split}
[\langle f^2\rangle] &=L^2T^{-3}\ ,\quad [\zeta_N ]=L^3T^{-4}\  ,\quad [Z_I ]=L^2T^{-1}. 
\end{split}
\end{equation}
\item Dimensions of Spectral functions :

\be\begin{split} [\rho[12]] =MT^{-1}\ ,\quad [\rho[12_+]] =M^2L^2T^{-2}\ ,\end{split}\ee

\be\begin{split}  [\rho[123]]=ML^{-1}T^{-1} \ ,\quad  [\rho[123_+]]=[\rho[12_+3]]=M^2LT^{-2} \ ,\quad   [\rho[12_+3_+]]=M^3L^3T^{-3}\ .\end{split}\ee
\item Dimensions of System Bath couplings

\be\begin{split} [g_x] =[g_y] =MT^{-2}\ ,\quad [g_{xy}] =ML^{-1}T^{-2}\   .\end{split}\ee
\item Dimensions of System Bath coupling distribution functions

\be\begin{split} [ \Big\langle \Big\langle \frac{g_x^2}{m_x}\Big\rangle\Big\rangle] =[\Big\langle \Big\langle \frac{g_y^2}{m_y}\Big\rangle\Big\rangle] =MT^{-3}\ ,\
 [ \Big\langle \Big\langle \frac{g_{xy}^2}{m_xm_y}\Big\rangle\Big\rangle ] =L^{-2}T^{-2}\  ,\\
[\Big\langle \Big\langle \frac{g_x g_y g_{xy}}{m_xm_y}\Big\rangle\Big\rangle] =ML^{-1}T^{-4} .\end{split}\ee

\item Dimensions of Spectral function parameters

\be\begin{split} [\gamma_x] =[\gamma_y] =[\Omega]=T^{-1}\ ,\quad [\Gamma_{xy}] =L^{-2}T^2\ ,\quad [\Gamma_3] =L^{-1}\ .\end{split}\ee
\item Dimensions of Thermal  parameters

\be\begin{split} [\beta] =T\ ,\quad [v_{th}^2] =L^2T^{-2}\  .\end{split}\ee
\end{itemize}

\section{Structure of 1PI effective action} \label{a1}

The dynamics of the combined system of particle and bath obeys unitary time evolution.
Consequently the 1-PI effective action of the Brownian particle has to satisfy the following conditions\cite{Haehl:2016pec, 2017arXiv170102820H,Avinash:2017asn,Chaudhuri:2018ihk,Chaudhuri:2018ymp}.

\begin{enumerate}
\item  \textbf{Collapse Conditions :} \\
The contour-ordered correlator of the particle picks up a sign when one slides an operator insertion  across a turning point of the time contour. This is true, provided as we move  from one leg to another, we return to the  same real time instant and there is no operator insertion in between. The sign change is due to our choice of putting an extra minus sign in the q's on even legs over the usual convention followed in discussions on the Schwinger-Keldysh formalism \cite{ Chou:1984es, 2009AdPhy..58..197K, Haehl:2016pec, 2017arXiv170102820H}.

At the level of the 1-PI effective action this implies that
under identification of the degrees of freedom on any two consecutive legs, the effective action becomes independent of the common degree of freedom of these two legs i.e.
the effective action becomes independent of $\tilde{q}$ under any one of the following identifications  \cite{Haehl:2016pec, 2017arXiv170102820H, Chaudhuri:2018ihk, Chaudhuri:2018ymp}:
\begin{enumerate}
\item \textbf{(1,2) collapse:} $q_1=-q_2=\tilde{q}$
\item \textbf{(2,3) collapse:} $q_2=-q_3=\tilde{q}$ 
\item \textbf{(3,4) collapse:} $q_3=-q_4=\tilde{q}$ .
\end{enumerate}
\label{Collapse rules}
Under any of these collapses, the 1-PI effective action reduces to the Schwinger-Keldysh 1-PI effective action for the degrees of freedom on the remaining two legs. These rules
further impose the following relations  between the couplings \cite{Avinash:2017asn, Chaudhuri:2018ihk}:
\begin{equation}
\begin{split}
& Z_\Delta=\text{Im}[Z],\ m_{\Delta}^2 = \text{Im}[m^2],\ \text{Im}[\lambda_3+3 \sigma_3]=0\ .
\end{split}
\end{equation}
\par
\item \textbf{Reality condition :} \\
A correlation function of hermitian operators gets complex conjugated when the operators are inserted in the reverse order within the correlation function. 
This is assured when the 1PI effective action picks up a negative sign under simultaneous complex conjugation of all the effective couplings and exchange of the degrees of freedom
in the following way:
\begin{equation*}
q_1 \leftrightarrow -q_4\ ,\ q_2 \leftrightarrow -q_3 \ .
\end{equation*}
\label{Reality condition}
This constraint is satisfied when the effective couplings $\widehat{F}$, $\widehat{\gamma}$ and $\kappa_3$ are real.
\end{enumerate}

\section{Contour integrals and Poles  for the Effective Couplings}\label{app:poles}
In this appendix, we will present some more details about the contour integrals that have to be evaluated to obtain the effective couplings in our model. 

For each coupling, we write down the explicit integrals over which double contour integration \footnote{The integral over $\omega_2$ is trivial due to the presence of a delta function coming from energy conservation. So, one has to integrate over the frequencies $\omega_1$ and $\omega_3$.} needs to be performed. In table 8, we tabulate the poles in each integrand whose residues add to give the final coupling as we perform the contour integral first over $\omega_1$ and then over $\omega_3$. We will write our integrands performing an integral over the frequency delta function:
\begin{equation}
\begin{split}
\int_{\mathcal{C}_3^{\prime}}\equiv \int_{\mathcal{C}_3} 2\pi\delta(\omega_1+\omega_2+\omega_3)\ .
\end{split}
\end{equation}

The explicit integrals for the cubic couplings are given by the following expressions :

\be
\begin{split}
\overline{\lambda}_{3}
&=-4\Gamma_3 \int_{\mathcal{C}_3^{\prime}} \frac{(\omega_1-\omega_2)}{\omega_1}\left(1-\frac{\omega_1\omega_2}{\Omega^2}\right)\times \prod_{k=1}^3\frac{ \Omega^2}{\omega_k^2+\Omega^2} 
=-\frac{3\Gamma_3 \Omega^2}{2}\ ,\\
\overline{\kappa}_{3} &=
-4\Gamma_3 \int_{\mathcal{C}_3^{\prime}} \frac{(\omega_3-\omega_2)}{\omega_3}\left(1-\frac{\omega_2\omega_3}{\Omega^2}\right)\times \prod_{k=1}^3\frac{ \Omega^2}{\omega_k^2+\Omega^2} 
=-\frac{3\Gamma_3 \Omega^2}{2}\ ,\\
\overline{\lambda}_{3\gamma}&=
2 i 
\Gamma_3 \int_{\mathcal{C}_3^{\prime}} \frac{(2\omega_3-\omega_1) (\omega_1-\omega_2)}{\omega_1^2 \omega_3}\left(1-\frac{\omega_1\omega_2}{\Omega^2}\right)\times \prod_{k=1}^3\frac{ \Omega^2}{\omega_k^2+\Omega^2} 
=-2\Gamma_3\Omega\ ,\\
\overline{\kappa}_{3\gamma}&=
2 i 
\Gamma_3 \int_{\mathcal{C}_3^{\prime}} \frac{(\omega_3-2\omega_1) (\omega_3-\omega_2)}{\omega_1 \omega_3^2}\left(1-\frac{\omega_2\omega_3}{\Omega^2}\right)\times \prod_{k=1}^3\frac{ \Omega^2}{\omega_k^2+\Omega^2} 
=-2\Gamma_3\Omega\ ,\\
\end{split}
\label{rekappa3gamma}
\ee

\begin{equation}
\begin{split}
\zeta_{\gamma}&=-
\Gamma_3 v^2_{th}\int_{\mathcal{C}_3^{\prime}} \Big[\left(\frac{2} {\omega_1^2}+\frac{3} {\omega_1 \omega_3}\right) (1+\frac{\omega_3^2}{ \Omega^2}) 
-\left(\frac{2} {\omega_3^2}+\frac{3} {\omega_1 \omega_3}\right) (1+\frac{\omega_1^2}{ \Omega^2})\\
& \qquad \qquad \qquad -\frac{3}{\omega_1 \omega_3} \left(3+2 \frac{\omega_3^2+\omega_1 \omega_3+\omega_1^2}{ \Omega^2} \right) \Big]\times \prod_{k=1}^3\frac{ \Omega^2}{\omega_k^2+\Omega^2} = \frac{3}{2}\Gamma_3 v^2_{th} \ ,\\
\widehat{\kappa}_{3\gamma}&=6 \Gamma_3 v^2_{th}\int_{\mathcal{C}_3^{\prime}} \frac{1}{\omega_1 \omega_3} 
\left(3+2 \frac{\omega_3^2+\omega_1 \omega_3+\omega_1^2}{\Omega^2} \right) \times \prod_{k=1}^3\frac{ \Omega^2}{\omega_k^2+\Omega^2} =3\Gamma_3 v^2_{th} \ ,\\
\end{split}
\label{imsigma3ga}
\end{equation}
\begin{equation}
\begin{split}
\zeta_N&=2 \Gamma_3 v^4_{th}\int_{\mathcal{C}_3^{\prime}} \frac{1}{\omega_1 \omega_3} 
\left(3+2 \frac{\omega_3^2+\omega_1 \omega_3+\omega_1^2}{\Omega^2} \right) \times \prod_{k=1}^3\frac{ \Omega^2}{\omega_k^2+\Omega^2} =\Gamma_3 v^4_{th} \ ,\\
\zeta_\mu&=\frac{2 \Gamma_3 v^2_{th}}{i m_p \Omega^2}\int_{\mathcal{C}_3^{\prime}} \frac{(\omega_1-\omega_3)}{ \omega_1 \omega_3} 
\left(2\Omega^2+\omega_1 \omega_3+2 (\omega_1^2+\omega_1 \omega_3+\omega_3^2)\right) \times \prod_{k=1}^3\frac{ \Omega^2}{\omega_k^2+\Omega^2} \\
&=-2\Gamma_3 \Omega v^2_{th} \ .
\end{split}
\label{zetamu}
\end{equation}

\begin{table}[ht]
\begin{center}
\begin{tabular}{|p{0.5cm}|p{3cm}|p{2cm}||p{0.5cm}|p{3cm}|p{2cm}|}
    \hline
     & \textbf{Poles in $\omega_1$} & \textbf{Poles in $\omega_3$}& & \textbf{Poles in $\omega_1$} & \textbf{Poles in $\omega_3$} \\
     \hline
    $\bar\kappa_3$ & $i\Omega+i\epsilon_1$ & $i\Omega-i\epsilon_2$ &$\overline\lambda_3$ & $i\Omega+i\epsilon_1$ & $i\Omega-i\epsilon_2$ \\
    \hline   
    & $-\omega_3+i\Omega+i\epsilon_1-i\epsilon_2$ & $2i\Omega-i\epsilon_2$ & & $-\omega_3+i\Omega+i\epsilon_1-i\epsilon_2$ & $i\Omega-i\epsilon_2$  \\
    \hline
    \hline 
    $\overline{\kappa}_{3\gamma}$& $i\epsilon_1$ & $i\Omega-i\epsilon_2$ &  $\widehat{\kappa}_{3\gamma} $ & $i\epsilon_1$ & $i\Omega-i\epsilon_2$  \\ 
\hline
 & $i\Omega+i\epsilon_1$ & $i\Omega-i\epsilon_2$ & & $i\Omega+i\epsilon_1$ & $i\Omega-i\epsilon_2$    \\
 \hline
& $-\omega_3+i\Omega+i\epsilon_1-i\epsilon_2$ & $i\Omega-i\epsilon_2$  & & $-\omega_3+i\Omega+i\epsilon_1-i\epsilon_2$ & $i\Omega-i\epsilon_2$  \\
 \hline
 & $-\omega_3+i\Omega+i\epsilon_1-i\epsilon_2$ & $2i\Omega-i\epsilon_2$   & & $-\omega_3+i\Omega+i\epsilon_1-i\epsilon_2$ & $2i\Omega-i\epsilon_2$   \\
 \hline
 \hline
 $\overline{\lambda}_{3\gamma}$& $i\epsilon_1$ & $i\Omega-i\epsilon_2$ &  $\zeta_{\mu}$ & $i\epsilon_1$ & $i\Omega-i\epsilon_2$ \\
 \hline
  & $i\Omega+i\epsilon_1$ & $i\Omega-i\epsilon_2$ & & $-\omega_3+i\Omega+i\epsilon_1-i\epsilon_2$ & $i\Omega-i\epsilon_2$  \\
  \hline
   & $-\omega_3+i\Omega+i\epsilon_1-i\epsilon_2$ & $i\Omega-i\epsilon_2$ & & $-\omega_3+i\Omega+i\epsilon_1-i\epsilon_2$ & $2i\Omega-i\epsilon_2$  \\
  \hline
  \hline
   $\zeta_{N}$ & $i\epsilon_1$ & $i\Omega-i\epsilon_2$ &  $\zeta_{\gamma}$& $i\epsilon_1$ & $i\Omega-i\epsilon_2$ \\
  \hline
  & $i\Omega+i\epsilon_1$ & $i\Omega-i\epsilon_2$ & & $i\Omega+i\epsilon_1$ & $i\Omega-i\epsilon_2$  \\
  \hline
  & $-\omega_3+i\Omega+i\epsilon_1-i\epsilon_2$ & $i\Omega-i\epsilon_2$ &  & $-\omega_3+i\Omega+i\epsilon_1-i\epsilon_2$ & $i\Omega-i\epsilon_2$  \\
  \hline
  & $-\omega_3+i\Omega+i\epsilon_1-i\epsilon_2$ & $2i\Omega-i\epsilon_2$  & &  $-\omega_3+i\Omega+i\epsilon_1-i\epsilon_2$ & $2i\Omega-i\epsilon_2$  \\
  \hline
\end{tabular}
\end{center}
\caption{Poles  for determining cubic couplings}
\label{tabmulticol}
\end{table}

\end{appendix}

\vspace{4cm}

\bibliographystyle{JHEP}
\bibliography{NonLinLangevinRefs}
\end{document}

%% file: oto-macros.tex
\definecolor{rust}{rgb}{0.8,0.2,0.2}

\newcommand{\be}{\begin{equation}}
\newcommand{\ee}{\end{equation}}
\newcommand{\ba}{\begin{eqnarray}}
\newcommand{\ea}{\end{eqnarray}}
\newcommand{\bi}{\begin{itemize}}
\newcommand{\ei}{\end{itemize}}
\newcommand{\bse}{\begin{subequations}}
\newcommand{\ese}{\end{subequations}}

\newcommand{\bose}{\mathfrak{f}}









%% file: OTOQuantumDissv4.bbl
\providecommand{\href}[2]{#2}\begingroup\raggedright\begin{thebibliography}{10}

\bibitem{Schwinger:1960qe}
J.~S. Schwinger, {\it {Brownian motion of a quantum oscillator}},  {\em J.
  Math. Phys.} {\bf 2} (1961) 407--432.

\bibitem{Feynman:1963fq}
R.~P. Feynman and F.~L. Vernon, Jr., {\it {The Theory of a general quantum
  system interacting with a linear dissipative system}},  {\em Annals Phys.}
  {\bf 24} (1963) 118--173. [,257(1963)].

\bibitem{Caldeira:1982iu}
A.~O. Caldeira and A.~J. Leggett, {\it {Path integral approach to quantum
  Brownian motion}},  {\em Physica} {\bf 121A} (1983) 587--616.

\bibitem{PhysRevD.45.2843}
B.~L. Hu, J.~P. Paz, and Y.~Zhang, {\it Quantum brownian motion in a general
  environment: Exact master equation with nonlocal dissipation and colored
  noise},  {\em Phys. Rev. D} {\bf 45} (Apr, 1992) 2843--2861.

\bibitem{PhysRevD.47.1576}
B.~L. Hu, J.~P. Paz, and Y.~Zhang, {\it Quantum brownian motion in a general
  environment. ii. nonlinear coupling and perturbative approach},  {\em Phys.
  Rev. D} {\bf 47} (Feb, 1993) 1576--1594.

\bibitem{BRE02}
H.~P. Breuer and F.~Petruccione, {\em The theory of open quantum systems}.
\newblock Oxford University Press, Great Clarendon Street, 2002.

\bibitem{weiss2012quantum}
U.~Weiss, {\em Quantum Dissipative Systems}.
\newblock Series in modern condensed matter physics. World Scientific, 2012.

\bibitem{schlosshauer2007decoherence}
M.~Schlosshauer, {\em Decoherence: And the Quantum-To-Classical Transition}.
\newblock The Frontiers Collection. Springer, 2007.

\bibitem{Chaudhuri:2018ihk}
S.~Chaudhuri and R.~Loganayagam, {\it {Probing Out-of-Time-Order Correlators}},
   {\em JHEP} {\bf 07} (2019) 006, [\href{http://arxiv.org/abs/1807.09731}{{\tt
  arXiv:1807.09731}}].

\bibitem{2017arXiv170102820H}
F.~M. {Haehl}, R.~{Loganayagam}, P.~{Narayan}, and M.~{Rangamani}, {\it
  {Classification of out-of-time-order correlators}},  {\em ArXiv e-prints}
  (Jan., 2017) [\href{http://arxiv.org/abs/1701.02820}{{\tt
  arXiv:1701.02820}}].

\bibitem{Haehl:2017eob}
F.~M. Haehl, R.~Loganayagam, P.~Narayan, A.~A. Nizami, and M.~Rangamani, {\it
  {Thermal out-of-time-order correlators, KMS relations, and spectral
  functions}},  {\em JHEP} {\bf 12} (2017) 154,
  [\href{http://arxiv.org/abs/1706.08956}{{\tt arXiv:1706.08956}}].

\bibitem{Chaudhuri:2018ymp}
S.~Chaudhuri, C.~Chowdhury, and R.~Loganayagam, {\it {Spectral Representation
  of Thermal OTO Correlators}},  \href{http://arxiv.org/abs/1810.03118}{{\tt
  arXiv:1810.03118}}.

\bibitem{PhysRev.32.97}
J.~B. Johnson, {\it Thermal agitation of electricity in conductors},  {\em
  Phys. Rev.} {\bf 32} (Jul, 1928) 97--109.

\bibitem{PhysRev.32.110}
H.~Nyquist, {\it Thermal agitation of electric charge in conductors},  {\em
  Phys. Rev.} {\bf 32} (Jul, 1928) 110--113.

\bibitem{PhysRev.83.34}
H.~B. Callen and T.~A. Welton, {\it Irreversibility and generalized noise},
  {\em Phys. Rev.} {\bf 83} (Jul, 1951) 34--40.

\bibitem{BERNARD:1959zz}
W.~Bernard and H.~B. Callen, {\it {Irreversible Thermodynamics of Nonlinear
  Processes and Noise in Driven Systems}},  {\em Rev. Mod. Phys.} {\bf 31}
  (1959) 1017--1044.

\bibitem{Kubo:1957mj}
R.~Kubo, {\it {Statistical mechanical theory of irreversible processes. 1.
  General theory and simple applications in magnetic and conduction problems}},
   {\em J. Phys. Soc. Jap.} {\bf 12} (1957) 570--586.

\bibitem{Martin:1959jp}
P.~C. Martin and J.~S. Schwinger, {\it {Theory of many particle systems. 1.}},
  {\em Phys. Rev.} {\bf 115} (1959) 1342--1373. [,427(1959)].

\bibitem{1973PhRvA...8..423M}
P.~C. {Martin}, E.~D. {Siggia}, and H.~A. {Rose}, {\it {Statistical Dynamics of
  Classical Systems}},  {\em pra} {\bf 8} (July, 1973) 423--437.

\bibitem{1978PhRvB..18..353D}
C.~{de Dominicis} and L.~{Peliti}, {\it {Field-theory renormalization and
  critical dynamics above T$_{c}$: Helium, antiferromagnets, and liquid-gas
  systems}},  {\em prb} {\bf 18} (July, 1978) 353--376.

\bibitem{1976ZPhyB..23..377J}
H.-K. {Janssen}, {\it {On a Lagrangean for classical field dynamics and
  renormalization group calculations of dynamical critical properties}},  {\em
  Zeitschrift fur Physik B Condensed Matter} {\bf 23} (Dec., 1976) 377--380.

\bibitem{2017JPhA...50c3001H}
J.~A. {Hertz}, Y.~{Roudi}, and P.~{Sollich}, {\it {Path integral methods for
  the dynamics of stochastic and disordered systems}},  {\em Journal of Physics
  A Mathematical General} {\bf 50} (Jan., 2017) 033001,
  [\href{http://arxiv.org/abs/1604.05775}{{\tt arXiv:1604.05775}}].

\bibitem{2011PhRvL.106w3601M}
K.~H. {Madsen}, S.~{Ates}, T.~{Lund-Hansen}, A.~{L{\"o}ffler},
  S.~{Reitzenstein}, A.~{Forchel}, and P.~{Lodahl}, {\it {Observation of
  Non-Markovian Dynamics of a Single Quantum Dot in a Micropillar Cavity}},
  {\em Physical Review Letters} {\bf 106} (June, 2011) 233601,
  [\href{http://arxiv.org/abs/1012.0740}{{\tt arXiv:1012.0740}}].

\bibitem{2013arXiv1305.6942G}
S.~{Groeblacher}, A.~{Trubarov}, N.~{Prigge}, G.~D. {Cole}, M.~{Aspelmeyer},
  and J.~{Eisert}, {\it {Observation of non-Markovian micro-mechanical Brownian
  motion}},  {\em ArXiv e-prints} (May, 2013)
  [\href{http://arxiv.org/abs/1305.6942}{{\tt arXiv:1305.6942}}].

\bibitem{2017Sci...355..156M}
X.~{Mi}, J.~V. {Cady}, D.~M. {Zajac}, P.~W. {Deelman}, and J.~R. {Petta}, {\it
  {Strong coupling of a single electron in silicon to a microwave photon}},
  {\em Science} {\bf 355} (Jan., 2017) 156--158,
  [\href{http://arxiv.org/abs/1703.03047}{{\tt arXiv:1703.03047}}].

\bibitem{2018NatCo...9..904P}
A.~{Poto{\v c}nik}, A.~{Bargerbos}, F.~A.~Y.~N. {Schr{\"o}der}, S.~A. {Khan},
  M.~C. {Collodo}, S.~{Gasparinetti}, Y.~{Salath{\'e}}, C.~{Creatore},
  C.~{Eichler}, H.~E. {T{\"u}reci}, A.~W. {Chin}, and A.~{Wallraff}, {\it
  {Studying light-harvesting models with superconducting circuits}},  {\em
  Nature Communications} {\bf 9} (Mar., 2018) 904,
  [\href{http://arxiv.org/abs/1710.07466}{{\tt arXiv:1710.07466}}].

\bibitem{stratonovich2012nonlinear}
R.~L. Stratonovich, {\em Nonlinear nonequilibrium thermodynamics I: linear and
  nonlinear fluctuation-dissipation theorems}, vol.~57.
\newblock Springer Science \& Business Media, 2012.

\bibitem{stratonovich2013nonlinear}
R.~Stratonovich, {\em Nonlinear Nonequilibrium Thermodynamics II: Advanced
  Theory}.
\newblock Springer Series in Synergetics. Springer Berlin Heidelberg, 2013.

\bibitem{PhysRev.37.405}
L.~Onsager, {\it Reciprocal relations in irreversible processes. 1.},  {\em
  Phys. Rev.} {\bf 37} (Feb, 1931) 405--426.

\bibitem{PhysRev.38.2265}
L.~Onsager, {\it Reciprocal relations in irreversible processes. 2.},  {\em
  Phys. Rev.} {\bf 38} (Dec, 1931) 2265--2279.

\bibitem{RevModPhys.17.343}
H.~B.~G. Casimir, {\it On onsager's principle of microscopic reversibility},
  {\em Rev. Mod. Phys.} {\bf 17} (Apr, 1945) 343--350.

\bibitem{2018PhRvB..97j4306C}
A.~{Chakraborty} and R.~{Sensarma}, {\it {Power-law tails and non-Markovian
  dynamics in open quantum systems: An exact solution from Keldysh field
  theory}},  {\em prb} {\bf 97} (Mar., 2018) 104306,
  [\href{http://arxiv.org/abs/1709.04472}{{\tt arXiv:1709.04472}}].

\bibitem{1971JSP.....3..245B}
M.~{Bixon} and R.~{Zwanzig}, {\it {Brownian motion of a nonlinear oscillator}},
   {\em Journal of Statistical Physics} {\bf 3} (Sept., 1971) 245--260.

\bibitem{1973JSP.....9..215Z}
R.~{Zwanzig}, {\it {Nonlinear generalized Langevin equations}},  {\em Journal
  of Statistical Physics} {\bf 9} (Nov., 1973) 215--220.

\bibitem{1979PhLA...69..313S}
A.~{Schenzle} and H.~{Brand}, {\it {Multiplicative stochastic processes in
  statistical physics}},  {\em Physics Letters A} {\bf 69} (Jan., 1979)
  313--315.

\bibitem{Brun:1993qj}
T.~A. Brun, {\it {Quasiclassical equations of motion for nonlinear Brownian
  systems}},  {\em Phys. Rev.} {\bf D47} (1993) 3383--3393,
  [\href{http://arxiv.org/abs/gr-qc/9306013}{{\tt gr-qc/9306013}}].

\bibitem{efremov1969fluctuation}
G.~Efremov, {\it A fluctuation dissipation theorem for nonlinear media},  {\em
  Sov. Phys. JETP} {\bf 28} (1969) 1232.

\bibitem{stratonovich1970izv}
R.~Stratonovich, {\it Thermal noise in nonlinear resistors},  {\em VUZ
  Radiofizika} {\bf 13} (1970) 1512--1522.

\bibitem{stratonovich1970contribution}
R.~Stratonovich, {\it Contribution to the quantum nonlinear theory of thermal
  fluctuations},  {\em SOVIET PHYSICS JETP} {\bf 31} (1970), no.~5.

\bibitem{efremov1972gf}
G.~Efremov, {\it Contribution to theory of heat fluctuations in the
  nonequilibrium media},  {\em Izv. Vuzov (Radiofizika)} {\bf 15} (1972) 1207.

\bibitem{bochkov1977general}
G.~Bochkov and Y.~E. Kuzovlev, {\it General theory of thermal fluctuations in
  nonlinear systems},  {\em Zh. Eksp. Teor. Fiz} {\bf 72} (1977) 238--243.

\bibitem{sitenko1978sov}
A.~Sitenko, {\it The fluctuation-dissipation relation in nonlinear
  electrodynamics},  {\em Sov. Phys. JETP} {\bf 48} (1978) 51.

\bibitem{gordon1978fluctuation}
J.~Gordon, {\it The fluctuation-dissipation relation in nonlinear
  electrodynamics},  {\em Zh. Eksp. Teor. Fiz} {\bf 75} (1978) 104--115.

\bibitem{klyshko2018photons}
D.~Klyshko, {\em Photons Nonlinear Optics}.
\newblock CRC Press, 2018.

\bibitem{Chou:1984es}
K.-c. Chou, Z.-b. Su, B.-l. Hao, and L.~Yu, {\it {Equilibrium and
  Nonequilibrium Formalisms Made Unified}},  {\em Phys. Rept.} {\bf 118} (1985)
  1--131.

\bibitem{Haag:1967sg}
R.~Haag, N.~M. Hugenholtz, and M.~Winnink, {\it {On the Equilibrium states in
  quantum statistical mechanics}},  {\em Commun. Math. Phys.} {\bf 5} (1967)
  215--236.

\bibitem{Hou:1998yc}
D.-f. Hou, E.~Wang, and U.~W. Heinz, {\it {n point functions at finite
  temperature}},  {\em J. Phys.} {\bf G24} (1998) 1861--1868,
  [\href{http://arxiv.org/abs/hep-th/9807118}{{\tt hep-th/9807118}}].

\bibitem{Wang:1998wg}
E.~Wang and U.~W. Heinz, {\it {A Generalized fluctuation dissipation theorem
  for nonlinear response functions}},  {\em Phys. Rev.} {\bf D66} (2002)
  025008, [\href{http://arxiv.org/abs/hep-th/9809016}{{\tt hep-th/9809016}}].

\bibitem{Weldon:2005nr}
H.~A. Weldon, {\it {Two sum rules for the thermal n-point functions}},  {\em
  Phys. Rev.} {\bf D72} (2005) 117901.

\bibitem{Evans:1991ky}
T.~S. Evans, {\it {N point finite temperature expectation values at real
  times}},  {\em Nucl. Phys.} {\bf B374} (1992) 340--370.

\bibitem{Guerin:1993ik}
F.~Guerin, {\it {Retarded - advanced N point Green functions in thermal field
  theories}},  {\em Nucl. Phys.} {\bf B432} (1994) 281--314,
  [\href{http://arxiv.org/abs/hep-ph/9306210}{{\tt hep-ph/9306210}}].

\bibitem{Tsuji:2016kep}
N.~Tsuji, T.~Shitara, and M.~Ueda, {\it {Out-of-time-order
  fluctuation-dissipation theorem}},  {\em Phys. Rev.} {\bf E97} (2018), no.~1
  012101, [\href{http://arxiv.org/abs/1612.08781}{{\tt arXiv:1612.08781}}].

\bibitem{Avinash:2017asn}
A.~Baidya, C.~Jana, R.~Loganayagam, and A.~Rudra, {\it {Renormalization in open
  quantum field theory. Part I. Scalar field theory}},  {\em JHEP} {\bf 11}
  (2017) 204, [\href{http://arxiv.org/abs/1704.08335}{{\tt arXiv:1704.08335}}].

\bibitem{kamenev_2011}
A.~Kamenev, {\em Field Theory of Non-Equilibrium Systems}.
\newblock Cambridge University Press, 2011.

\bibitem{SC}
S.~Chaudhuri and R.~Loganayagam, {\it {Simplifying OTO Diagrammatics}},  {\em
  to appear}.

\bibitem{weinberg1995quantum}
S.~Weinberg, {\em The Quantum Theory of Fields}.
\newblock Quantum Theory of Fields. Cambridge University Press, 1995.

\bibitem{Chakrabarty:2019qcp}
B.~Chakrabarty and S.~Chaudhuri, {\it {Out of time ordered effective dynamics
  of a quartic oscillator}},  \href{http://arxiv.org/abs/1905.08307}{{\tt
  arXiv:1905.08307}}.

\bibitem{klyshko1988photons}
D.~Klyshko, {\em Photons Nonlinear Optics}.
\newblock Taylor \& Francis, 1988.

\bibitem{kippenberg2007cavity}
T.~J. Kippenberg and K.~J. Vahala, {\it Cavity opto-mechanics},  {\em Optics
  express} {\bf 15} (2007), no.~25 17172--17205.

\bibitem{PhysRevA.45.5056}
A.~Joshi and R.~R. Puri, {\it Dynamical evolution of the two-photon
  jaynes-cummings model in a kerr-like medium},  {\em Phys. Rev. A} {\bf 45}
  (Apr, 1992) 5056--5060.

\bibitem{PhysRevA.45.6816}
P.~G\'ora and C.~Jedrzejek, {\it Nonlinear jaynes-cummings model},  {\em Phys.
  Rev. A} {\bf 45} (May, 1992) 6816--6828.

\bibitem{PhysRevA.52.4214}
W.~Vogel and R.~L. d.~M. Filho, {\it Nonlinear jaynes-cummings dynamics of a
  trapped ion},  {\em Phys. Rev. A} {\bf 52} (Nov, 1995) 4214--4217.

\bibitem{Liu2018}
X.-J. Liu, J.-B. Lu, S.-Q. Zhang, J.-P. Liu, H.~Li, Y.~Liang, J.~Ma, Y.-J.
  Weng, Q.-R. Zhang, H.~Liu, X.-R. Zhang, and X.-Y. Wu, {\it The nonlinear
  jaynes-cummings model for the multiphoton transition},  {\em International
  Journal of Theoretical Physics} {\bf 57} (Jan, 2018) 290--298.

\bibitem{2017NatPh..13..781G}
M.~{G{\"a}rttner}, J.~G. {Bohnet}, A.~{Safavi-Naini}, M.~L. {Wall}, J.~J.
  {Bollinger}, and A.~M. {Rey}, {\it {Measuring out-of-time-order correlations
  and multiple quantum spectra in a trapped-ion quantum magnet}},  {\em Nature
  Physics} {\bf 13} (Aug., 2017) 781--786,
  [\href{http://arxiv.org/abs/1608.08938}{{\tt arXiv:1608.08938}}].

\bibitem{PhysRevX.7.031011}
J.~Li, R.~Fan, H.~Wang, B.~Ye, B.~Zeng, H.~Zhai, X.~Peng, and J.~Du, {\it
  Measuring out-of-time-order correlators on a nuclear magnetic resonance
  quantum simulator},  {\em Phys. Rev. X} {\bf 7} (Jul, 2017) 031011.

\bibitem{Zhu:2016uws}
G.~Zhu, M.~Hafezi, and T.~Grover, {\it {Measurement of many-body chaos using a
  quantum clock}},  {\em Phys. Rev.} {\bf A94} (2016), no.~6 062329,
  [\href{http://arxiv.org/abs/1607.00079}{{\tt arXiv:1607.00079}}].

\bibitem{Swingle:2016var}
B.~Swingle, G.~Bentsen, M.~Schleier-Smith, and P.~Hayden, {\it {Measuring the
  scrambling of quantum information}},  {\em Phys. Rev.} {\bf A94} (2016),
  no.~4 040302, [\href{http://arxiv.org/abs/1602.06271}{{\tt
  arXiv:1602.06271}}].

\bibitem{2016arXiv160701801Y}
N.~Y. {Yao}, F.~{Grusdt}, B.~{Swingle}, M.~D. {Lukin}, D.~M. {Stamper-Kurn},
  J.~E. {Moore}, and E.~A. {Demler}, {\it {Interferometric Approach to Probing
  Fast Scrambling}},  {\em ArXiv e-prints} (July, 2016)
  [\href{http://arxiv.org/abs/1607.01801}{{\tt arXiv:1607.01801}}].

\bibitem{2017PhRvL.119d0501G}
L.~{Garc{\'{\i}}a-{\'A}lvarez}, I.~L. {Egusquiza}, L.~{Lamata}, A.~{del Campo},
  J.~{Sonner}, and E.~{Solano}, {\it {Digital Quantum Simulation of Minimal AdS
  /CFT}},  {\em Physical Review Letters} {\bf 119} (July, 2017) 040501,
  [\href{http://arxiv.org/abs/1607.08560}{{\tt arXiv:1607.08560}}].

\bibitem{2015NatSR...514873B}
F.~{Barra}, {\it {The thermodynamic cost of driving quantum systems by their
  boundaries}},  {\em Scientific Reports} {\bf 5} (Oct., 2015) 14873,
  [\href{http://arxiv.org/abs/1509.04223}{{\tt arXiv:1509.04223}}].

\bibitem{PhysRevA.94.012341}
G.~Sadiek and S.~Almalki, {\it Entanglement dynamics in heisenberg spin chains
  coupled to a dissipative environment at finite temperature},  {\em Phys. Rev.
  A} {\bf 94} (Jul, 2016) 012341.

\bibitem{Ilievski:2014gra}
E.~Ilievski, {\em {Exact solutions of open integrable quantum spin chains}}.
\newblock PhD thesis, Ljubljana U., 2014.
\newblock \href{http://arxiv.org/abs/1410.1446}{{\tt arXiv:1410.1446}}.

\bibitem{PhysRevLett.121.024101}
A.~Das, S.~Chakrabarty, A.~Dhar, A.~Kundu, D.~A. Huse, R.~Moessner, S.~S. Ray,
  and S.~Bhattacharjee, {\it Light-cone spreading of perturbations and the
  butterfly effect in a classical spin chain},  {\em Phys. Rev. Lett.} {\bf
  121} (Jul, 2018) 024101.

\bibitem{Scaffidi:2017ghs}
T.~Scaffidi and E.~Altman, {\it {Semiclassical Theory of Many-Body Quantum
  Chaos and its Bound}},  \href{http://arxiv.org/abs/1711.04768}{{\tt
  arXiv:1711.04768}}.

\bibitem{Khemani:2018sdn}
V.~Khemani, D.~A. Huse, and A.~Nahum, {\it {Velocity-dependent Lyapunov
  exponents in many-body quantum, semiclassical, and classical chaos}},  {\em
  Phys. Rev.} {\bf B98} (2018), no.~14 144304,
  [\href{http://arxiv.org/abs/1803.05902}{{\tt arXiv:1803.05902}}].

\bibitem{PhysRevLett.118.086801}
E.~B. Rozenbaum, S.~Ganeshan, and V.~Galitski, {\it Lyapunov exponent and
  out-of-time-ordered correlator's growth rate in a chaotic system},  {\em
  Phys. Rev. Lett.} {\bf 118} (Feb, 2017) 086801.

\bibitem{Herzog:2002pc}
C.~P. Herzog and D.~T. Son, {\it {Schwinger-Keldysh propagators from AdS/CFT
  correspondence}},  {\em JHEP} {\bf 03} (2003) 046,
  [\href{http://arxiv.org/abs/hep-th/0212072}{{\tt hep-th/0212072}}].

\bibitem{Skenderis:2008dh}
K.~Skenderis and B.~C. van Rees, {\it {Real-time gauge/gravity duality}},  {\em
  Phys. Rev. Lett.} {\bf 101} (2008) 081601,
  [\href{http://arxiv.org/abs/0805.0150}{{\tt arXiv:0805.0150}}].

\bibitem{deBoer:2017xdk}
J.~de~Boer, E.~Llabrés, J.~F. Pedraza, and D.~Vegh, {\it {Chaotic strings in
  AdS/CFT}},  {\em Phys. Rev. Lett.} {\bf 120} (2018), no.~20 201604,
  [\href{http://arxiv.org/abs/1709.01052}{{\tt arXiv:1709.01052}}].

\bibitem{Kobes:1986za}
R.~L. Kobes and G.~W. Semenoff, {\it {Discontinuities of Green Functions in
  Field Theory at Finite Temperature and Density. 2}},  {\em Nucl. Phys.} {\bf
  B272} (1986) 329--364.

\bibitem{Kobes:1990ua}
R.~Kobes, {\it {Retarded functions, dispersion relations, and Cutkosky rules at
  zero and finite temperature}},  {\em Phys. Rev.} {\bf D43} (1991) 1269--1282.

\bibitem{Haehl:2016pec}
F.~M. Haehl, R.~Loganayagam, and M.~Rangamani, {\it {Schwinger-Keldysh
  formalism. Part I: BRST symmetries and superspace}},  {\em JHEP} {\bf 06}
  (2017) 069, [\href{http://arxiv.org/abs/1610.01940}{{\tt arXiv:1610.01940}}].

\bibitem{2009AdPhy..58..197K}
A.~{Kamenev} and A.~{Levchenko}, {\it {Keldysh technique and non-linear
  {\ensuremath{\sigma}}-model: basic principles and applications}},  {\em
  Advances in Physics} {\bf 58} (May, 2009) 197--319,
  [\href{http://arxiv.org/abs/0901.3586}{{\tt arXiv:0901.3586}}].

\end{thebibliography}\endgroup
